\documentclass[%
 reprint,
superscriptaddress,
nofootinbib,
 amsmath,amssymb,
 aps,onecolumn
]{revtex4-2}

\usepackage{color}
\usepackage{graphicx}
\usepackage{stmaryrd}

\usepackage{dcolumn}
\usepackage{bm}
\usepackage[normalem]{ulem}
\usepackage{amsmath}

\providecommand\mnras{Monthly Notices of the Royal Astronomical Society}
\providecommand\apjl{The Astrophysical Journal Letters}
\providecommand\baas{Bulletin of the American Astronomical Society}
\providecommand\physrep{Physics Reports}
\providecommand\jcap{Journal of Cosmology and Astroparticle Physics}
\providecommand\sovast{Soviet Astronomy}

\newcommand{\tj}[6]{ \begin{pmatrix}
  #1 & #2 & #3 \\
  #4 & #5 & #6 
 \end{pmatrix}}

\newcommand{\cmnt}[1]{}

\begin{document}
\title{Corrections to Hawking radiation from asteroid-mass primordial black holes: Numerical evaluation of dissipative effects}
\date{\today}

\author{Emily Koivu}
\email{koivu.1@osu.edu}
\affiliation{%
 Center for Cosmology and Astroparticle Physics, The Ohio State University,
 191 West Woodruff Avenue, Columbus, Ohio 43210, USA
}
\affiliation{%
 Department of Physics, The Ohio State University,
 191 West Woodruff Avenue, Columbus, Ohio 43210, USA
}

\author{John Kushan}
\email{john.kushan@case.edu}
\affiliation{
Case Western Reserve University,
Rockefeller Building,
2076 Adelbert Road,
Cleveland, Ohio 44106, USA}
\author{Makana Silva}
\email{makanas@lanl.gov}
\affiliation{%
 Computational Physics and Methods Group (CCS-2), Los Alamos National Laboratory, Los Alamos, New Mexico 87544, USA
}
\author{Gabriel Vasquez}
\email{vasquez.119@osu.edu}
\affiliation{%
 Center for Cosmology and Astroparticle Physics, The Ohio State University,
 191 West Woodruff Avenue, Columbus, Ohio 43210, USA
}
\affiliation{%
 Department of Physics, The Ohio State University,
 191 West Woodruff Avenue, Columbus, Ohio 43210, USA
}
\author{Arijit Das}
\email{das.241@osu.edu}
\affiliation{%
 Center for Cosmology and Astroparticle Physics, The Ohio State University,
 191 West Woodruff Avenue, Columbus, Ohio 43210, USA
}
\affiliation{%
 Department of Physics, The Ohio State University,
 191 West Woodruff Avenue, Columbus, Ohio 43210, USA
}
\author{Christopher M. Hirata}
\email{hirata.10@osu.edu}
\affiliation{%
 Center for Cosmology and Astroparticle Physics, The Ohio State University,
 191 West Woodruff Avenue, Columbus, Ohio 43210, USA
}
\affiliation{%
 Department of Physics, The Ohio State University,
 191 West Woodruff Avenue, Columbus, Ohio 43210, USA
}
\affiliation{%
 Department of Astronomy, The Ohio State University,
 140 West 18th Avenue, Columbus, Ohio 43210, USA
}

\begin{abstract}
    Primordial black holes (PBHs) are theorized objects that may make up some --- or all --- of the dark matter in the universe. At the lowest allowed masses, Hawking radiation (in the form of photons or electrons and positrons) is the primary tool to search for PBHs. This paper is part of an ongoing series in which we aim to calculate the ${\cal O}(\alpha)$ corrections to Hawking radiation from asteroid-mass primordial black holes, based on a perturbative quantum electrodymanics (QED) calculation on Schwarzschild background.
    Silva \textit{et.~al.}\ [{\slshape Phys. Rev. D}, {\bfseries 107}:045004 (2023)] divided the corrections into dissipative and conservative parts; this work focuses on the numerical computation of the dissipative ${\cal O}(\alpha)$ corrections to the photon spectrum. We generate spectra for primordial black holes of mass $M=1$--$8 \times 10^{21} m_{\rm planck}$. This calculation confirms the expectation that at low energies, the inner bremsstrahlung radiation is the dominant contribution to the Hawking radiation spectrum. At high energies, the main ${\cal O}(\alpha)$ effect is a suppression of the photon spectrum due to pair production (emitted $\gamma\rightarrow e^+e^-$), but this is small compared to the overall spectrum.
    We compare the low-energy tail in our curved spacetime QED calculation to several approximation schemes in the literature, and find deviations that could have important implications for constraints from Hawking radiation on primordial black holes as dark matter.
\end{abstract}
\maketitle

\section{Introduction}
\label{sec:intro}

Primordial Black Holes (PBHs) \cite{1967SvA....10..602Z, 1971MNRAS.152...75H} in the asteroid mass range ($10^{17}$---$10^{23}$ g) are of current great interest as a dark matter candidate \cite{2019BAAS...51c..51K, 2021JPhG...48d3001G, 2021RPPh...84k6902C, 2023PDU....4101231B, 2024PhR..1054....1C, 2024NuPhB100316494G}. In this range, PBHs could account for some or all of the dark matter in our Universe \cite{green2020}, without invoking a new long-lived elementary particle. This ``asteroid mass window'' is bounded at the high-mass end by gravitational microlensing surveys \cite{2014ApJ...786..158G, 2019NatAs...3..524N, 2020PhRvD.101f3005S} (which provides a higher event rate for {\em higher} PBH masses). It is bounded at the low-mass end by Hawking radiation \cite{1975CMaPh..43..199H} constraints from gamma rays \cite{2016PhRvD..94d4029C}, positrons \cite{2019PhRvL.122d1104B}, and positron annihilation gamma rays \cite{2019PhRvL.123y1102D, 2019PhRvL.123y1101L} (which is a stronger signal for {\em lower} PBH masses). There is ongoing work on signatures of asteroid-mass PBHs, coming from stellar capture \cite{2009arXiv0901.1093R, 2020PhRvD.102h3004G, 2022MNRAS.517...28O, 2023PhRvD.107j3052E, 2024arXiv240603114T}, X-ray microlensing \cite{2019PhRvD..99l3019B, 2024arXiv240520365T}, and picolensing \cite{1995ApJ...452L.111N, 1999ApJ...512L..13M}. While difficult to detect from much larger black holes due to the inverse relationship between black hole mass and Hawking temperature ($T_{\rm H}\propto 1/M)$, Hawking radiation from sufficiently low-mass PBHs is intense (the emitted power per unit cosmological volume scales as $\propto 1/M^3$) and may be detectable in the gamma-ray regime \cite{2021PhRvD.104b3516R, 2023PrPNP.13104040A}.

This paper is part of a series of papers dedicated to the careful perturbative calculation of the Hawking radiation spectrum, including the effects of interacting particles on curved spacetime. While the free-particle emission spectrum is well-established \cite{1976PhRvD..13..198P, 1977PhRvD..16.2402P}, at sufficiently high temperatures charged leptons and hadrons can be produced, and the secondary particles from their decays and interactions are included in modern codes \cite{2019EPJC...79..693A, 2021PhRvD.103j4010A, 2021EPJC...81..910A}. The temperature of a non-rotating (Schwarzschild) black hole of mass $M$ is related to the electron mass $\mu$ via
\begin{equation}
\frac{T_{\rm H}}\mu = \frac {2.07\times 10^{16}\,\rm g}M.
\end{equation}
(We use natural units, $G=c=\hbar=k_{\rm B}=\epsilon_0=1$, to interconvert temperatures and masses.)
Thus at masses of order $10^{17}$\,g or less --- i.e., near the current lower bound of the asteroid-mass window --- electrons and positrons can be produced and thus Quantum Electrodymanics (QED) effects could be important.
Previous studies have mostly considered the generation of secondary particles using flat spacetime arguments derived from nuclear and particle physics \cite{1936Phy.....3..425K, 1977NuPhB.126..298A, 2020JCAP...01..056C}. They have revealed at least one interacting-particle QED effect that is relevant: the ``inner bremsstrahlung'' or final-state radiation from a charged particle such as an electron emitted from the black hole \cite{2008PhRvD..78f4044P, coogan2021}. At low photon frequencies $\omega$ ($\omega\lesssim T_{\rm H}$, so below $\sim 100$ keV for black holes in the $\sim 10^{17}$\,g mass range), the inner bremsstrahlung can even dominate over the primary particle emission. This tail may be relevant for X-ray constraints on PBHs \cite{2020PhLB..80835624B}.

The existence of inner bremsstrahlung, combined with the use of Hawking radiation for PBH constraints, motivates us to do the complete calculation of Hawking radiation, taking into account the fact that the ``particles'' emitted are interacting on a curved spacetime background as they climb away from the hole. The inner bremsstrahlung is a correction of order ${\cal O}(\alpha)$, where $\alpha = e^2/4\pi\approx 1/137$ is the fine structure constant, which may exceed the primary spectrum when all other numerical factors are included. As a matter of fundamental principle, it should be understood in the context of QED on curved spacetime, and we should be able to determine at order ${\cal O}(\alpha)$ whether there are any other significant corrections. Since QED is the simplest realistic interacting quantum field theory (in the sense of describing known particles rather than a toy model), we also expect that exploring QED on a Schwarzschild background (the simplest type of black hole) is also a good starting point for learning how to {\em numerically} treat interacting theories on black hole spacetimes.

This paper follows the results of a previous work from our group, \citet{Paper1} (hereafter Paper I), in which we worked out the formalism for the dissipative first-order QED corrections to the photon Hawking radiation spectra from a Schwarzschild black hole (the simplest case). Paper I began with the quantization of QED on the Schwarzschild spacetime, using canonical methods and a generalization of the Coulomb gauge (this gauge uses the vector potential for propagating waves, but the scalar potential to mediate electrostatic interactions; some ``loose ends'' on the latter, needed for corrections to the electron/positron spectrum but not the photon spectrum, are presented in \citet{2024arXiv240709724V}). We employed a perturbative approach, working to first order in $\alpha$. We also distinguished {\em dissipative effects}, where the number of particles is changing (electrons, positrons, and photons are created or destroyed), from {\em conservative effects}, in which the number of particles do not change (rather the transmission probability for a particle to escape to infinity instead of falling back into the black hole is modified by the plasma or by virtual particles). Conservative effects, while important, will be explored in a future paper in this series. They are inherently more challenging because their full treatment requires renormalization. Here we will focus on the numerical implementation of the dissipative effect formalism introduced in Paper I. 

This paper is organized as follows. We review the structure of the equation for the Hawking radiation spectrum that we need to evaluate (Paper I, Eq.~80) in Section~\ref{sec:outline}. We describe the numerical evaluation of Paper I Eq.~(80) in Section~\ref{sec:numerical}. The results for several black hole masses are presented in Section~\ref{sec:results}. We conclude and discuss directions for future work in Section~\ref{sec:discussion}.

\section{Outline of the calculation}
\label{sec:outline}

\begin{figure}
    \centering
    \includegraphics[scale=.5]{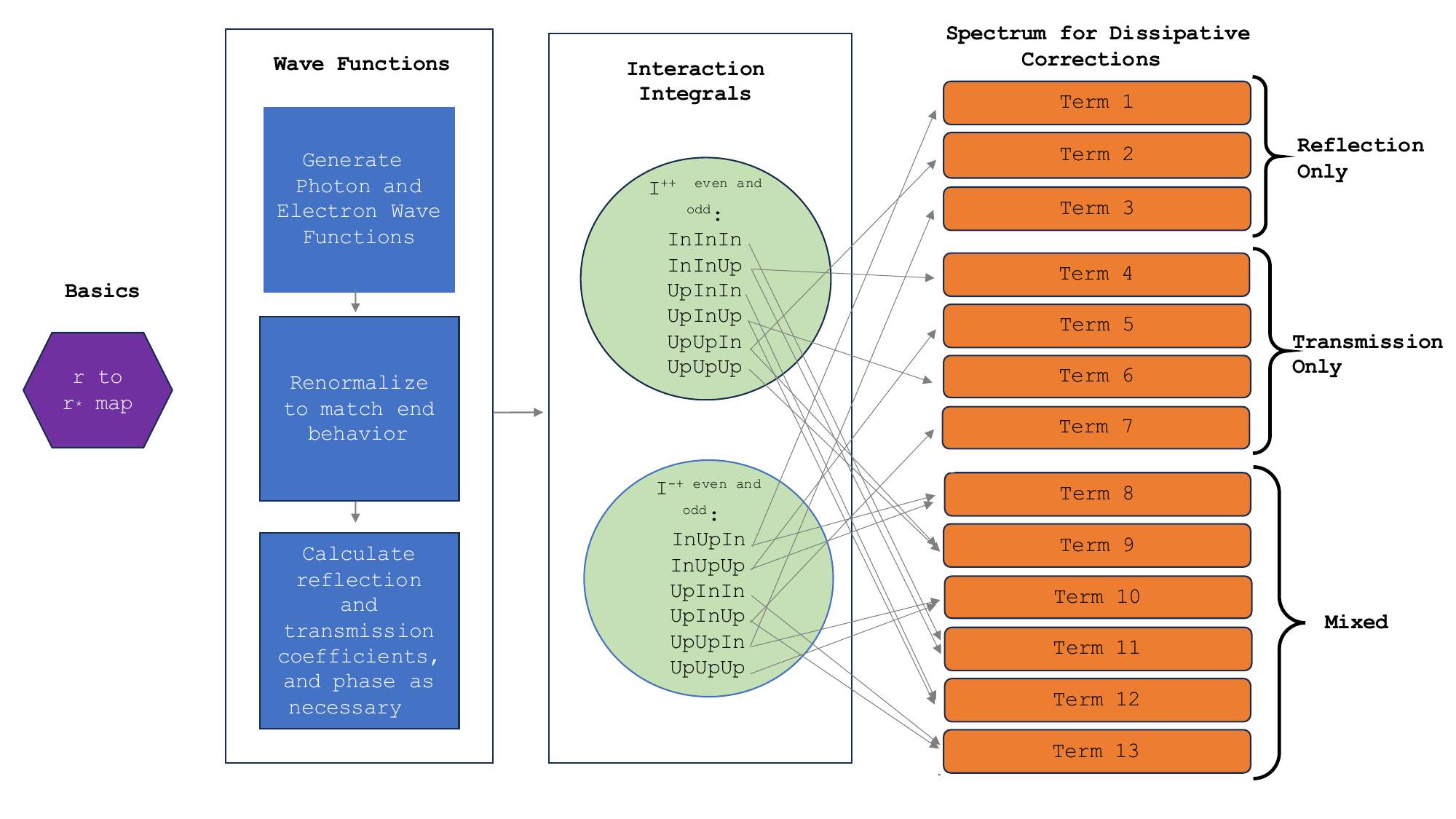}
    \caption{Flow chart demonstrating the numerical processes that have been built to calculate the dissipative spectrum. This includes the $r\leftrightarrow r_\star$ mapping (\S\ref{ss:mapping}); the photon (\S\ref{ss:pwf}) and electron (\S\ref{ss:ewf}) wave functions; the interaction integrals (\S\ref{ss:integral}); and the evaluation of the spectrum (\S\ref{ss:spec}).} 
    \label{fig: Numerical Flow Chart}
\end{figure}

The calculation of the corrections to the Hawking radiation spectrum proceeds in several steps, starting from the basics of the Schwarzschild geometry through to the matrix elements and ultimately the particle spectrum. See Figure~\ref{fig: Numerical Flow Chart} for a schematic. We work in natural units where the universal gravitation constant $G_{\rm N}$, the speed of light $c$, the reduced Planck's constant $\hbar$, Boltzmann's constant $k_{\rm B}$, and the permittivity of the vacuum $\epsilon_0$ are all equal to 1. The $-+++$ metric and Dirac equation conventions from Paper I are used.

\newcounter{cstep}
\begin{list}{\S III\Alph{cstep}:}{\usecounter{cstep}}
\item The first step --- and the lowest-level utility in our code --- is the mapping between the Schwarzschild radial coordinate $r$, and the tortoise coordinate $r_\star$ in which radial geodesics have $dr_\star/dt = \pm 1$.
\item Next, we compute the single-particle photon radial wave functions. The photon quantum numbers are $\{\ell,m,p,\omega,X\}$, where $\ell$ and $m$ are the total angular momentum and its projection on the $z$-axis; $p\in\{e,o\}$ is the parity (electric-type for $e$ and magnetic-type for $o$); the single-particle energy $\omega$ is the only continuous parameter; and $X\in\{{\rm in,up}\}$ selects the scattering basis state (coming ``in'' from spatial infinity or ``up'' from the horizon). Spherical symmetry implies that the solutions do not depend on $m$. Furthermore, the duality of electric and magnetic fields in Maxwell's equations implies that the same mode functions $\Psi_{X,\ell, \omega}(r_\star)$ can be used to describe both parities (see Paper I, Eq.~29).
\item The single-particle electron wave functions are described by quantum numbers $\{k,m,h,X\}$, where the fermion energies are described by $h$. The angular momentum is captured in the Schr\"odinger separation constant $k = \pm 1,\pm 2,\pm3...$, where the total angular momentum is $j=|k|-\frac12$. The sign of $k$ distinguishes between the two partial waves of a spin $\frac12$ particle with the same $j$: $k=1,2,3,...$ correspond to s$_{1/2}$, p$_{3/2}$, d$_{5/2}$ ..., whereas $k=-1,-2,-3,...$ correspond to p$_{1/2}$, d$_{3/2}$, f$_{5/2}$ ...\,. The radial wave functions are described by two complex functions $F_{kh}(r_\star)$ and $G_{kh}(r_\star)$, since once $k$ is fixed only 2 of the 4 Dirac spinor components are independent. (For completeness and as a numerical check, we present the results here, even though the radial functions are well-studied in the literature \cite[e.g.][]{1976RSPSA.348...39C, 1977PhRvD..16.2402P, 2021PhRvD.103j4010A, 2021PhRvD.104h4016A}.)
\item The matrix elements describing the photon-fermion-fermion vertex are described as $I$-integrals (Paper I, Eq.~56). Numerically, we work with the reduced version $\llbracket I\rrbracket$ in the sense of the Wigner-Eckart theorem, so that the $m$ quantum numbers do not need to be specified. The $\llbracket I^{\pm\pm}_{Xk,X'k',X_\gamma\ell}(h,h',\omega) \rrbracket$ integrals are the most computationally expensive part of our analysis, since they depend on many quantities: the angular momenta ($k,k',\ell)$ of the two fermions and the photon; the single-particle energies $(h,h',\omega)$ (although since in this paper the particles are all real, energy conservation is enforced so only 2 of the 3 energies are independent); the photon parity $p$; the electron vs.\ positron selections (the two $\pm$ superscripts, so $2^2=4$ choices, although 1 of the 4 turns out to not be needed due to energy conservation, and $I^{+-}$ can be related to $I^{-+}$ by charge conjugation); and the scattering states (in vs.\, up for each particle: $X,X',X_\gamma$, or $2^3=8$ choices total). 
\item Finally, we compute the individual terms in the correction to the photon spectrum, Paper I Eq.~(80).
\end{list}

\begin{table}
\caption{\label{tab:fd}Diagrammatic representations of the 13 terms in Paper I Eq.~(80). The first three represent the terms where the in-going photon is reflected. The next 4 represent processes where the up-going photon is transmitted through the momentum barrier. The last 6 terms individually represent the interference between 2 processes; some of these processes do not contribute overall to the spectra by themselves, but combined with the other process, they can create interference that have measurable impacts. For all diagrams, time flows vertically, the black hole horizon is on the left, and spatial infinity is on the right. Any "in" photon must be reflected off the angular momentum barrier of the black hole to contribute, which is demonstrated by the bend in the "in" photon paths.} 
\hfill
\includegraphics[width= 7.0in]{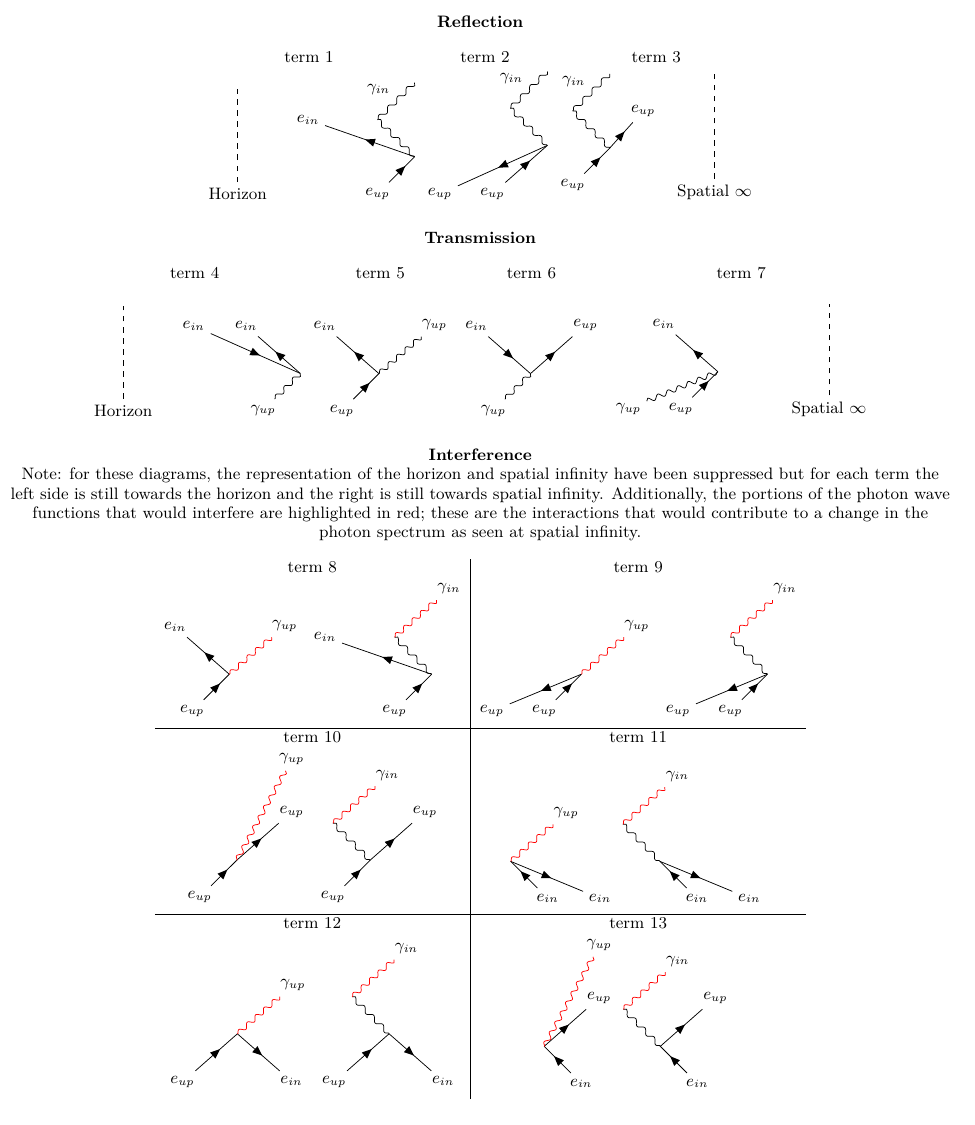}
\end{table}

\section{Numerical Methods}
\label{sec:numerical}

We begin with a description of our numerical methods. The organization of our code is shown in Fig.~\ref{fig: Numerical Flow Chart}; this section is organized around the modules of this code, progressing from left to right.

\subsection{$r\leftrightarrow r_\star$ mapping}
\label{ss:mapping}

In order to start our calculation, we need an accurate and numerically stable routine to map between $r$ and $r_\star$. This is almost trivial for the direction $r\rightarrow r_\star$ since there is an analytic mapping:
\begin{equation}
r_\star = r+ 2M \ln\frac{r-2M}{2M}.
\label{r to rstar}
\end{equation}
The one subtlety is that very close to the horizon, where $r-2M\ll 2M$, there is a loss of numerical precision in taking the difference $r-2M$. This issue is fundamental if $r$ is represented with finitely many bits, and our cure will be to take $r_\star$ as the independent variable for our calculations.

To compute the inverse of this interaction and map $r_\star$ to $r$, we recursively solve the function in both the near-horizon and far away limits. If we are trying to compute solutions for $r$ close to the black hole, in our case $r_\star<-3M$, we use
\begin{equation} 
    r_{i+1} = 2M \left( 1+ \exp\frac{r_\star -r_{i}}{2M} \right),
\end{equation}
where $i = 0, 1, 2...$ is the iteration number. We start with $r_0=2M$ and proceed until the fractional difference in $r$ is less than our set tolerance of $10^{-10}$. 

If we are trying to compute solutions for $r$ far from the black hole, in our case $r_\star>12M$, we instead use
\begin{equation} 
    r_{i+1} = r_\star - 2M \ln\frac{r_{i} - 2M}{2M}
\end{equation}
until convergence is met, starting with $r_0=(2+10^{-10})M$.

For $-3M<r_\star<12M$, we perform a root-finding solution using Brent's method.

\subsection{Photon Wave function}
\label{ss:pwf}

\begin{figure}
    \centering
    \includegraphics[scale=.6]{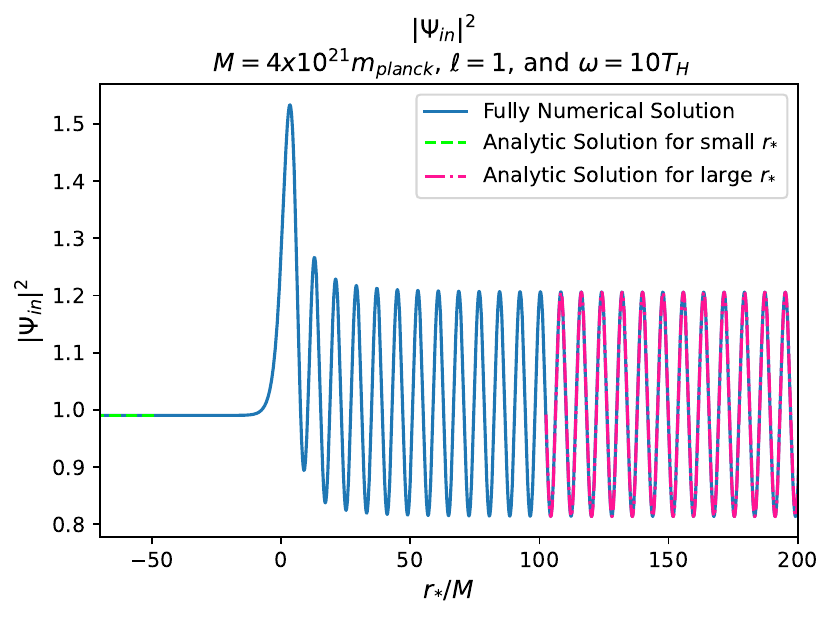}
    \includegraphics[scale=.6]{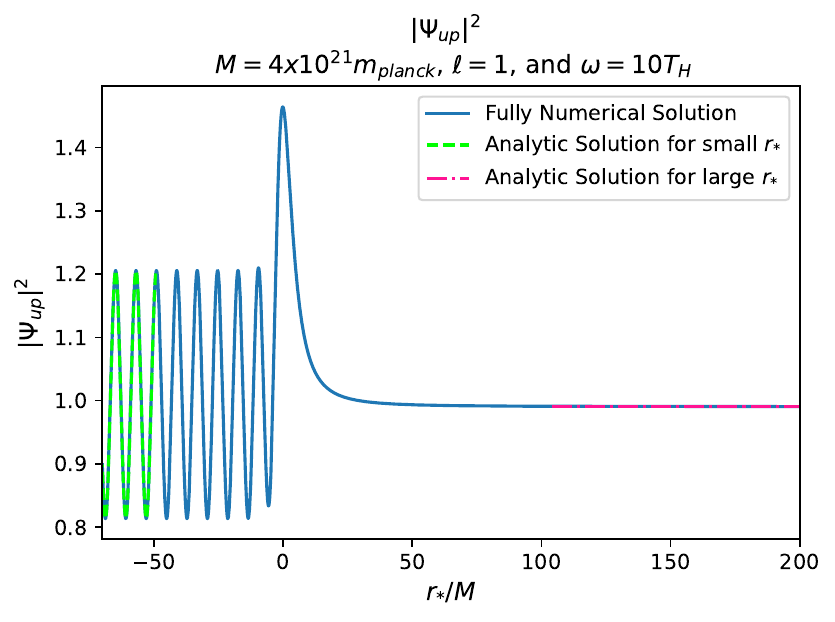}
    \caption{Example Photon Wave Function, for $M = 4\times10^{21}$, $\ell=1$, and $\omega = 10 T_{H}$. The left figure shows the 'in' solution, and the right shows the 'up' solution. The wave function found with our numerical approach is given in blue, and the expected analytic solutions are given in green (pink) for the small (large) $r_{*}$ limits. 
    }
    \label{fig:PhotonWF}
\end{figure}

The first step in our goal of computing the numerical solution for the density evolution is to numerically compute the unperturbed photon wave function around a Schwarzschild black hole. To begin, the differential equation ${\cal H}_{l}\Psi = \omega^{2}\Psi$ (Paper I, Eq. 16) for the photon wave function can be re-written as a system of coupled first-order ODEs:
\begin{equation}
    z = \Psi' ~~~{\rm and}~~~
    z' = \left[-\omega^{2} + \frac{l(l+1)(1-2M/r)}{r^{2}} \right] \Psi.
\label{eq:Psi-ODE}
\end{equation}
These two first order differential equations are solved simultaneously using the fourth-order Runge-Kutta (RK4) method.

We need different initial conditions for the ``in'' and ``up'' solutions, integrating in opposite directions. The ``in'' solution is characterized by having a wave propagating purely inward at $r_\star\rightarrow-\infty$, which can be accommodated by starting at $r_\star = r_{\star,\rm min}$ with $\Psi=1$ and $z=-i\omega$ and integrating outward. In order to ensure the normalization of these wave functions are consistent with our previous analytic calculations, we rescale the wave functions by matching the numerical solutions to the analytic forms at the boundary conditions.
For the ``in'' solution, we let the properly scaled wave function take the form
\begin{equation}
\Psi_{\rm in} = a \Psi_{\rm in}^{\rm old}   
\end{equation}
where $\Psi_{\rm in}^{\rm old}$ is the wave function taken directly from the RK4 integrator. To calculate $a$, we use Paper I, Eq.~(22) and take the combination $ -i \omega\Psi_{in} +  \Psi'_{in}$ in the $r_{\star} \rightarrow \infty$ limit, giving 
\begin{equation}
    a =\left[ \frac{- 2i\omega e^{-i\omega r_{\star}}}{-i\omega\Psi_{\rm in}^{\rm old} + \Psi_{\rm in}^{\rm old}{'}}  \right]_{r_{\star} = r_{\star,\rm max}}.
\label{eq:a-norm}
\end{equation}

A similar procedure is done for the ``up'' solution. This time, we start at $r_\star = r_{\star,\rm max}$ with $\Psi=1$ and $z=i\omega$, and integrate inward. To rescale the solution from the RK4 integrator, we let $ \Psi_{\rm up} = b \Psi_{\rm up}^{\rm old}$ and use the combination of $i \omega\Psi_{\rm up} +  \Psi'_{\rm up}$ in the $r_{\star} \rightarrow -\infty$ limit to find 
\begin{equation}
    b =\left[ \frac{ 2i\omega e^{i\omega r_{\star}}}{i\omega\Psi_{\rm up}^{\rm old} + \Psi_{\rm up}^{\rm old}{'}}\right]_{r_{\star} = r_{\star,\rm min} }.
\end{equation}

Now that we have appropriately scaled wave functions, we can calculate the  reflection and transmission coefficients, $R$ and $T$, of the photon wave functions, which are 
\begin{equation}
    R =\left[ \frac{i\omega \Psi_{\rm in} + \Psi'_{\rm in}}{2i\omega e^{i\omega r_{\star}}}\right]_{r_{\star} = r_{\star,\rm max}}
    ~~{\rm and}~~
    T = \left[ \Psi_{\rm up}e^{-i\omega r_{\star}} \right]_{r_{\star} = r_{\star,\rm max}}. 
\end{equation}

To generate our table of photon wave functions, we fix a grid of $r_{\star}$ from $-70M$ to $2000M$ with evaluation on 240,000 equally spaced points, and store data for every other point to ensure that our ODE solver is appropriately sampled. We compute a grid of photon energies $\omega$ in steps of $0.01T_{\rm H}$ from $0.01T_{\rm H}$ to $20.00T_{\rm H}$. Finally, we let $\ell$ vary from 1 to 5 (for $M=10^{21}$ and $2\times 10^{21}M_{\rm Planck}$; convergence is faster for larger black hole masses, and we sum through $\ell=4$ for $M=4\times 10^{21}M_{\rm Planck}$ and $\ell=3$ for $M=8\times 10^{21}M_{\rm Planck}$).

Figure \ref{fig:PhotonWF} shows example photon wave functions for $M = 2\times 10^{21}$, $\ell=1$, and $\omega = 10T_{H} $. The three different regimes for the wave functions are clearly demonstrated in this figure: $r_{*}\ll -M$, $r_{*}\sim 0$, and $r_{*}\gg M$. From Figure~\ref{fig:PhotonWF}, we can see our wavefunction is behaving as expected in Paper I Eqs.~(22,23). For the ``in'' solution we have a purely ingoing wave near the horizon ($r_{*}\rightarrow-\infty$), indicated by a nearly constant line, and a superposition of incoming and outgoing waves far from infinity ($r_{*}\rightarrow\infty$), indicated by an oscillating behavior as we go further from the horizon. Likewise for the ``up'' solution, we have a purely outgoing wave far from the horizon (indicated by a near constant line) and a superposition of ingoing and outgoing waves near the horizon (oscillations near the horizon). In addition to complete numerical solutions for the wave functions, the analytic limits for small and large $r_{*}$ are plotted to confirm our methods. We also see that the photon wave functions are of order unity as expected from our normalization conditions.

\begin{figure}
    \centering
    \includegraphics[width=5in]{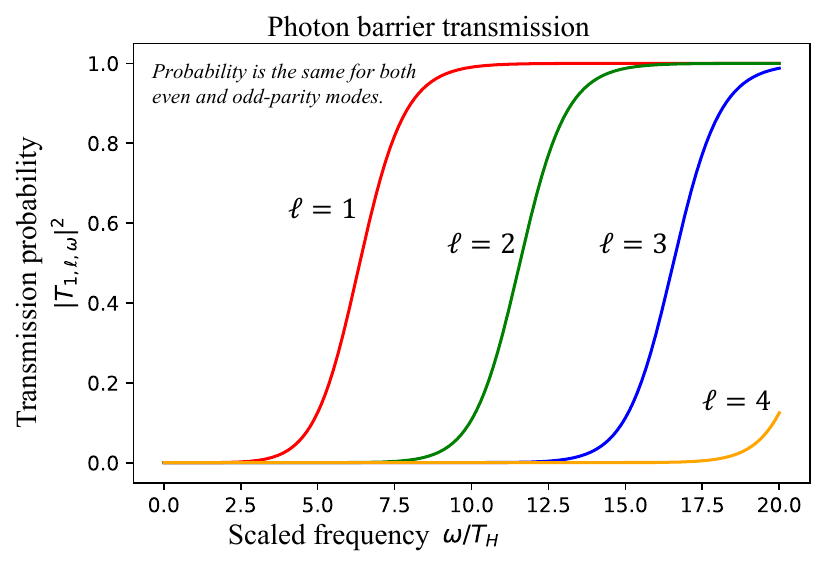}
    \caption{Photon transmission probability $|T_{1,\ell,\omega}|^{2}$ for $\ell =1-4$. Each color represents a different $\ell$, and the solid lines represent the reflection coefficients while the dashed represent the transmission coefficients. All the different masses are identical display identical behavior. As the value of $\ell$ increases, we see the probability of transmission through the barrier staying negligible for higher energies, until the $\ell=5$ case where there is a vanishing transmission coefficient over our entire photon energy range. This figure is independent of the black hole mass since the photon is massless.}
    \label{fig:TCoeffsPhoton}
\end{figure}

In order to fully understand the following discussions where the reflection and transmission coefficients for the photon play an important role, we also show the full set of coefficients in Figure~\ref{fig:TCoeffsPhoton}. At low $\omega$, the transmission coefficients go to zero (the black hole is a poor absorber or emitter), but at large $\omega$ the transmission probabilities go to 1. As we increase $\ell$, the frequency that corresponds to the transition between mostly reflection ($|T_{1,\ell\omega|^2}<\frac12$) and  mostly transmission ($|T_{1,\ell\omega|^2}>\frac12$) increases. The geometric optics prediction is that this transition should occur when the impact parameter (which is the angular momentum-to-energy ratio for a massless particle) is $\sqrt{\ell(\ell+1)}/\omega = 3\sqrt3\,M$ (e.g., \cite{1973grav.book.....M}, Ch.~23), or
\begin{equation}
\frac\omega{T_{\rm H}} = \frac{8\pi M}{\sqrt{\ell(\ell+1)}/\omega} \sqrt{\ell(\ell+1)}
= \frac{8\pi}{3\sqrt3} \sqrt{\ell(\ell+1)} = 4.84\sqrt{\ell(\ell+1)} = \left\{\begin{array}{rcl}
6.84 & & \ell=1 \\
11.85 & & \ell=2 \\
16.76 & & \ell=3 \\
21.63 & & \ell=4; \\
\end{array}\right.
\label{eq:50-1}
\end{equation}
this expectation does indeed correspond to the results in Fig.~\ref{fig:TCoeffsPhoton}.

\subsection{Electron Wave function}
\label{ss:ewf}

We now turn to the electron wave functions. As before, we construct an RK4 solver for the differential equations of $F$ and $G$ as shown in Paper I Eq.~(35), which can be slightly rewritten as
\begin{equation} 
    \frac{dF}{dr_{\star}} = \bigg(-\mu \sqrt {1 - \frac {2M}{r}} - h\bigg) G + \frac{k}{r} \sqrt {1 - \frac {2M}{r}}\, F
       ~~~ {\rm and} ~~~
    \frac{dG}{dr_{\star}} = \bigg(h - \mu \sqrt {1 - \frac {2M}{r}}\bigg) F - \frac{k}{r} \sqrt {1 - \frac {2M}{r}}\, G .
\end{equation}
Again, we will need to rescale the ``up'' and ``in'' solutions to match the appropriate boundary conditions. Unlike the photon, the electron is massive and so it has different behavior for the bound ($h<\mu$) and unbound ($h>\mu$) solutions: the bound case has no ``in'' solution, and must have a reflection probability of unity.

We first consider the ``in'' solution for the unbound case, i.e. $h>\mu$. We start with $(F_{\rm in},G_{\rm in}) = (1,i)$ at $r_\star=r_{\star,\rm min}$ and integrate outward.
To rescale these solutions, we take $(F_{\rm in},G_{\rm in}) = b (F_{\rm in}^{\rm old},G_{\rm in}^{\rm old})$ and use the combination of $\sqrt{h+\mu}\,F_{\rm in} - i\sqrt{h + \mu}\,G_{\rm in}$ from Paper I Eq.~(39) in the $r_{\star} \rightarrow \infty$ limit to find
\begin{equation}
    b =\left[ \frac{ 2h\sqrt{v}\,e^{-i\zeta \ln(r_\star/2M)}e^{-i\sqrt{h^{2}-\mu^{2}}r_\star}}{\sqrt{h-\mu}F^{\rm old}_{\rm in} -i\sqrt{h+\mu}G^{\rm old}_{\rm in}} \right]_{r_\star = r_{\star,\rm max}},
\end{equation}
where $\zeta$ and $v$ are as defined in Paper I Eqs.~(37) and (40).

For the unbound ``up'' solution, we start with $(F_{\rm up},G_{\rm up}) = (\sqrt{h+\mu},-i\sqrt{h-\mu})$ at $r_\star=r_{\star,\rm max}$ and integrate inward.
The rescaling is $(F_{\rm up},G_{\rm up}) = a (F_{\rm up}^{\rm old}, G_{\rm up}^{\rm old})$; we use the combination $F_{\rm up}+ iG_{\rm up}$ from Paper I Eq.~(41) in the $r_\star \rightarrow -\infty$ limit to arrive at 
\begin{equation}
    a = \left[ \frac{ 2\sqrt{h}e^{ihr_{\star}}}{F^{\rm old}_{\rm up} + iG^{\rm old}_{\rm up}} \right]_{r_{\star} =r_{\star,\rm min}}.
\end{equation}
Using this combined knowledge, we can define the reflection and transmission coefficients for our scaled wave functions as  
\begin{equation}
    R_{\frac12,k,h} = \left[ \frac{(\sqrt{h-\mu}F_{\rm in}+ i\sqrt{h+\mu}G_{\rm in})e^{-i\zeta \ln (r_\star/2M)}e^{-i\sqrt{h^{2}-\mu^{2}}r_{\star}}}{2h\sqrt{v}} \right]_{r_{\star} = r_{\star,\rm max}} 
\end{equation}
and
\begin{equation}
    T_{\frac12,k,h} = \left[ \frac{ G_{\rm up}\sqrt{v}e^{-i\zeta\ln (r_\star/2M)}e^{-i\sqrt{h^{2}-\mu^{2}}r_{\star}}}{-i\sqrt{h-\mu}}\right]_{r_{\star} = r_{\star,\rm max}}.
\end{equation}

This process must now be repeated for the bound case, where $0<h<\mu$. Here we may formally take $(F_{\rm in},G_{\rm in})=0$, so we just need to consider the ``up'' solutions. We find these by setting $(F_{\rm up}, G_{\rm up}) = ((\mu+h)/\sqrt{\mu^2-h^2}, 1)$ at $r_\star=r_{\star,\rm max}$ (the ratio corresponds to the exponentially decaying solution at large $r_\star$) and integrating inward.
Let us define $(F_{\rm up},G_{\rm up}) = a (F_{up}^{old}, G_{up}^{old})$. Using Paper I Eq.~(42), we can look at the combination $-iF_{up} + G_{\rm up}$ and find
\begin{equation}
a = \left[ \frac{-2i\sqrt{h}e^{ihr_{\star}}}{-iF_{up}^{old} + G_{up}^{old}} \right]_{r_{\star} =r_{\star,\rm min}}.
\label{eq:up-a}
\end{equation}
For this solution set, we know $|R_{\frac12,k,h}|=1$ and $T_{\frac12,k,h}=0$, so the only piece left to fully define our solutions is the phase $\delta_{1/2,k,h}$:
\begin{equation}
\delta_{\frac12,k,h} = \frac{-i}{2} \ln\left[ \frac{iF_{\rm up}+G_{\rm up}}{2i\sqrt{h}e^{-ihr_{\star}}} \right]_{r_\star=r_{\star,\rm min}}.
\end{equation}
The normalization of Eq.~(\ref{eq:up-a}) guarantees that the quantity in brackets has absolute value unity, and hence that this phase is real. This leaves an ambiguity in which $\delta_{1/2,k,h}$ could be incremented by $\pi$, but this is not important for the numerical evaluation in this paper.

Again, we fix a grid of $r_\star$ from $-70M$ to $2000M$ with evaluation on 240,000 equally spaced points, and store data for every other point. We also use the same grid of energies where $h$ is spaced from $0.01T_{\rm H}$ to $20.00T_{\rm H}$ with a spacing of $\Delta h = 0.01T$. Finally, we let $k$ vary from $-10$ to $10$, not including 0 (which is not an allowed value of the Schr\"odinger separation constant).

\begin{figure}
    \centering
    \includegraphics[scale=.6]{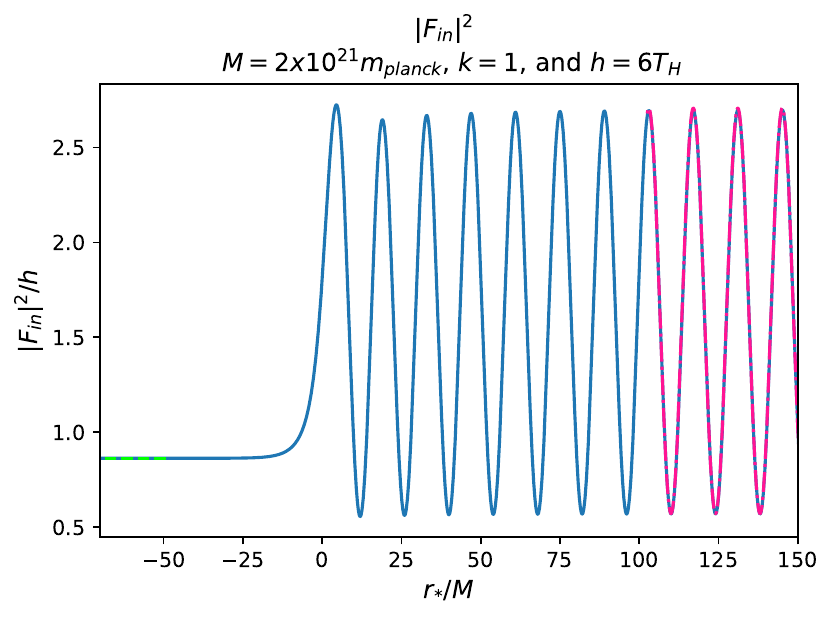}
    \includegraphics[scale=.6]{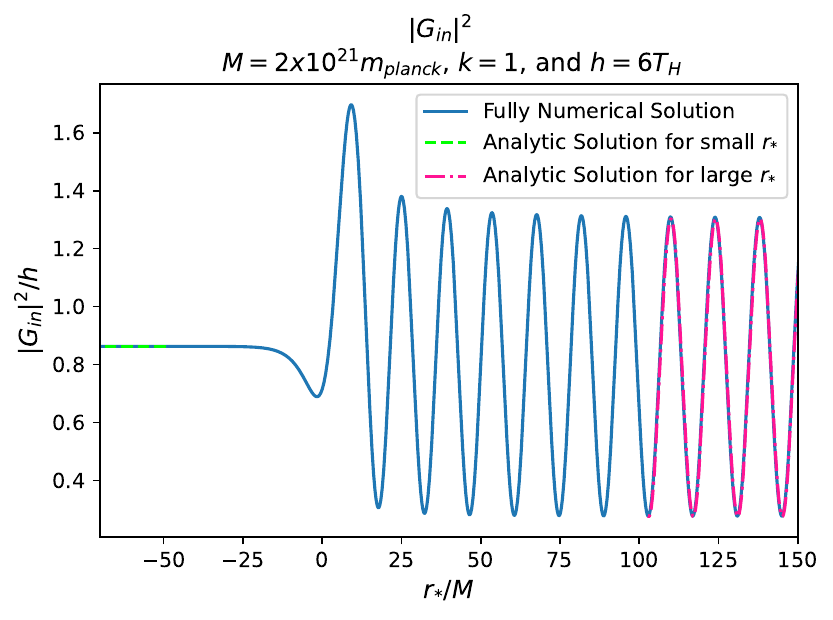}
    \includegraphics[scale=.6]{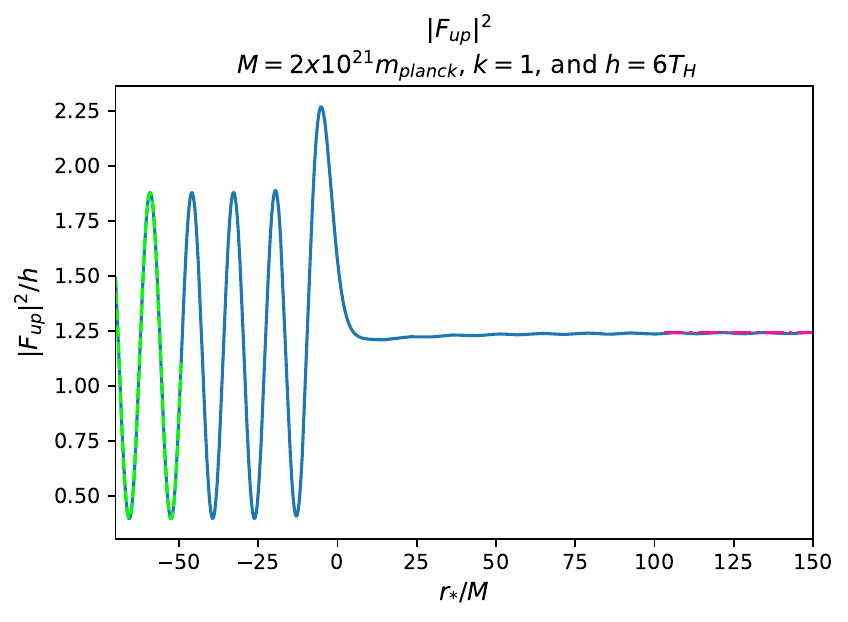}
    \includegraphics[scale=.6]{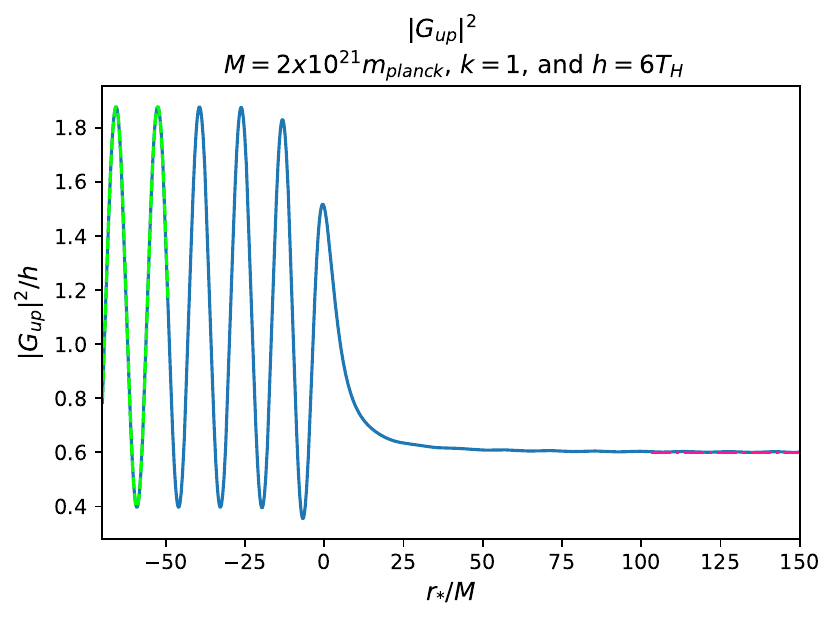}
    \caption{Example Electron Wave Functions for $M = 2\times10^{21}$, $k=1$, and $h = 6T_{H}$. The top (bottom) row represents fermionic wave functions in the 'in' ('up') basis, and the left (right) figures represent F (G) solutions. For each figure, the numerical solution is shown in blue, the analytic solution for small $r_{*}$ is shown in green, and the analytic solution for large $r_{*}$ is in pink. The $|F|^{2}$ and  $|G|^{2}$ are of the $\mathcal{O}(1/h)$, as expected.}
    \label{fig:ElectronWF}
\end{figure}

Figure \ref{fig:ElectronWF} shows an example fermion wave function, specifically the solution for $M = 2\times10^{21}$, $k=1$, and $h = 6T_{H}$. These wave functions scale as $\mathcal{O}(1/\sqrt{h})$, which is seen in this figure. We also see the limiting behavior for small and large $r_{*}$ numerically match our analytic expectations. For "in" solutions, we expect the behavior for the electron to be highly oscillatory far from the black hole due to the superposition of the in-going and outgoing waves, and a close to the black hole we have a purely in-going wave which has flat asymptotic behavior close to the black hole; the opposite behavior is true for the "up" solution. 

\begin{figure}
    \centering
    \includegraphics[width=6.5in]{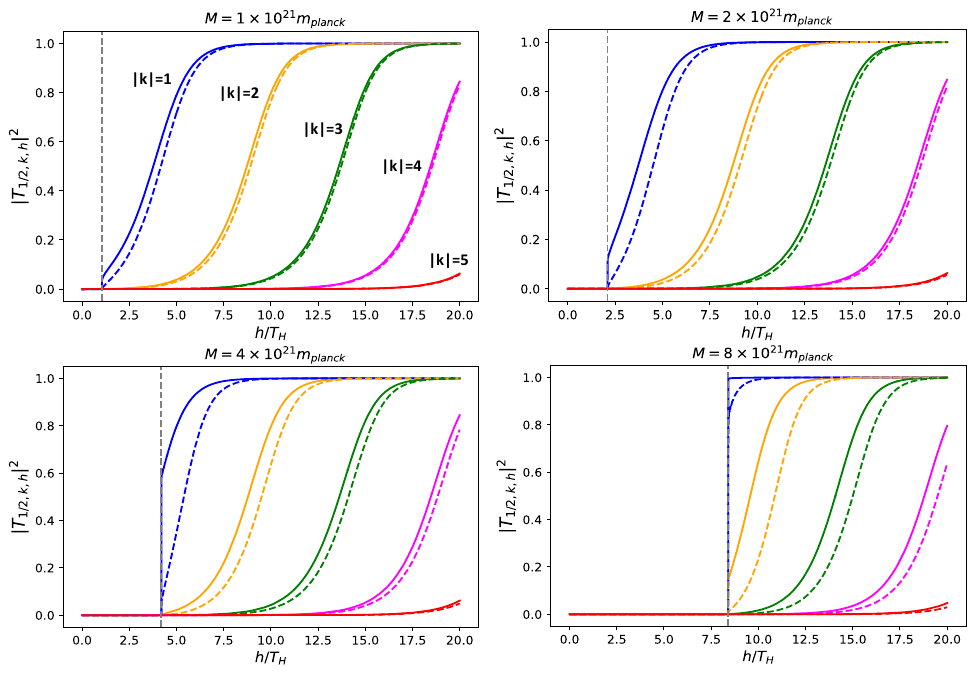}
    \caption{Electron transmission coefficients $|T_{\frac12,\ell,\omega}|$ for $\lvert k\rvert =1-5$. Each panel shows a different PBH mass. Each color represents a different $|k|$, and the solid lines represent the $k>0$ coefficients while the dashed represent the $k<0$ coefficients. There is a boundary at $h = \mu$ (expressed at different $T_{H}$ values) where lower energies are not allowed to transmit through the barrier, which is shown as the gray vertical dashed line.}  
    \label{fig:RTCoeffsElectron}
\end{figure}

Figure~\ref{fig:RTCoeffsElectron} shows the transmission probabilities for the electrons. Because the electron is massive, this depends on the black hole mass, even after scaling by the Hawking temperature. Once again, there is a behavior that the transmission probability grows with $h$. For $h\gg\mu$ or $h/T_H\gg 8\pi \mu M$, arguments similar to Eq.~(\ref{eq:50-1}) should apply, and the energy at which there is a 50\% transmission probability should be
\begin{equation}
\frac h{T_{\rm H}} =
4.84\sqrt{k^2-\frac14} = \left\{\begin{array}{rcl}
4.19 & & |k|=1 \\
9.37 & & |k|=2 \\
14.31 & & |k|=3 \\
19.20 & & |k|=4. \\
\end{array}\right.
\label{eq:50-0.5}
\end{equation}
Recall that the squared angular momentum is $j(j+1)=k^2-\frac14$.) This is good for large $|k|$, but the critical energy is larger than this for large black hole masses and low $|k|$ (the regime where the electron is non-relativistic). In the non-relativistic regime, at a given $|k|$, the case with positive $k$ has a larger transmission probability than negative $k$ because it has smaller angular momentum: positive $k$ corresponds to $\ell =|k|-\frac12$ whereas negative $k$ corresponds to $\ell = |k|+\frac12$. On the other hand, in the ultrarelativistic regime, we can neglect the electron mass ($\mu\rightarrow 0$) and then the left- and right-handed electrons do not mix: this leads to two modes for each $h,j,m$ with the same transmission coefficients.\footnote{A similar result even applies in the Kerr case due to the mapping between $s=\frac12$ and $s=-\frac12$ solutions to the radial equation --- see \citet{1973ApJ...185..635T}, Eqs.~(B2, B3) --- even though the degeneracy of different values of $m$ is lifted and the angular momentum quantum number $j$ is replaced by a counting index.}

\subsection{Interaction integrals}
\label{ss:integral}

In order to arrive at the dissipative  evolution of the photon density matrix, we must construct the interaction integrals of Paper I Eq.~(56). These terms describe the 3-particle (electron-electron-photon) interactions. We separate the even and odd contributions since they have different forms and allow for different transitions. This calculation involved a numerical integration of the analytic equations using the midpoint method. For the Wigner $3j$ symbols, we use the {\sc SymPy} package \cite{SymPy}.

\begin{figure}
    \centering
    \includegraphics[width=3.0in]{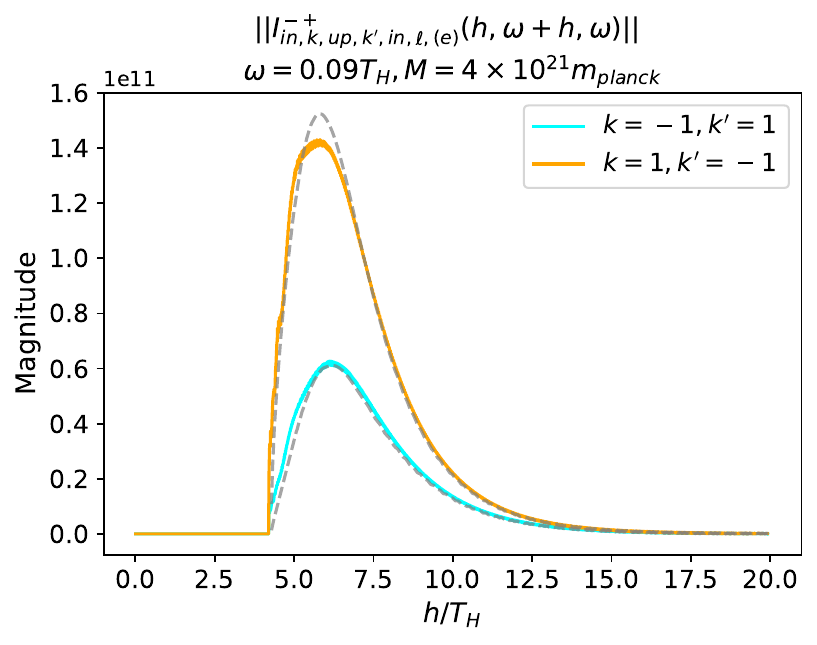}
    \includegraphics[width=3.0in]{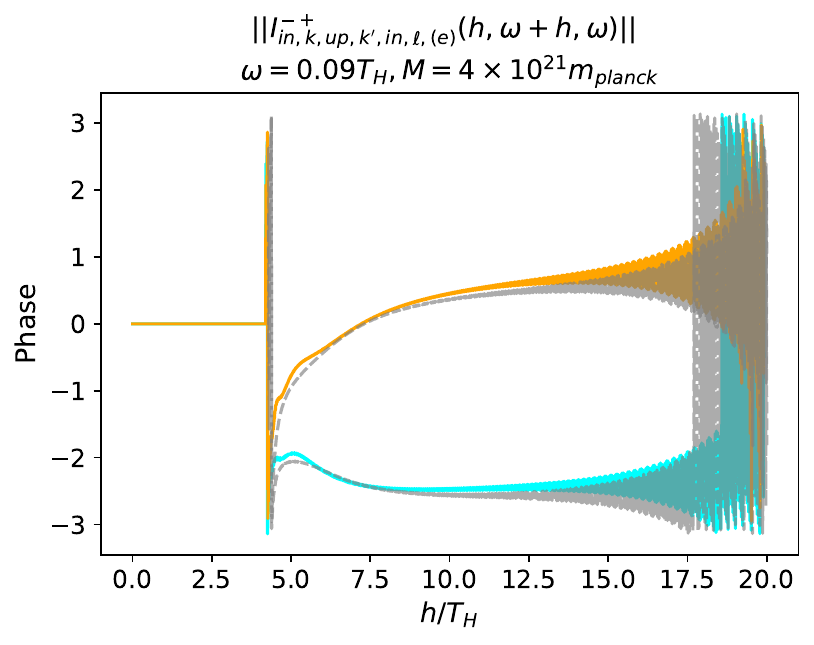}
    \includegraphics[width=3.0in]{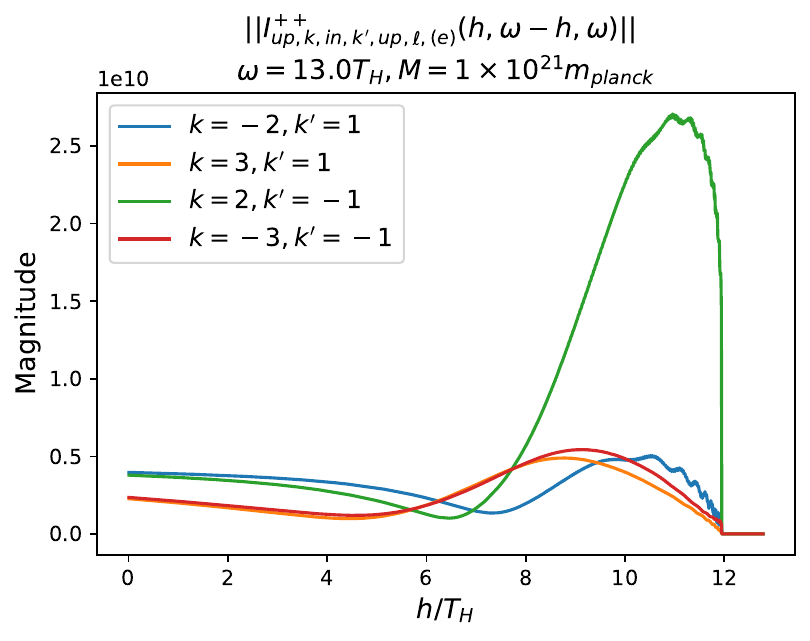}
    \includegraphics[width=3.1in]{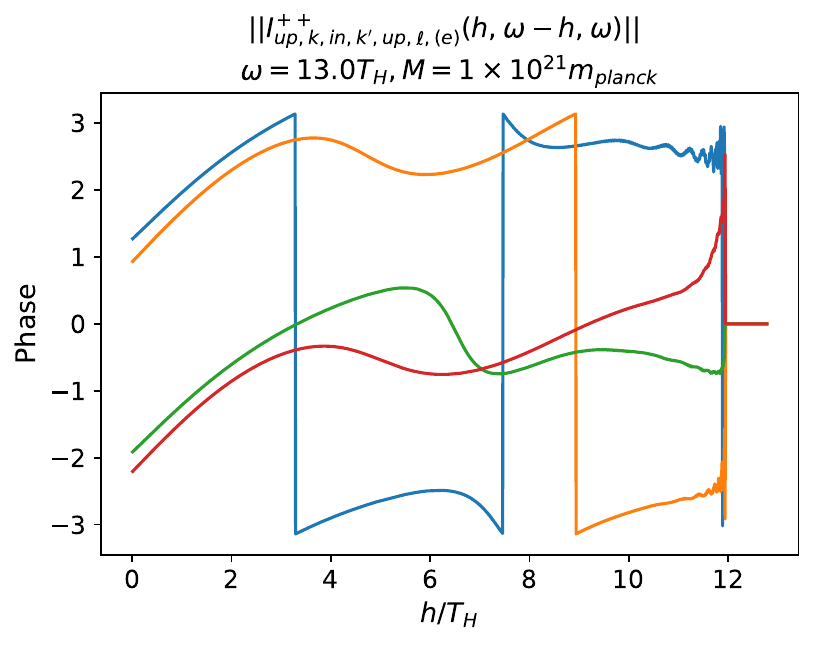}
    \caption{Representation of the magnitude (left) and phase (right) of the selected $I$-integrals. The top row shows the behavior of the $\llbracket I^{-+}_{{\rm in},k,{\rm up},k',{\rm in},\ell,(e)}(h,\omega - h,\omega) \rrbracket$ integral as a function of fermion energy with the $k$ and $k' =\pm 1$ for $M=4\times10^{21}m_{planck}$ and $\ell =1$ at $\omega= 0.09 T_{H}$. For this interaction integral, we compare our fully numerical results with the classical approximation in eq.~\ref{B7}, and show that the two methods of calculation converge. The second row shows the  $\llbracket I^{++}_{{\rm up},k,{\rm in},k',{\rm in},\ell,(e)}(h,\omega - h,\omega) \rrbracket$  for $M=1\times10^{21}m_{planck}$ and $\ell =2$ at $\omega= 13.0 T_{H}$ for various $k$ and $k'$ combinations. This $I$-integral is the dominant contribution to the spectrum at the high  photon energy. Again, magnitude is plotted on the left, and phase on the right. Note that the phase exhibits numerical ringing when the magnitude of the integral approaches zero.}
    \label{fig:IIntegrals}
\end{figure}

It is difficult to display the full space of $I$-integrals since they depend on 2 continuous energies ($\omega$ and $h$), three angular momenta ($k,k',\ell$), a parity $p$, and three binary indices (in vs.\ up). In Figure~\ref{fig:IIntegrals}, we show some slices through this space. Specifically, we investigate the behavior of different $I$-integrals as a function of $h$ for different $k$ values evaluated at two specific $\omega$s - one in the low and one in the high frequency limits. First, we look at $\llbracket I^{-+}_{{\rm in},k,{\rm up},k',{\rm in},\ell,(e)}(h,\omega - h,\omega) \rrbracket$ which contributes to the expected inner bremsstrahlung process of the spectrum below at low frequencies. This fermion  energy dependent investigation was critical in our understanding of the comparison to the classical results for the electric dipole behavior. A detailed discussion of this specific check can be found in Appendix~\ref{app:corr}, and the classical result expectation from Eq.~(\ref{B7}) is also shown in Figure ~\ref{fig:IIntegrals} for the magnitude and phase of the $I$-integral. 

In the higher photon energy limit, the $\llbracket I^{++}_{{\rm up},k,{\rm in},k',{\rm in},\ell,(e)}(h,\omega - h,\omega) \rrbracket$ has a sizeable contribution to the photon spectrum, so we also show this in Figure~\ref{fig:IIntegrals}, though there is no expected classical result in this region (it corresponds to a pair production or pair annihilation process). The majority of the contribution for this process comes from regions where the fermion energy is close to $\omega-\mu$, whereas the other $I$-integral we show has most of its contribution around $h\approx \mu$. We also see that the low $k$ and $k'$ states are more consequential. 
After investigating the behavior of our integrals and ensuring their behavior is correct, we generate a spectrum. 

\subsection{Spectrum}
\label{ss:spec}

The final calculation comes from numerically integrating Paper I, Eq.~(80). This calculation is also done via midpoint integration method. Each term is separately calculated, and also subdivided by their even and odd contributions for further analysis. Additionally, each $\ell$ contribution is calculated separately, and the cumulative $\ell$ spectrum is calculated after all terms and integrals are separately considered.

In order to reduce the accumulated error of our calculation and balance computational load, we evaluate the photon spectra only at some values of $\omega$, rather than every $0.01T_{\rm H}$. We choose the sampling rate based on the expected interpolation error. There are two regimes:

\begin{list}{$\bullet$}{}
\item
For a power law $\omega ^{n}$, the fractional error induced by linear interpolation to a point halfway in between the samples is
\begin{equation}
    \text{frac.~error} = 
    \frac{[(\omega+\Delta\omega)^n+\omega^n]/2}{(\omega+\Delta\omega/2)^n} - 1 \approx
     \frac{(n-1)n}{8} \left(\frac{\Delta \omega}{\omega} \right)^2.
\end{equation}
In the low frequency limit, we expect the spectrum to behave as  $\omega^{-1}$ ($n=-1$), so to have a 1 \% error, we have to restrict our spacings such that 
    $\Delta \omega < 0.2\omega$.
 \item  At some point, we expect the behavior of the spectrum to deviate and behave such that the behavior is $\omega e^{-\omega/T_{\rm H}} $, for which the fractional error is
 \begin{equation}
    \text{frac.~error} = \frac{[e^{-(\omega+\Delta\omega)/T_{\rm H}}+e^{-\omega/T_{\rm H}}]/2}{e^{-(\omega+\Delta\omega/2)/T_{\rm H}}} - 1
    \approx \frac18 \left(\frac{\Delta \omega}{T_{\rm H}}\right)^2.
\end{equation}
For 1\% interpolation error, this suggests that we use
$\Delta \omega < 0.283 T_{\rm H}$.
\end{list}

The two spacing formulae cross each other at 
$ h\sim 1.4 T_{H}$. At the very highest energies, we sample every $0.5T_{\rm H}$ so that we cut down on computational load since we expect the spectrum to be subdominant in this regime and thus we accept the larger ($\sim 3\%$) error. 

That is,
\begin{equation}
\begin{array}{cccl}
    \Delta \omega &<& 0.2\omega & \text{,      }\omega<1.25T_{H},\\
    \Delta \omega &=& 0.25T_{H} &\text{, }1.25T_{H}<\omega<8.0T_{H},~~~{\rm and}\\
    \Delta \omega &=& 0.5T_{H} &\text{, } \omega>8.0T_{H}.
\end{array}
\end{equation}

As a consequence of computing our spectrum for a finite range of $r_{*}$, a correction factor is needed in order to calculate results for the analytically infinite $r_{*}$ integration range. A detailed discussion of this correction factor can be found in Appendix~\ref{app:outer}. We divide our results by the correction factor in post-processing of the spectra.

\section{Results}
\label{sec:results}

\begin{figure}
    \centering
    \includegraphics[width=7.0in]{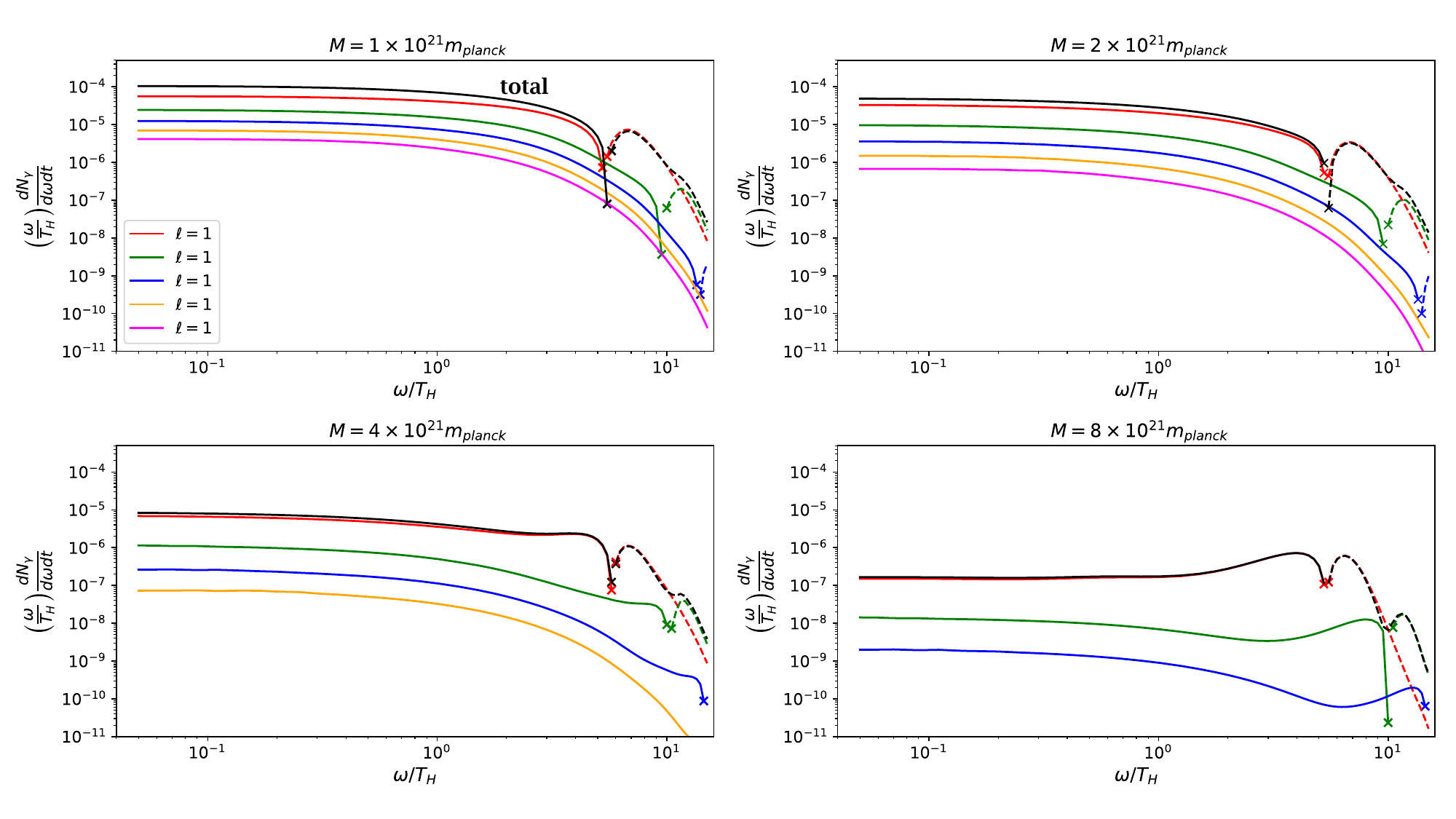}
    \caption{Results for first order correction to the photon energy spectrum. Here we show the total spectrum summed over $\ell$ (top black line), along with the different $\ell$ contributions (colors, with the highest $\ell$ generally at the bottom). The solid lines represent where the correction contributes positively to the overall spectrum, and the dashed lines show the regions where the contributions are negative. We indicate the frequencies where there is a zero-crossing in the spectrum contributions by an 'x' marker on the highest frequency where the contribution is positive and on the lowest frequency contribution that is negative.}
    \label{fig:ResultsPerL}
\end{figure}

\begin{figure}
    \centering
    \includegraphics[width=5.5in]{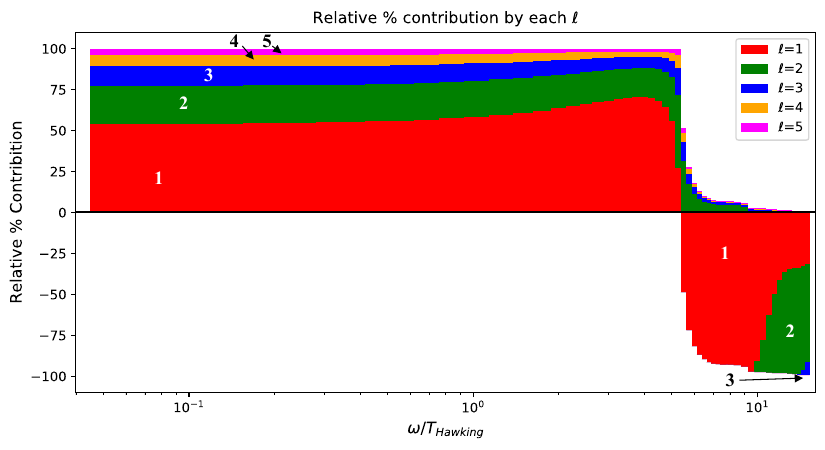}
    \caption{Convergence of results as more multipole modes are added for $M=10^{21}M_{\rm Planck}$. The $\ell=5$ term contributes 4\% at low $\omega/T_{\rm H}$. This is the most extreme of the 4 cases; the higher masses exhibit faster convergence with $\ell$.}
    \label{fig:M1LConvergence}
\end{figure}

The results for our first order dissipative corrections to the Hawking radiation photon spectrum for each of the four masses of interest can be found in Figure~\ref{fig:ResultsPerL}. This data is broken down by $\ell$ value, and differentiates which contributions are contribute positively or negatively to the overall spectra. At low frequencies, the inner bremsstrahlung terms are most important; these are distributed among the values of $\ell$, up to values of order the Lorentz factor of the emitted electron (see Appendix~\ref{app:classical}). At larger PBH masses, the Hawking temperature is lower, and so the peak Lorentz factor of the electrons is reduced; by the time we get to $M=8\times 10^{21}M_{\rm Planck}$, most of the emitted electrons are nonrelativistic. In this case, $\ell=1$ accounts for almost all of the inner bremsstrahlung, since the multipole expansion is also an expansion in powers of $R/\lambda \sim v/c$, where $R$ is the scale of the emission region, $\lambda$ is the wavelength of the emitted radiation, and $v$ is the velocity of the charged particle. However, at the highest frequencies (still in units of $T_{\rm H}$), larger multipoles can still contribute. We also find that the ${\cal O}(\alpha)$ dissipative correction changes to absorption.

This behavior can also be seen in Fig.~\ref{fig:M1LConvergence}, where the contributions from each $\ell$ are broken down in the case of $M=10^{21}M_{\rm Planck}$. Here we see the rate of convergence of the spectrum with $\ell$, with the higher $\ell$s being most important at low frequency; $\ell=5$, for example, contributes 4\%. This rate of convergence makes sense in the context of Appendix~\ref{app:classical}: the peak of the electron spectrum in this case is at $h \approx 4\mu$ (Lorentz factor 4), where Appendix~\ref{app:classical} predicts that 3.3\% of the contribution should come from $\ell=5$. 

\begin{figure}
    \centering
    \includegraphics[height=8in]{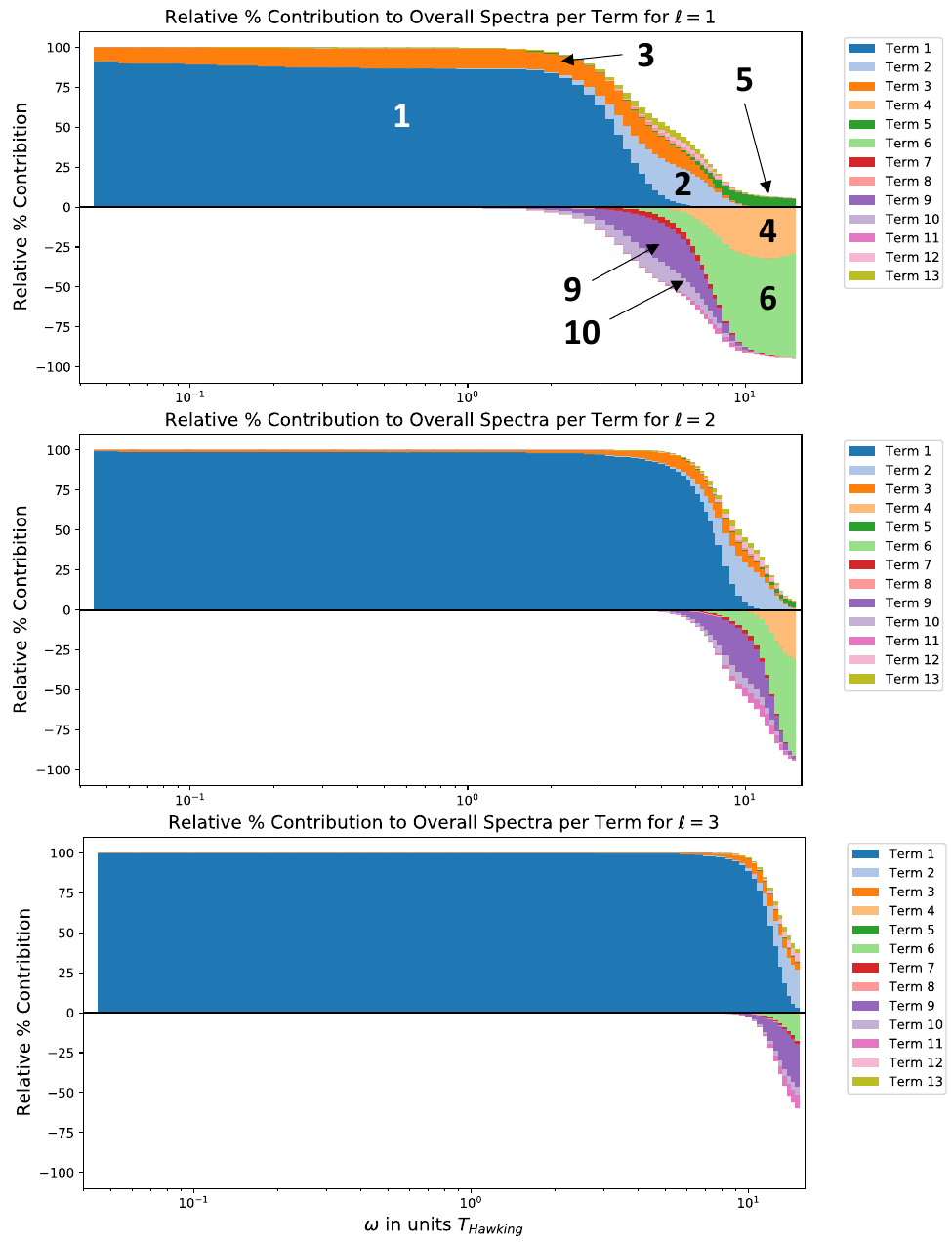}
    \caption{The breakdown of the contributions to the ${\cal O}(\alpha)$ dissipative corrections for $M=10^{21}M_{\rm Pl}$ into the different terms. The panels show $\ell=1$, 2, and 3 (top to bottom). The percentage contributions to the absolute value are shown, with positive contributions above the horizontal axis and negative contributions below. At low frequencies, the inner bremsstrahlung terms (1 and 3) dominate, while at high frequencies the negative contribution from pair production (terms 4 and 6) is most important. The transition between the different contributions moves to higher frequencies as $\ell$ is increased.}
    \label{fig:TermPerL}
\end{figure}

To understand the transition from positive to negative contributions better, we looked at the contributions of each term to the overall spectra for each $\ell$ value for each mass. This is displayed in Fig.~\ref{fig:TermPerL}. At the lowest frequencies, terms 1 and 3 dominate: these are the inner bremsstrahlung terms,
\begin{equation}
e^\pm_{\rm up} \rightarrow e^\pm_{\rm in,up} + \gamma.
\end{equation}
(The excess of term \#1 to \#3 indicates that the final electron is most likely to be in the ``in'' mode, but we should recall that ``in/up'' is a different basis choice than ``down/out,'' the latter determining the fate of the outgoing electron.) At the highest frequencies, terms 4 and 6 dominate: these are the pair production terms,
\begin{equation}
\gamma_{\rm up} \rightarrow e^+ + e^-.
\end{equation}
(The electrons may be in the ``in+in'' or ``in+up'' modes; recall that ``up+up'' is already in thermal equilibrium since all the ``up'' particles have the same Hawking temperature.) Near the crossover point ($\omega \approx 5.5T_{\rm H}$), terms 2 and 9 are larger: these involve pair annihilation,
\begin{equation}
e^-_{\rm up} + e^+_{\rm up} \leftrightarrow \gamma_{\rm in~or~up}.
\end{equation}
It matters to the emitted spectrum (photons in the ``out'' state) that the photon is emitted into a superposition of ``in'' and ``up:'' via the interference term, the $\gamma_{\rm up}$ diagram can contribute even though $e^-_{\rm up} + e^+_{\rm up} \leftrightarrow \gamma_{\rm up}$ is in equilibrium. But the fact that terms 2 and 9 have opposite sign and similar magnitude leads to a reduced overall effect on the spectrum.

\begin{figure}
    \centering
    \includegraphics[height=7in]{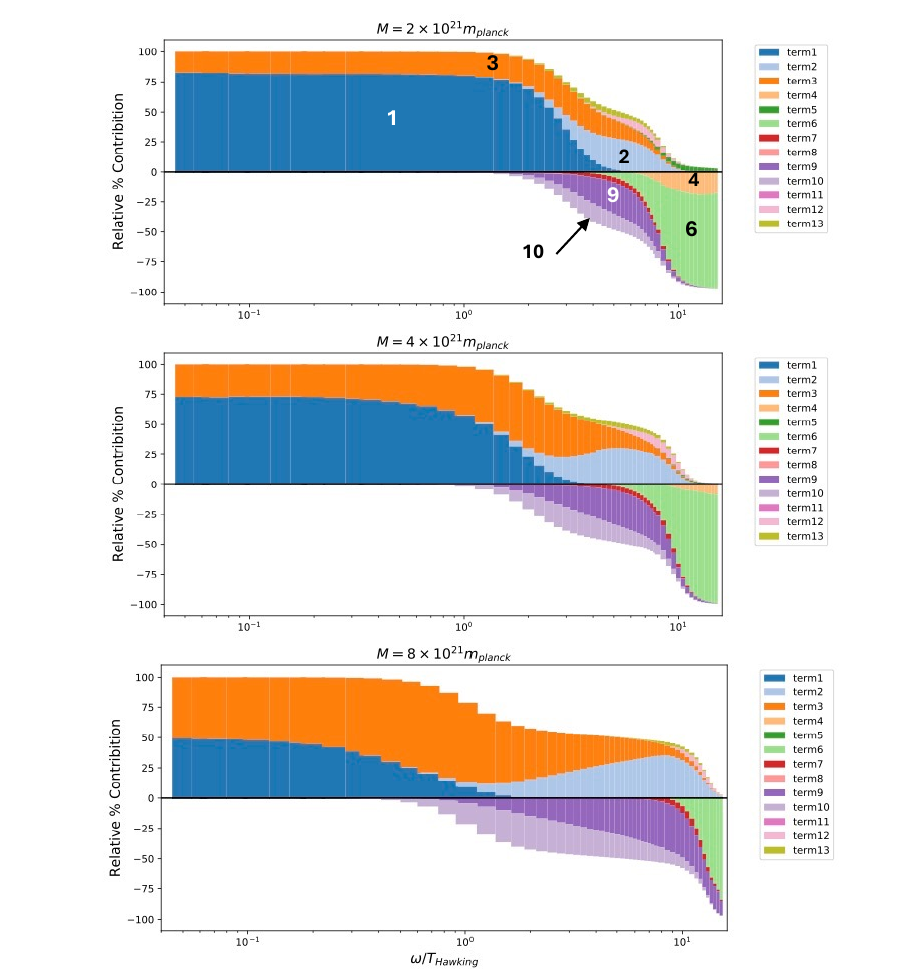}
    \caption{The breakdown of the contributions to the ${\cal O}(\alpha)$ dissipative corrections for $\ell =1$ in different masses into the different terms. The panels show $M=2\times 10^{21} m_{planck}$, $4\times 10^{21} m_{planck}$, and $8\times 10^{21} m_{planck}$ (top to bottom). The percentage contributions to the absolute value are shown, with positive contributions above the horizontal axis and negative contributions below. As the PBH mass increases, we see a decrease in the dominance of term 1 at the low photon energy range, as well as an extension of the region where terms 2, 9 and 10 are important.}
    \label{fig:TermPerM}
\end{figure}

We also see the behavior of the term contributions is PBH mass dependent, as demonstrated in Fig.~\ref{fig:TermPerM}. In particular, for the inner bremsstrahlung contribution, Term \#1 corresponds to the final electron in the ``in'' mode, whereas Term \#3 corresponds to the final electron in the ``up'' mode. Inner bremsstrahlung comes from an escaping electron in the ``out'' mode, but emission of even a soft photon can change the angular momentum $j$ (the length scale over which the photon is emitted gets longer as the photon gets softer). The ``out'' electron mode is a superposition of ``in'' and ``up,'' so the ratio of Term \#3 to Term \#1 should be $\sim |T_{\frac12,k,h}|^2/|R_{\frac12,k,h}|^2$ (see also Eq.~\ref{eq:C9}). The transmission coefficient is reduced as $\ell$ increases, so we expect the behavior from Fig.~\ref{fig:TermPerL} that Term \#3 is less important for high $\ell$. But we also expect the ratio of Term \#3 to Term \#1 to vary with $M$. At large $M$, we have $T_{\rm H}\ll\mu$, and most of the electrons that escape to $\infty$ are non-relativistic (near threshold: $h-\mu$ is of order $T_{\rm H}$). Also in this case, the transmission coefficients are close to 1 for small $j$ (see the final panel of Fig.~\ref{fig:RTCoeffsElectron}). This means that after emitting one unit of angular momentum, the electrons are still in a partial wave that has a large transmission coefficient. (In the large-$M$ limit, we could even make this statement semi-classically: an electron on a geodesic from the past horizon to future infinity is still on such a geodesic after emitting an angular momentum $\sim\hbar$.) We ascribe the trend of Term \#3 having a larger contribution for larger $M$ to this effect.

\begin{figure}
    \centering
    \includegraphics[width=6.75in]{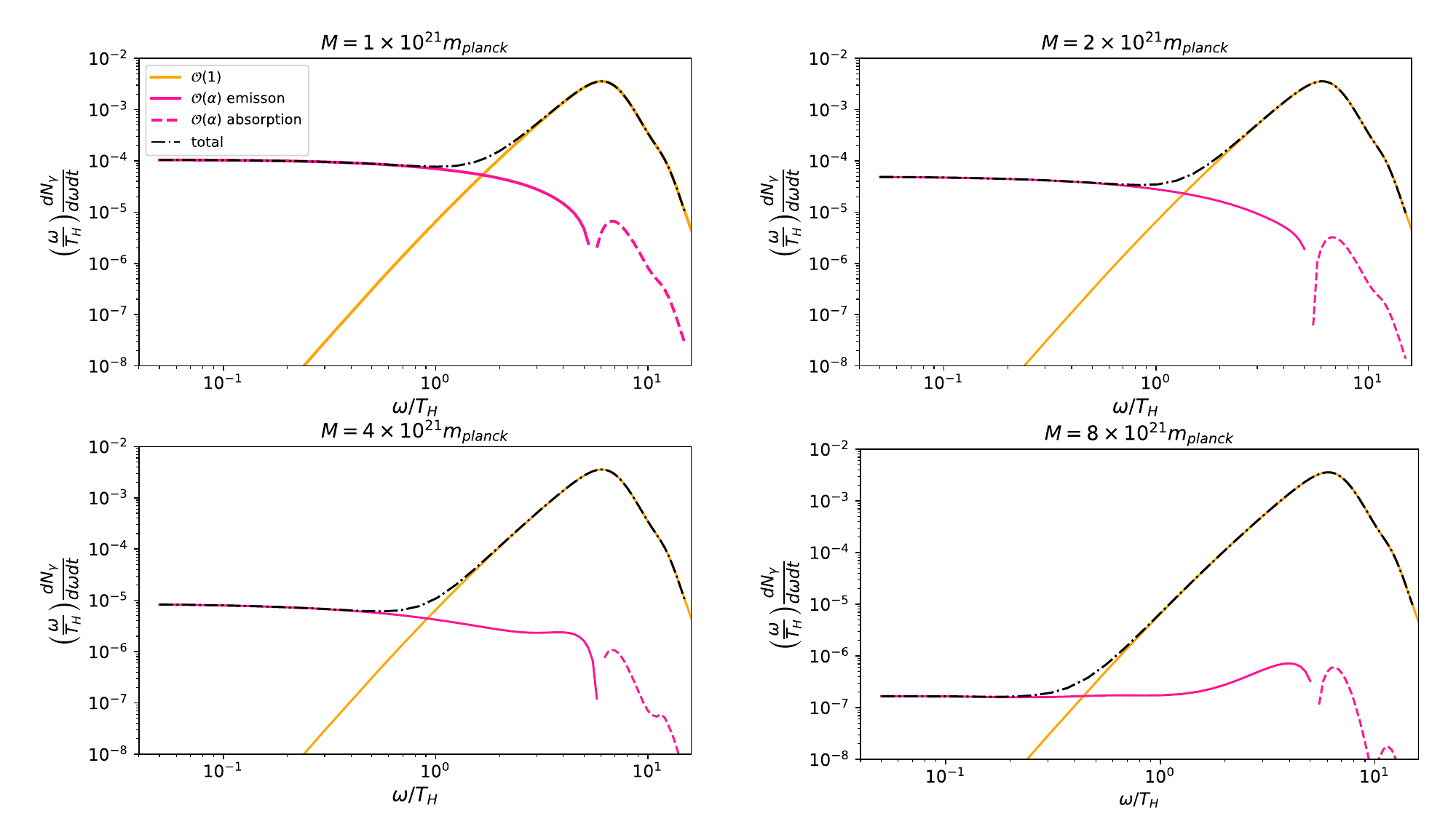}
    \caption{Results for zeroth and first order correction to the photon spectrum. At low frequencies, the dominant contribution to the spectrum is the ${\cal O}(\alpha)$ correction (pink) we compute here. The ${\cal O}(1)$ contribution (yellow) is dominant at high photon energies, and is more dominant as PBH mass increases. The region where the ${\cal O}(\alpha)$ correction is negative (indicated by dashed pink line) is in the region where the ${\cal O}(1)$ term is most significant.}
    \label{fig:Results}
\end{figure}

Figure~\ref{fig:Results} highlights the overall results from this study: the ${\cal O}(\alpha)$ dissipative correction to the photon emission spectrum from a Schwarzschild black hole. The free-field or ${\cal O}(1)$ contribution has the familiar graybody peak at $\omega \sim 6T_{\rm H}$. At low frequencies, the inner bremsstrahlung tail dominates, with the familiar power law (equal number of photons per logarithmic range in frequencies): even with the factor of $\alpha$, this exceeds what is possible with a graybody (where the small ``size'' of the black hole limits its ability to radiate as a dipole). Absorption terms appear at high frequency, but these are suppressed by $\sim 2$ orders of magnitude (due to the factor of $\alpha$), or even more at high masses (where electrons are Boltzmann-suppressed).

\begin{figure}
    \centering
    \includegraphics[width=3.78in]{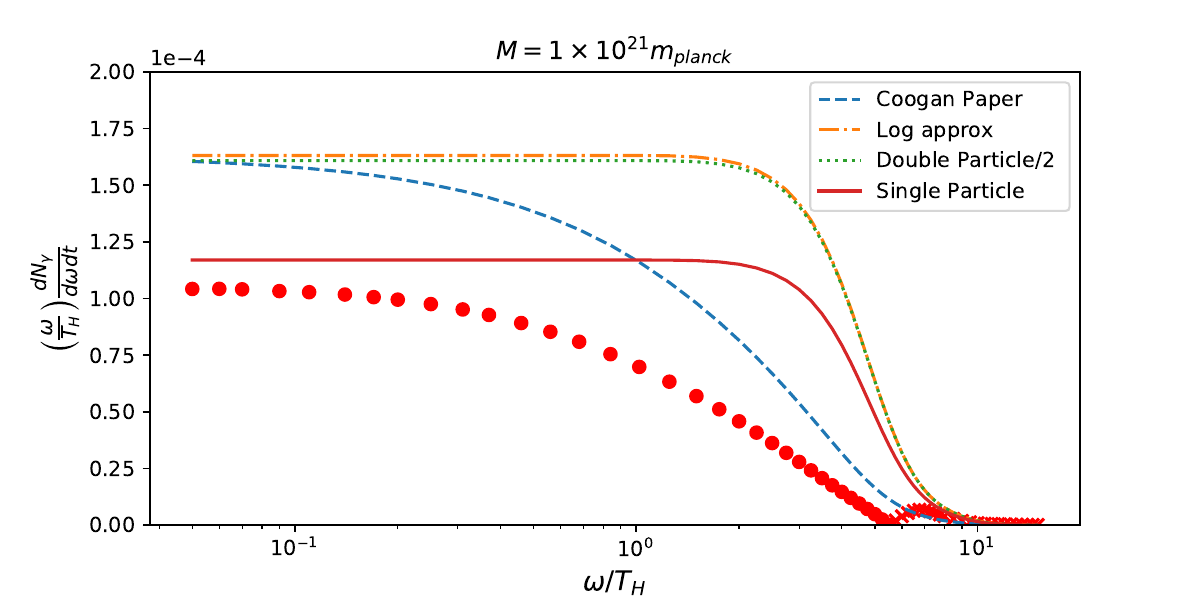}
    \hskip-0.55in
    \includegraphics[width=3.78in]{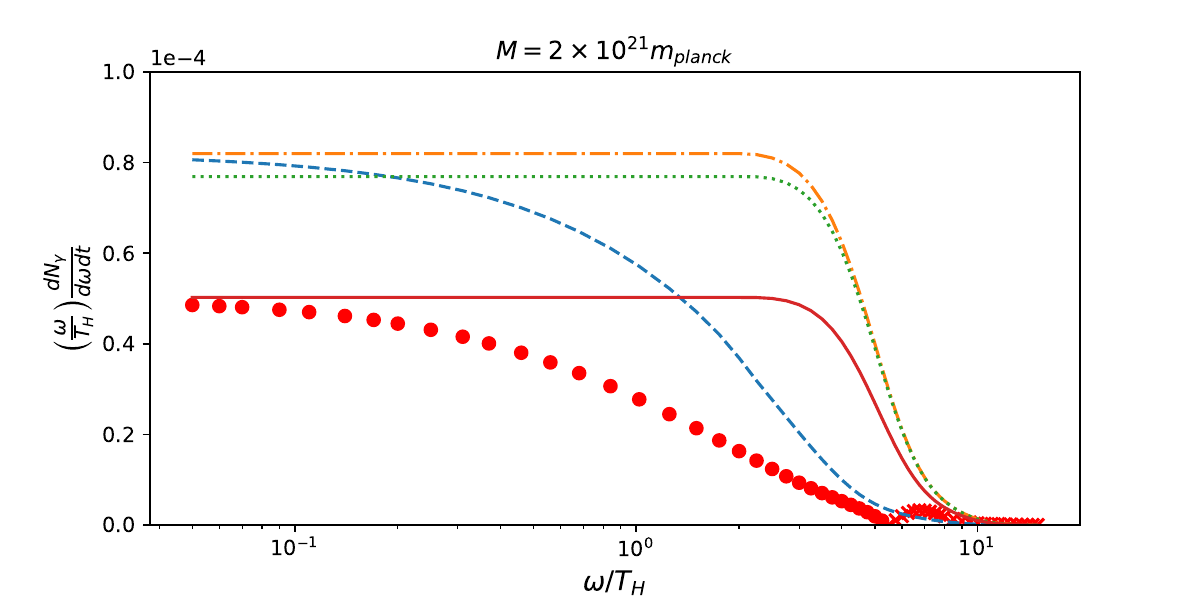}
    \\
    \includegraphics[width=3.78in]{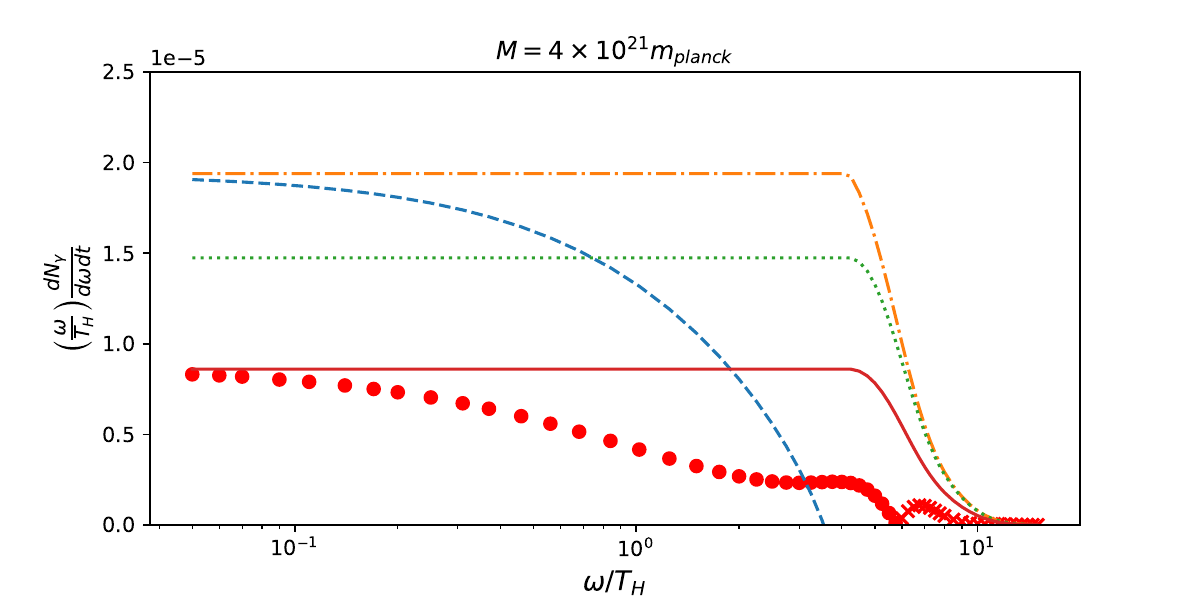}
    \hskip-0.55in
    \includegraphics[width=3.78in]{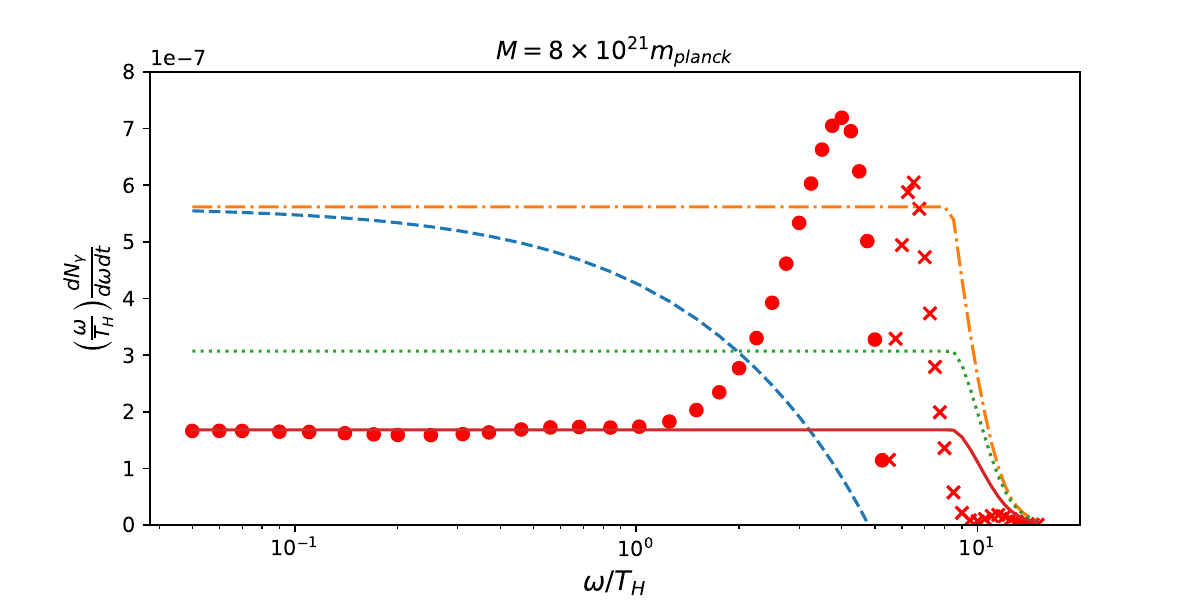}
    \caption{Comparison of different low frequency approximation schemes for the energy spectrum relative to the fully numerical calculation. There the red markers show our fully numerical results calculated in this paper, where the round dots are positive and the x markers are negative contributions. The red solid line shows the approximation in Eq.~(\ref{eq:S-single}) which is most similar to the calculation we perform. Shown in a green dotted line is the double particle/2 approximation from Eq.~(\ref{eq:S-coogan}). The blue dashed line shows the energy spectra using the full expression in Eq.~(6) of \citet{coogan2021}, and the green dotted-dashed line shows the further approximation of the Coogan spectrum which we calculated in Eq.~(\ref{eq:S-double}). 
    }
    \label{fig:LowOmega}
\end{figure}

\section{Discussion}
\label{sec:discussion}

This work aimed to numerically compute the dissipative ${\cal O}(\alpha)$ correction to the the Hawking radiation spectrum for a range of asteroid-mass PBHs, the results of which are in Fig.~\ref{fig:Results}. This is the first step to a full computation of the ${\cal O}(\alpha)$ Hawking radiation calculation based on a perturbative QED calculation on a Schwarzschild background. In performing this calculation, we also gained an understanding of the types of interactive photon and electron/positron processes are most important at different energy ranges. We specifically see the importance of inner-bremsstrahlung interactions at low photon energy, and pair-production type processes at higher energies, though these processes' contributions are overshadowed by the ${\cal O}(1)$ greybody spectrum. We also see that the soft inner bremsstrahlung photons are more prominent for lower mass PBHs.

Figure~\ref{fig:LowOmega} puts this work in the context of the other approximation schemes in the literature for the inner bremsstrahlung from PBHs. A more detailed discussion of these approximation schemes can be found in Appendix~\ref{app:classical}, but a common theme we find is the low energy photon spectrum is typically overestimated in previous treatments. We also see that even amongst approximation schemes previously used, there are inconsistent asymptotic behaviors which are more prominent in higher mass PBHs. 
Our fully numerical calculation for all ${\cal O}(\alpha)$ dissipative corrections is shown at the red points.
The classical low-frequency approximation to the inner bremsstrahlung is shown in the red solid line. 
We do see that this approximation is closer to our numerical results for higher masses. At the lowest mass, there is a 11\% discrepancy between the single particle approximation and our data. We can account for 1.5\% difference from not being truly in the low $\omega$ limit (based on a linear fit of the percentage difference versus $\omega$), and have an expected 4.5\% error from multipoles $\ell>5$ (based on the expansions in Appendix~\ref{app:classical} at $h/\mu=4$, close to the peak of the electron spectrum). The remaining 5\% difference is likely a combination of numerical errors and contributions from the high-energy tail of the electron distribution (which have larger contributions from higher $\ell$).

This spectrum deviation between our work and other calculations at low photon energies could have some important implications for PBH mass bounds and forecasting of constraints of future surveys \cite{coogan2021, 
2022PhRvD.105j3026S, 2023PrPNP.13104040A}. While we do not revisit existing constraints in this work, we are interested in investigating the consequences of the ${\cal O}(\alpha)$ spectrum calculation for future constraints, especially given that the X-ray emission has been revised downward by factors of a few (depending on the PBH mass). 

We have split the ${\cal O}(\alpha)$ Hawking radiation calculation into the dissipative terms (which are the best studied and the subject of this paper) and the conservative terms. The conservative corrections include the plasma effects and the vacuum polarization effects which could have a significant contribution, but these conservative effects require implementation of renormalization techniques and therefore is still a work in progress.
We also are interested in examining the corrections to the electron Hawking radiation spectrum at ${\cal O}(\alpha)$ both to try to provide a thorough treatment and because the $e^\pm$ spectra are relevant to upcoming PBH searches from low-energy positrons and the 511 keV line \cite{2023JCAP...02..006C}.

Much work awaits to complete the calculation of the ${\cal O}(\alpha)$ corrections to Hawking radiation --- and this is for the simplest interacting field theory that is actually realized in Nature (QED), and is being carried out for the simplest black hole (Schwarzschild). Nevertheless, the numerical evaluation of the dissipative terms here is an important step toward understanding how quantum fields interact with each other near a black hole. 

\section*{Acknowledgements}

We thank Bowen Chen, Aditi Fulsundar, and Cara Nel for comments on the draft and presentation. Computations were performed on the Pitzer cluster at the Ohio Supercomputer Center \cite{Pitzer2018}. This project was supported by the David \& Lucile Packard Foundation award 2021-72096. C.H.\ additionally received support from the National Aeronautics and Space Administration, under subaward AWP-10019534 from the Jet Propulsion Laboratory. M.S. received support from the Los Alamos National Laboratory (LANL), operated by Triad National Security, LLC, under the Laboratory Directed Research and Development program of LANL project number 20230863PRD LA~-~UR: PENDING. This material is based upon work supported by the U.S. Department of Energy, Office of Science, Office of Workforce Development for Teachers and Scientists, Office of Science Graduate Student Research (SCGSR) program for E.K. The SCGSR program is administered by the Oak Ridge Institute for Science and Education for the DOE under contract number DE‐SC0014664.

\appendix 

\section{Comparison to classical soft inner bremsstrahlung in special relativity}
\label{app:classical}

This appendix reviews the classical special relativistic approach to inner bremsstrahlung. We work in spherical harmonic space, since one of our main goals is to predict the rate of convergence of the sum over multipoles ($\sum_{\ell=1}^\infty$) in the inner bremsstrahlung case. This has a long history of investigation in nuclear physics, particularly the photons radiated in beta decay \cite{1936Phy.....3..425K, 1936PhRv...50..272B}. A similar multipolar decomposition applied to a classical particle around a black hole can be found in Ref.~\cite{2003PhRvD..68h4011C} (in that case the particle falls into the hole rather than being emitted, but the soft radiation spectrum is the same).

\subsection{Classical calculation}

We consider the multipole distribution of the low-frequency inner bremsstrahlung emitted when a particle of charge $e$ is accelerated instantaneously from rest to velocity ${\bf v}$ at time $t=0$ in flat spacetime. At a distance $R$ from the source, the electric field is
\begin{equation}
\tilde{\bf E}_{\rm rad}(\omega, R\hat{\bf n}) = \int_{-\infty}^\infty {\bf E}(t)\,e^{i\omega t}\,dt
= \frac{e}{4\pi R} e^{i\omega R} \frac{\hat{\bf n} \times (\hat{\bf n}\times{\bf v})}{1-{\bf v}\cdot\hat{\bf n}}
\end{equation}
(this follows from Eqs.~14.64 and 14.66 of \citet{1998clel.book.....J}, but with a factor of $\sqrt{2\pi} / 4\pi$ due to our choice of units and conventions for Fourier transforms). If ${\bf v}$ is placed on the $z$-axis, and $\hat{\bf n}$ is expressed in polar coordinates (a distance $\theta$ from the $z$-axis), then this becomes
\begin{equation}
\tilde{\bf E}_{\rm rad}(\omega, R,\theta,\phi) 
= \frac{e}{4\pi R} e^{i\omega R} \frac{v\sin\theta\,\hat{\bf e}_\theta}{1-v\cos\theta}.
\label{eq:Erad}
\end{equation}
The number of photons radiated per unit frequency is then\footnote{The prefactor contains a $1/(2\pi)$ from Parseval's theorem; a 2 from considering both positive- and negative-frequency contributions; a $1/2$ from the electric field energy density $E^2/2$; a 2 from the fact that there is an equal density in magnetic field.}
\begin{equation}
\frac{dN_\gamma}{d\omega} = \frac1{\pi\omega} \oint_{S^2} |R{\bf E}_{\rm rad}(\omega, R,\theta,\phi)|^2\,d\Omega
\label{eq:dNdomega}
\end{equation}
where $d\Omega = \sin\theta\,d\theta\,d\phi$ is the solid angle element.
(again, this is Eq.~14.53 of \citet{1998clel.book.....J} but re-written according to our units, and with a factor of $\omega$ since we are counting photons instead of energy).
The integral is straightforward to evaluate and gives (using the substitution $\bar\mu = \cos\theta$ followed by partial fractions):
\begin{equation}
\frac{dN_\gamma}{d\omega}
= \frac{e^2v^2}{8\pi^2\omega} \int_{-1}^1 \frac{1-\bar\mu^2}{(1-v\bar\mu)^2}\,d\bar\mu
= \frac{e^2}{4\pi^2\omega} \left[ \frac1v\ln \frac{1+v}{1-v} - 2\right].
\label{eq:DB}
\end{equation}

The decomposition of the emitted radiation in multipoles is obtained by decomposing $R{\bf E}_{\rm rad}(\omega, R,\theta,\phi)$ in vector spherical harmonics. Since our problem is axisymmetric and ${\bf E}_{\rm rad}$ is in the $\theta$-direction, only the $m=0$ harmonics with electric parity contribute. The orthonormal vector harmonics are
\begin{equation}
{\bf X}_{\ell 0}(\theta,\phi) \equiv \frac{\nabla_{\rm ang}Y_{\ell 0}(\theta,\phi)}{\sqrt{\ell(\ell+1)}}
= -\sqrt{\frac{2\ell+1}{4\pi\ell(\ell+1)}} ~P'_\ell(\cos\theta)\,\sin\theta\,\hat{\bf e}_\theta,
\end{equation}
(where $\nabla_{\rm ang} = \hat{\bf e}_\theta \partial_\theta + \csc\theta \hat{\bf e}_\phi \partial_\phi$ denotes the gradient on the unit sphere)
and the corresponding decomposition is $R{\bf E}_{\rm rad}(\omega, R,\theta,\phi) = \sum_{\ell=1}^\infty a_\ell {\bf X}_{\ell 0}(\theta,\phi)$ with
\begin{equation}
a_\ell = \oint_{S^2} {\bf X}^\ast_{\ell 0}(\theta,\phi) R {\bf E}_{\rm rad}(\omega, R,\theta,\phi) \,d\Omega
= - \sqrt{\frac{2\ell+1}{16\pi\ell(\ell+1)}} ~ev e^{i\omega R} \int_{-1}^1
 P'_\ell(\bar\mu)\frac{1-\bar\mu^2}{1-v\bar\mu}\,d\bar\mu,
\end{equation}
where as usual we substituted $\bar\mu = \cos\theta$. The contribution of the $\ell$ multipole to the integral in Eq.~(\ref{eq:dNdomega}) is
$|a_\ell|^2$; then the contribution to $dN_\gamma/d\omega$ is
\begin{equation}
\left.\frac{dN_\gamma}{d\omega}\right|_\ell =
\frac{2\ell+1}{16\pi^2\ell(\ell+1)} \frac{e^2v^2}{\omega} \left[ \int_{-1}^1
 P'_\ell(\bar\mu)\frac{1-\bar\mu^2}{1-v\bar\mu}\,d\bar\mu \right]^2
 \equiv \frac{2\ell+1}{16\pi^2\ell(\ell+1)} \frac{e^2v^2}{\omega} \varsigma^2_\ell,
\label{eq:dnl1}
\end{equation}
where we denote the integral in brackets by $\varsigma_\ell(v)$.

Analytic forms are possible for the first few such functions, e.g.:
\begin{eqnarray}
\varsigma_1(v) \!\! &=& \!\!\frac{2}{v^2} - \frac{1-v^2}{v^3}\ln\frac{1+v}{1-v} = \sum_{\sigma=0}^\infty \frac{4}{(2\sigma+1)(2\sigma+3)} v^{2\sigma}
~~~{\rm and}
\nonumber \\
\varsigma_2(v) \!\! &=& \!\! 
\frac{6}{v^3} - \frac4v - \frac{3(1-v^2)}{v^4}\ln\frac{1+v}{1-v}
= \sum_{\sigma=1}^\infty \frac{12}{(2\sigma+1)(2\sigma+3)} v^{2\sigma-1}.
\label{eq:zeta-12}
\end{eqnarray}
For higher functions it is more convenient {\em not} to use partial fractions, but rather to derive a recursion relation by using the associated Legendre polynomials:
\begin{equation}
\varsigma_\ell(v) = - \int_{-1}^1 \frac{\sqrt{1-\bar\mu^2} P_\ell^{1}(\bar\mu)
}{1-v\bar\mu}\,d\bar\mu.
\end{equation}
Using the recursion relation
\begin{equation}
(\ell+1) P_{\ell-1}^1(\bar\mu) + \ell P_{\ell+1}^1(\bar\mu)
=(2\ell+1)\bar\mu P_\ell^1(\bar\mu) ,
\end{equation}
(where it is understood that $P_0^1(\bar\mu)=0$) we see that
\begin{equation}
v[(\ell+1) \varsigma_{\ell-1}(v) + \ell \varsigma_{\ell+1}(v)]
= -(2\ell+1) \int_{-1}^1 \frac{\sqrt{1-\bar\mu^2} \,v\bar\mu P_\ell^{1}(\bar\mu)
}{1-v\bar\mu}\,d\bar\mu.
\end{equation}
Subtracting $(2\ell+1)\varsigma_\ell(v)$ from both sides gives
\begin{equation}
-(2\ell+1)\varsigma_\ell(v) +
v[(\ell+1) \varsigma_{\ell-1}(v) + \ell \varsigma_{\ell+1}(v)]
= (2\ell+1) \int_{-1}^1 \sqrt{1-\bar\mu^2} \,P_\ell^{1}(\bar\mu) \,d\bar\mu
= -4\delta_{\ell,1}.
\end{equation}
Thus we find that
\begin{equation}
\varsigma_{\ell+1}(v) = \frac{2\ell+1}{\ell v}\varsigma_\ell(v) - \frac{\ell+1}{\ell} \varsigma_{\ell-1}(v)~~~{\rm for}~\ell\ge 2.
\label{eq:zeta-rec}
\end{equation}
Use of these relations, initialized with $\varsigma_1(v)$ from Eq.~(\ref{eq:zeta-12}), becomes numerically unstable for small $v$ and large $\ell$; however, the use of the Taylor series at $v<0.1$ and double precision leads to stability up through $\ell=8$ at all velocities.

\begin{figure}
\includegraphics[width=5in]{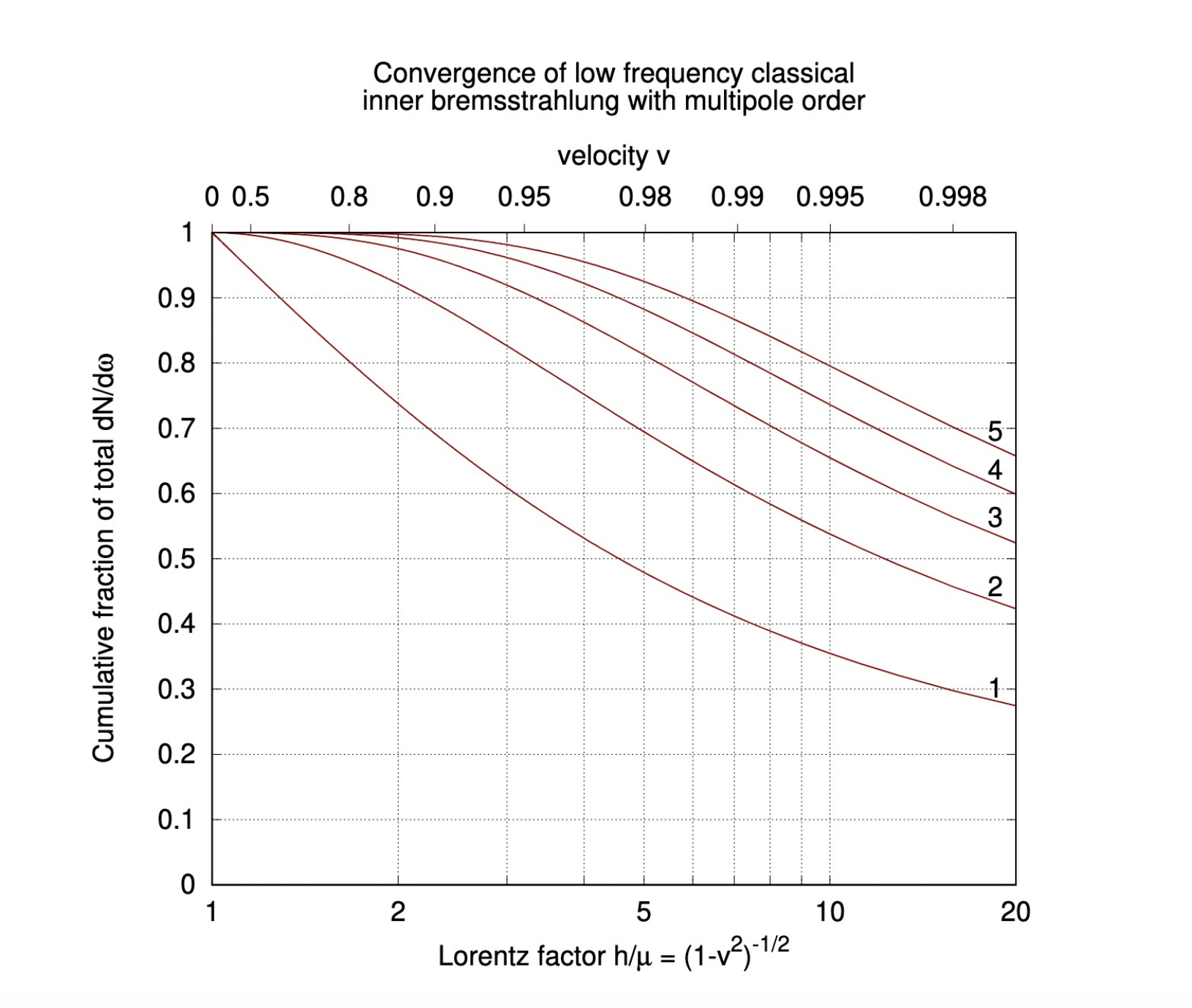}
\caption{\label{fig:icont}The fraction of the total $dN_\gamma/d\omega$ (Eq.~\ref{eq:dNdomega}) contributed by multipoles up to each order, according to the classical soft inner bremsstrahlung calculation (small $\omega$). For the non-relativistic case (small Lorentz factors), essentially the whole contribution comes from the dipole term ($\ell=1$): this is 90\% at $v=0.63$ or Lorentz factor 1.29. The first 5 terms (i.e., up through $\ell=5$) contribute 90\% of the total at $v=0.984$ or Lorentz factor 5.6. 
}
\end{figure}

The fraction of the contribution to $dN_\gamma/d\omega$ coming from the first few $\ell$s is shown in Fig.~\ref{fig:icont}.

\subsection{Behavior at large Lorentz factor}

We may also examine the limiting case where $v$ is close to 1: $v = (1-\gamma^{-2})^{-1/2}$ where $\gamma\gg 1$. To begin, we use polynomial division to re-write the defining integral as:
\begin{equation}
\varsigma_\ell = \int_{-1}^1 \left[ 1 + v\bar\mu + \frac{\bar\mu^2}{\gamma^2(1-v\bar\mu)} \right] P'_\ell(\bar\mu)\,d\bar\mu
= \int_{-1}^1 \frac{\bar\mu^2}{\gamma^2(1-v\bar\mu)} P\prime_\ell(\bar\mu)\,d\bar\mu.
\label{eq:a-transform}
\end{equation}
In the second equality, we eliminated the $1+v\bar\mu$ term since we can integrate by parts and use Legendre polynomial orthogonality to show that its contribution is zero. The remaining integral is dominated by the region where $\bar\mu\approx 1$, since then the denominator $1-v\bar\mu$ is close to zero and (after the transformation leading to Eq.~\ref{eq:a-transform}) the numerator has no zero at $\bar\mu=1$. Writing $v\approx 1 - \frac12\gamma^{-2}$, we find that to lowest order in $\gamma^{-1}$ and $\theta$ we have $1-v\bar\mu \approx \frac12\gamma^{-2} + \frac12\theta^2$. Then:
\begin{equation}
\varsigma_\ell \approx 2 \int_0^\pi \frac{1}{1+\gamma^2\theta^2} P'_\ell(\cos\theta)\, \theta\,d\theta.
\end{equation}
The large-$\ell$ expansion of the Legendre polynomial is $P_\ell(\cos\theta) \approx J_0(\ell\theta)$ (Ref.~\cite{1972hmfw.book.....A}, Eq.~9.1.71), implying that $-\theta P'_\ell(\cos\theta) \approx -\ell J_1(\ell\theta)$ (using the chain rule on both sides). Then with the substitution $x = \ell\theta$, we find
\begin{equation}
\varsigma_\ell \approx 2 \int_0^\infty \frac{1}{1 + (\gamma/\ell)^2 x^2} J_1(x) \,dx
= -2 \frac\ell\gamma K_1\left(\frac\ell\gamma\right),
\label{eq:ze}
\end{equation}
where we used the Hankel-Nicholson integral for the Bessel functions (Ref.~\cite{1972hmfw.book.....A}, Eq.~11.4.44) and $K_1$ is the modified Bessel function. This has a universal shape at large $\gamma$, being roughly constant at $\varsigma_\ell \approx -2$ for $\ell\ll \gamma$ but then approaching zero exponentially at $\ell>\gamma$. The number of photons emitted per unit frequency then goes to
\begin{equation}
\left. \frac{dN_\gamma}{d\omega}\right|_\ell \rightarrow \frac{e^2\ell}{2\pi^2\gamma^2\omega}
\left[K_1\left(\frac\ell\gamma\right) \right]^2
,~~~~
\gamma = \frac1{\sqrt{1-v^2}}\gg 1, ~~~\ell\gg 1.
\end{equation}
If one sums over $\ell$, one encounters a logarithmically large contribution since at $1\ll \ell\ll \gamma$ the contribution of each $\ell$ is $\approx e^2/(2\pi^2\omega\ell)$ and hence the sum is $\sim (e^2/2\pi^2\omega) \ln \gamma$. This is the expected limiting form (see Eq.~\ref{eq:DB}). The logarithmic behavior in summing over $\ell$ and the logarithmic divergence if we integrate over $\omega$ are the classical multipole-space description of the well-known double logarithmic asymptotics in gauge theory \cite{Sudakov56, Abrikosov56}.

The corollary to this discussion is that as one goes to high Hawking temperatures (small black hole masses), one must take many more terms in $\ell$ to achieve convergence, up through $\ell \sim 4T_{\rm H}/\mu$ (the Lorentz factor of electrons at the peak of the graybody distribution). For this reason, in this paper we have limited our numerical studies to masses $M \ge 10^{21}$ Planck masses.

\subsection{Comparison to other inner bremsstrahlung formulae}
\label{ss:comparison}

\begin{figure}
\includegraphics[width=6in]{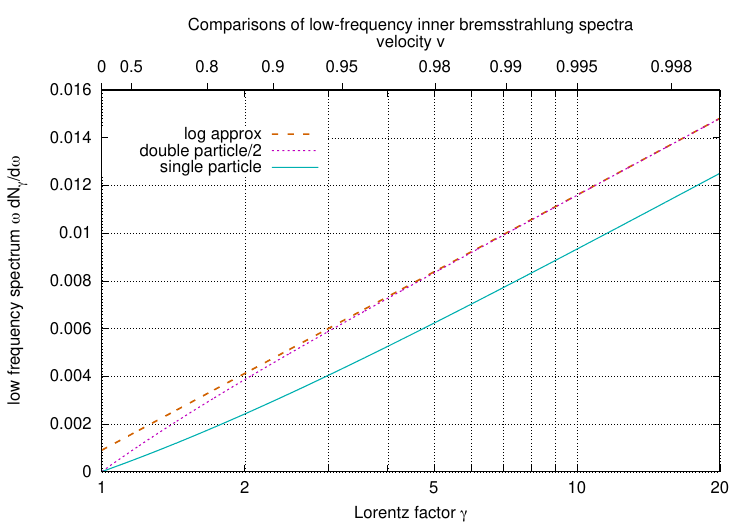}
\caption{\label{fig:doublelog}A comparison of three inner bremsstrahlung formulae in the low frequency limit. The ``single particle'' curve, Eq.~(\ref{eq:S-single}), is for emission from a single particle and is most appropriate for a PBH or for nuclear beta decay. The ``double particle/2'' curve, Eq.~(\ref{eq:S-double}), corresponds to emission from a particle-antiparticle pair emitted back-to-back, as in a neutral$\rightarrow \ell^+\ell^-$ process. The ``log approx'' formula, Eq.~(\ref{eq:S-coogan}), is that used in \cite{coogan2021}; it is the leading approximation to the double particle case at large Lorentz factor, and is quite good down to $\gamma\approx 2$.}
\end{figure}

We have derived the inner bremsstrahlung formula for a single particle,
\begin{equation}
\lim_{\omega\rightarrow 0^+} \omega \frac{dN_\gamma}{d\omega}
= \frac{e^2}{4\pi^2} \left[ \frac1v \ln \frac{1+v}{1-v} - 2 \right].
\label{eq:S-single}
\end{equation}
However, other treatments exist in the literature. The splitting function used \citet{coogan2021} is appropriate for large Lorentz factors ($v\rightarrow 1$) because it was derived in the limit of the electron having small mass; it corresponds to
\begin{equation}
\lim_{\omega\rightarrow 0^+} \omega \frac{dN_\gamma}{d\omega}
= \frac{\alpha}{\pi} \left[ 2 \ln \frac{2h}{\mu} - 1\right]
= \frac{e^2}{4\pi^2} \left[ \ln \frac{4}{1-v^2} - 1\right],
\label{eq:S-coogan}
\end{equation}
where $h$ is the final energy per fermion. Finally, one has the classical formula for inner bremsstrahlung for back-to-back electron and positron of the same velocity. This is not relevant for Hawking radiation, but it is appropriate for dark matter annihilation $\chi\chi\rightarrow\ell^+\ell^-$ and thus has attracted much attention in investigations of final state radiation for indirect dark matter detection \cite{2020JCAP...01..056C}. In this case, the radiated electric field becomes (with a particle of charge $e$ emitted along the $+z$ axis and a particle of charge $-e$ emitted along the $-z$ axis), instead of Eq.~(\ref{eq:Erad}),
\begin{equation}
\tilde{\bf E}_{\rm rad}(\omega,R,\theta,\phi) = \frac{e}{4\pi R} e^{i\omega R} \left( \frac{1}{1-v\cos\theta} + \frac{1}{1+v\cos\theta} \right) v\sin\theta\,\hat{\bf e}_\theta.
\label{eq:Erad-backtoback}
\end{equation}
Propagating this through to the emitted photon spectrum, and dividing by 2 to get the emitted photons per fermion, gives
\begin{align}
\lim_{\omega\rightarrow 0^+} \omega \frac{dN_\gamma}{d\omega}
&= \frac{1}{2\pi} \oint_{S^2} |R\tilde{\bf E}_{\rm rad}(\omega,R,\theta,\phi)|^2\,d\Omega & {\rm see~Eq.~(\ref{eq:dNdomega})} \nonumber \\
& = \frac{e^2v^2}{16\pi^2} \int_{-1}^1 \left( \frac1{1-v\bar\mu} + \frac1{1+v\bar\mu} \right)^2 (1-\bar\mu^2)\,d\bar\mu & {\rm see~Eq.~(\ref{eq:DB})} \nonumber \\
& = \frac{e^2}{4\pi^2} \left[
\frac{1+v^2}{2v} \ln \frac{1+v}{1-v} - 1
\right]. & {\rm (partial~fractions)}
\label{eq:S-double}
\end{align}
These results are compared in Fig.~\ref{fig:doublelog}. The log-divergent behaviors ($\ln\gamma$) all agree. However, the double particle case has a slightly larger constant offset, because of interference of the final state radiation from the two charged leptons. At high Lorentz factors, most of the radiation from each particle is forward-beamed in a cone of opening angle $\sim\gamma$, but this radiation is only logarithmically enhanced relative to the dipole component, and the dipole components add coherently. In the non-relativisitic limit, $v\ll 1$, the ``double particle/2'' curve is twice the ``single particle'' curve because in that case there is only a dipole component to the emission: the emission of two particles of opposite sign back-to-back leads to twice the amplitude and $4\times$ the energy, so the emitted number of photons per lepton is twice the single-particle case. The logarithmic approximation of Eq.~(\ref{eq:S-coogan}) contains a constant subtracted term that makes it asymptotically very accurate for large Lorentz factor, but it is still an overestimate in the non-relativistic regime.

For low black hole masses, the peak of the Hawking radiation corresponds to relativistic particles, $4T_{\rm H}\gg \mu$, and all the variants of the inner bremsstrahlung formulae are equivalent. But at higher black hole masses, $4T_{\rm H}\lesssim \mu$, there are more non-relativistic electrons, and the other variants of the formulae will overestimate the inner bremsstrahlung (by a factor of 2 for ``double particle/2'' and by a factor that diverges for ``log approx'').

\section{Correspondence of the quantum and classical results}
\label{app:corr}

This appendix considers how the QED treatment of inner bremsstrahlung can be reduced to the classical result of Appendix~\ref{app:classical} by taking appropriate limits. The key is to consider soft photons, where $\omega$ is small compared to both $T_{\rm H}$ and the energy of the electron involved ($h$ or $h'$).

For photons with $\omega\ll T_{\rm H}$, the ``in'' photon mode function essentially lives in flat spacetime and becomes a spherical Bessel function with unit reflection probability. We may also take $h+\omega = h'\approx h$ (except in the phase of a complex exponential). Finally, the overlap integrals $I^{-+}_{X,{\rm up},{\rm in}}$ are dominated by the region far from the black hole. This leads to a series of simplifications, showing that in an appropriate limit (soft photons and neglecting Pauli blocking) the classical result is recovered --- including the multipole distribution seen in Fig.~\ref{fig:icont}.

Let's consider the limit of soft photons and work in regions far from the black hole. In this case, we may take
\begin{equation}
\Psi_{{\rm in},\ell,\omega}(r_\star) = 2 (-i)^{\ell+1} \omega r j_\ell(\omega r_\star)
\end{equation}
and $|R_{1,\ell,\omega}|\rightarrow 1$ due to the enormous angular momentum barrier.

The electron wave functions are
\begin{equation}
\left( \begin{array}{c} F_{\rm up} \\ G_{\rm up} \end{array} \right)
= T_{\frac12,k,h} v^{-1/2} \left( \begin{array}c \sqrt{h+\mu} \\ -i \sqrt{h-\mu} \end{array} \right)
\left( \frac{r_\star}{2M}\right)^{i\zeta} e^{ihv r_\star}
\end{equation}
and
\begin{equation}
\left( \begin{array}{c} F_{\rm in} \\ G_{\rm in} \end{array} \right)
= v^{-1/2} \left( \begin{array}c \sqrt{h+\mu} \\ i \sqrt{h-\mu} \end{array} \right)
\left( \frac{r_\star}{2M}\right)^{-i\zeta} e^{-ihv r_\star}
+ R_{\frac12,k,h} v^{-1/2} \left( \begin{array}c \sqrt{h+\mu} \\ -i \sqrt{h-\mu} \end{array} \right)
\left( \frac{r_\star}{2M}\right)^{i\zeta} e^{ihv r_\star},
\end{equation} where $\zeta = \mu^2M/(hv)$ and $v=\sqrt{h^2-\mu^2}\,/h$.

Now we take the part of the $I$-integral that is slowly oscillating (i.e., keeping $e^{\pm i(hv-h'v')r_\star}$ but not $e^{\pm i(hv+h'v')r_\star}$). We have $h'=h+\omega$ (exactly), so we approximate $h\approx h'$ except in the oscillatory parts and transmission/reflection coefficients. We further use the fact that $|hv-h'v'|>\omega$. This leads to:
\begin{eqnarray}
\!\!\!\!\!\!\!\!
&&
\llbracket I^{-+}_{{\rm in}k,{\rm up}k',{\rm in}\ell(e)}(h,h',\omega)\rrbracket \rightarrow
-2 (-i)^{\ell+1} \llbracket\Delta^{kk'\ell}\rrbracket R_{\frac12,-k,h} T^\ast_{\frac12,-k',h'} \int_{{\rm few}\times M}^\infty
e^{i(hv-h'v')r_\star}
 j_\ell(\omega r_\star) \frac{\sqrt{\ell(\ell+1)}}{r\sqrt{2\omega}}
\,dr_\star
\nonumber \\ \!\!\!\!\!\!\!\!
 && ~~~\approx 
i (-1)^{\ell-j'+1/2}
\sqrt{\frac{(2j+1)(2j'+1)(2\ell+1)\ell(\ell+1)}{2\pi\omega}} \tj{j}{j'}{\ell}{\frac12}{-\frac12}{0} \delta_{ss',(-1)^{j-j'+\ell}} R_{\frac12,-k,h} T^\ast_{\frac12,-k',h'} 
\Xi_{\ell}\left( \frac{h'v'-hv}{\omega} \right),
\nonumber \\  \!\!\!\!\!\!\!\! &&
\end{eqnarray}
where we define the integral (for $y>1$):
\begin{align}
\Xi_{\ell}(y) ~& \equiv (-i)^\ell \int_0^\infty e^{-iy x} j_\ell(x)\,\frac{dx}{x}
\nonumber \\
&= \frac12 \lim_{\varepsilon\rightarrow 0^+} \int_\varepsilon^\infty e^{-iy x} \int_{-1}^1 e^{-ix\bar\mu} P_\ell(\bar\mu)\,d\bar\mu\,\frac1x\,dx
& {\rm Ref.~\left[53\right],~Eq.~(10.1.14)~with~}\theta=\cos\bar\mu
\nonumber \\
&= \frac12 \lim_{\varepsilon\rightarrow 0^+} \int_{-1}^1 E_1(i(y+\bar\mu)\varepsilon) P_\ell(\bar\mu)\,d\bar\mu
& {\rm definition,~Ref.~\left[53\right],~Eq.~(5.1.1)}
\nonumber \\
&= \frac12 \lim_{\varepsilon\rightarrow 0^+} \int_{-1}^1 \left[-\gamma_{\rm E} - \frac{i\pi}2 - \ln(y+\bar\mu) - \ln\varepsilon\right] P_\ell(\bar\mu)\,d\bar\mu & {\rm Ref.~\left[53\right],~Eq.~(5.1.11)}
\nonumber \\
&= -\frac12  \int_{-1}^1 \ln(y+\bar\mu)\, P_\ell(\bar\mu)\,d\bar\mu
\nonumber \\
&= \frac1{2\ell(\ell+1)}  \int_{-1}^1 \ln(y+\bar\mu)\,\frac{d}{d\bar\mu}[ (1-\bar\mu^2) P'_\ell(\bar\mu) ] \,d\bar\mu & {\rm defining~diff.~eq.,~Ref.~\left[53\right],~Eq.~(8.1.1)}
\nonumber \\
&= -\frac1{2\ell(\ell+1)}  \int_{-1}^1 \frac{1-\bar\mu^2}{y+\bar\mu} P'_\ell(\bar\mu) \,d\bar\mu & {\rm int.~by~parts,~}1-\bar\mu^2=0{\rm ~at~}\bar\mu=\pm1
\nonumber \\
&= \frac{(-1)^{\ell}}{2\ell(\ell+1)}  \int_{-1}^1 \frac{1-\bar\mu^2}{y-\bar\mu} P'_\ell(\bar\mu) \,d\bar\mu & P'_\ell(\bar\mu){\rm~is~odd~}(\ell~{\rm even)~or~even~}(\ell~{\rm odd})
\nonumber \\
&= \frac{(-1)^{\ell}}{2\ell(\ell+1)} y^{-1} \varsigma_\ell(y^{-1})
& {\rm def.~of~\varsigma_\ell,~Eq.~(\ref{eq:dnl1})},
\label{eq:Xi-def}
\end{align}
where $E_1$ is the exponential integral; $\gamma_{\rm E}$ is Euler's constant; and $\varsigma_\ell$ is the function defined in Appendix~\ref{app:classical}. With the further approximation that in the soft limit
\begin{equation}
\frac{h'v'-hv}{\omega} \approx \frac{h'v'-hv}{h'-h}
\approx \frac{d(hv)}{dh} = \frac1v,
\end{equation}
we arrive at
\begin{equation}
\llbracket I^{-+}_{{\rm in}k,{\rm up}k',{\rm in}\ell(e)}(h,h',\omega)\rrbracket \rightarrow
i (-1)^{j'-1/2}
\sqrt{\frac{(2j+1)(2j'+1)(2\ell+1)}{8\pi \ell(\ell+1)\omega }} \tj{j}{j'}{\ell}{\frac12}{-\frac12}{0} \delta_{ss',(-1)^{j-j'+\ell}} R_{\frac12,-k,h} T^\ast_{\frac12,-k',h'} v\varsigma_\ell(v).
\label{B7}
\end{equation}

Then term 1 from Paper I becomes:
\begin{eqnarray}
\left.\frac{dN^{(1)}}{d\omega\,dt}\right|_{{\rm term}~1,\ell,e} &\approx& \frac{e^2}{2\pi} \int \frac{dh}{2\pi} \sum_{kk'}
\Delta(j,j',\ell) \delta_{ss'(-1)^{k+k'+\ell},1} \frac{2}{e^{8\pi Mh}+1}
\nonumber \\ && \times
\frac{(2j+1)(2j'+1)(2\ell+1)}{8\pi \ell(\ell+1)\omega } \tj{j}{j'}{\ell}{\frac12}{-\frac12}{0}^2 |R_{\frac12,-k,h}|^2 |T_{\frac12,-k',h'}|^2 v^2\varsigma^2_\ell(v).
\label{eq:C8}
\end{eqnarray}
A similar result from term 3 can be combined to find --- in the soft limit ---
\begin{eqnarray}
\left.\frac{dN^{(1)}}{d\omega\,dt}\right|_{{\rm term}~1+3,\ell,e} &\approx& \frac{e^2}{2\pi} \int \frac{dh}{2\pi} \sum_{kk'}
\Delta(j,j',\ell) \delta_{ss'(-1)^{k+k'+\ell},1}
\frac{(2j+1)(2j'+1)(2\ell+1)}{4\pi \ell(\ell+1)\omega (e^{8\pi Mh}+1)}
\nonumber \\ && \times
 \tj{j}{j'}{\ell}{\frac12}{-\frac12}{0}^2 \left[ |R_{\frac12,-k,h}|^2 +
\frac{e^{8\pi Mh}}{e^{8\pi Mh}+1} |T_{\frac12,-k,h}|^2 \right] |T_{\frac12,-k',h}|^2 v^2\varsigma^2_\ell(v).
\label{eq:C9}
\end{eqnarray}
(Note that we kept $k$ and $k'$ separate, but approximated $h'\approx h$.)

A convenient next approximation is to also neglect Pauli blocking, i.e., take $e^{8\pi Mh}/(e^{8\pi Mh}+1) \rightarrow 1$ in the term in brackets. This is a good approximation since the particles emitted in Hawking radiation are typically at several times the Hawking temperature, $8\pi Mh=h/T_{\rm H} \sim 4$, but it is not a true limiting approximation in the sense that there is no small expansion parameter in the problem. But if we take it, then the term in brackets becomes 1. The sum over $k$ then collapses: we may split it into a sum over $j$ and a sum over $s$. The sum over $s$ has exactly one allowed value (according to the Kronecker delta), and then $j$ appears only via the combination
\begin{equation}
\sum_j \Delta(j,j',\ell) (2j+1) \tj{j}{j'}{\ell}{\frac12}{-\frac12}{0}^2 = 1.
\end{equation}
The consequence is a mass simplification:
\begin{equation}
\left.\frac{dN^{(1)}}{d\omega\,dt}\right|_{{\rm term}~1+3,\ell,e}\!\!({\rm soft,~no~Pauli}) \approx \frac{e^2}{8\pi^2}  
\frac{2\ell+1}{ \ell(\ell+1)\omega } \int \frac{dh}{2\pi}
 v^2\varsigma^2_\ell(v)  \sum_{k'} \frac{(2j'+1) |T_{\frac12,-k',h}|^2}{e^{8\pi Mh}+1}.
\end{equation}
Now the unperturbed rate of emission of electrons and positrons is
\begin{equation}
\frac{dN^{(0)}_{e^\pm}}{dh\,dt} = \frac{2}{2\pi} \sum_{k'} \frac{(2j'+1) |T_{\frac12,-k',h}|^2}{e^{8\pi Mh}+1},
\end{equation}
where the factor of 2 results from having both electron and positron degrees of freedom, so
\begin{equation}
\left.\frac{dN^{(1)}}{d\omega\,dt}\right|_{{\rm term}~1+3,\ell,e}\!\!({\rm soft,~no~Pauli}) \approx \frac{e^2}{16\pi^2}  
\frac{2\ell+1}{ \ell(\ell+1)\omega } \int dh\,
 v^2\varsigma^2_\ell(v) \frac{dN^{(0)}_{e^\pm}}{dh\,dt} .
\label{eq:C13}
\end{equation}
This is in agreement with the semiclassical result, Eq.~(\ref{eq:dnl1}).

To assess both the extent of the classical regime, and to further our understanding of the fully numerical results, we show the spectra in as calculated in Eqs.~(\ref{eq:C8}), (\ref{eq:C9}), and (\ref{eq:C13}) relative to the fully numerical results for a range of $\omega$ values in Table~\ref{Table:classical results l=1 M=1e21}.

\begin{table}[h!]
\begin{center}
\caption{Comparison of classical results and numerical results for $\ell$=1 of $M=1\times 10^{21} M_{planck}$ PBH, as well as the numerical results after implementing the rescaling factors. }
\label{Table:classical results l=1 M=1e21}
\begin{tabular}{||c || c c c || c c c c ||} 
 \hline
 $\omega$ & \multicolumn3c{analytic/numerical}  &\multicolumn3c{analytic/rescaled numerical} & \\ 
 & Eq.~(\ref{eq:C8}) & Eq.~(\ref{eq:C9}) & Eq.~(\ref{eq:C13})
 & Eq.~(\ref{eq:C8}) & Eq.~(\ref{eq:C9}) & Eq.~(\ref{eq:C13})
 & \\
 \hline\hline
$ 0.05T_{H}$ & 0.86 & 0.92 & 0.93 & 1.003 & 1.07 & 1.09 & \\ 
 \hline
 $ 0.07T_{H}$ & 0.88 & 0.93 & 0.93 & 1.015 & 1.07 & 1.07 & \\ 
 \hline
 $0.11T_{H}$ & 0.93 & 0.97 & 0.98  & 1.035 & 1.08 & 1.09 & \\ 
 \hline
 $0.46T_{H}$ & 1.14 & 1.17 & 1.17 & 1.164 &1.19 & 1.19 & \\ [1ex] 
 \hline
\end{tabular}
\end{center}
\end{table}

\section{Outer boundary effects}
\label{app:outer}

The main calculation in the paper computes electron and photon wave functions and integrals out to some maximum radius $r_{\star,\rm max}$. In this appendix, we want to derive an expected ``correction factor'' describing the difference between our calculation and the ideal case where we would set $r_{\star,\rm max}=\infty$.

We first note that the outer boundary is most likely to be important at low photon frequencies. Specifically, we define $\xi$ to be the distance from the black hole to the outer boundary in units of the reduced wavelength $\lambdabar = \lambda/2\pi = 1/\omega$. That is,
\begin{equation}
\xi = \frac{r_{\star,\rm max}}\lambdabar = \omega r_{\star,\rm max} = \frac{\omega}{T_{\rm H}} \frac{r_{\star,\rm max}}{8\pi M}
\approx 79.6 \frac{\omega}{T_{\rm H}}\frac{r_{\star,\rm max}}{2000M}.
\end{equation}
For $\omega / T_{\rm H} \gtrsim 1$, this means that the outer boundary is many wavelengths away from the black hole; but for small $\omega$ (we have computed down to $0.05T_{\rm H}$), the outer boundary may be only of order one wavelength from the hole. It is in this limit that a correction is necessary. At these low frequencies, the largest contribution by far is from terms 1 and 3, and from the even parity (``electric type'') photon mode; so we will restrict our attention here to the correction for these terms, drawing on Eq.~(\ref{eq:C13}). In particular, we may write a correction factor
\begin{equation}
f_{{\rm corr},\ell}(\xi) = \frac{[{\rm emitted~photon~rate~at~}r_{\star,\rm max}{\rm~actually~used}]}{[{\rm emitted~photon~rate~at~}r_{\star,\rm max}=\infty]}.
\end{equation}
This may in principle be written as a correction factor $f_{{\rm corr},\ell}(\xi|v)$ conditioned on the electron velocity (i.e., in the integrand of Eq.~\ref{eq:C13}), and then in accordance with Eq.~(\ref{eq:C13}) the overall factor is
\begin{equation}
f_{{\rm corr},\ell}(\xi) = \frac{\int_\mu^\infty v^2 \varsigma_\ell^2(v)\, [dN^{(0)}_{e^\pm}/dh\,dt] f_{{\rm corr},\ell}(\xi|v) \,dh}{\int_\mu^\infty v^2 \varsigma_\ell^2(v)\, [dN^{(0)}_{e^\pm}/dh\,dt] \,dh}\,.
\label{eq:f-int}
\end{equation}

\begin{figure}
\includegraphics[width=6.5in]{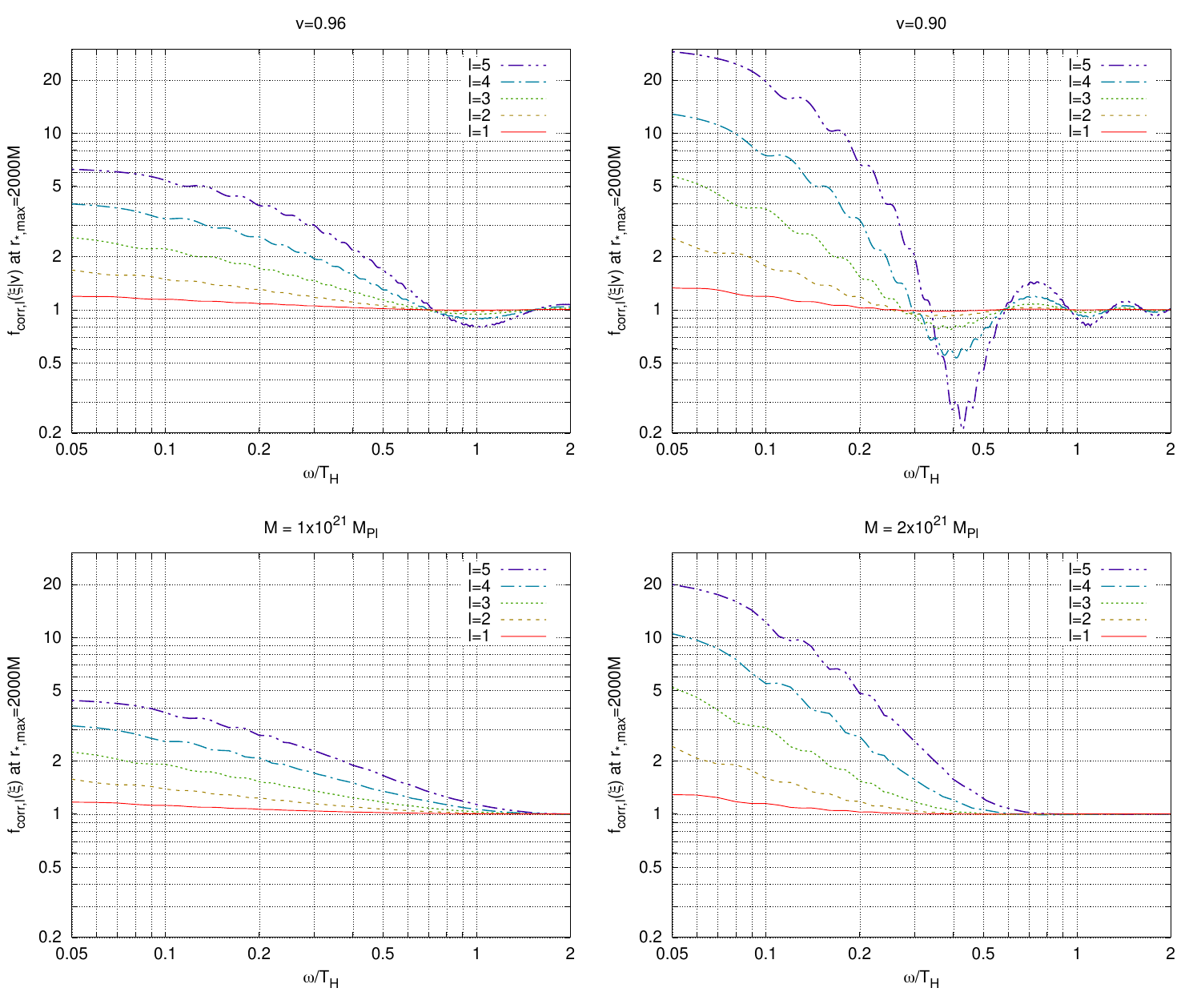}
\caption{\label{fig:corrfactor}{\em Top row}: The correction factors $f_{{\rm corr},\ell}(\xi|v)$, for $r_{\star,\rm max}=2000 M$, for two velocities. Note that the factors converge to 1 at large $\xi$ (or large $\omega/T_{\rm H}$, but with some oscillations as expected from integrating a spherical Bessel function. {\em Bottom row}: The total correction factors averaged over the emitted electron spectrum using Eq.~(\ref{eq:f-int}) for two primordial black hole masses, $M = 10^{21}$ and $2\times 10^{21}M_{\rm Pl}$.}
\end{figure}

The correction factor $f_{{\rm corr},\ell}(\xi)$ has two contributions. The first comes from the normalization of the photon wave functions. In the low-$\omega$ limit, the wave functions at large $r_\star$ become spherical Bessel functions, since they satisfy the spherical Bessel function equation (Eq.~\ref{eq:Psi-ODE}) when $r/2M\gg 1$ and hence are a linear combination of $xj_\ell(x)$ and $y_\ell(x)$ where $x=\omega r_\star$; and there is a large ``potential'' barrier at $\omega r_\star\ll 1$ that suppresses the $y_\ell$ solution. The true solution (with outer boundary at $r_\star\rightarrow\infty$) is given in Paper I as
\begin{equation}
\Psi_{\rm in}\rightarrow e^{-ix} + R e^{ix}
\approx 2(-i)^{\ell+1} xj_\ell(x),
~~~ R = (-1)^{\ell-1},  ~~~
{\rm at}~~~ \omega\ll M^{-1}, ~~ x=\omega r_\star\rightarrow\infty.
\end{equation}
Using this as $\Psi_{\rm in}^{\rm old}$ in Eq.~(\ref{eq:a-norm}), we see that if we were to normalize the radial wave function at some $r_{\star,\rm max}$, we would scale up the photon wave function normalization by $a$, where
\begin{equation}
|a|^2 = \left. \frac{4\omega^2}{|-i\omega\Psi_{\rm in}^{\rm old} + \Psi_{\rm in}^{\rm old}{}'|^2} \right|_{r_\star=r_{\star,\rm max}}
= \left. \frac1{[xj_\ell(x)]^2 + \{\partial_x [xj_\ell(x)]\}^2}
\right|_{x=\xi}.
\label{eq:a2}
\end{equation}

The other contribution comes from the finite range of the integrals. In evaluating $\llbracket I^{-+}_{{\rm in}\,k,{\rm up}\,k',{\rm in}\,\ell(e)}(h,h',\omega)\rrbracket$, we have cut off the integration at $r_{\star,\rm max}$ instead of the correct upper limit of $\infty$. From Eq.~(\ref{eq:Xi-def}), and propagating the limits of the integral through the remaining steps to Eq.~(\ref{eq:C13}), we have
\begin{equation}
f_{{\rm corr},\ell}(\xi|v) = \frac1{|a|^2} \left|\frac{\int_0^\xi x^{-1} e^{-ix/v} j_\ell(x)\,dx}{\int_0^\infty x^{-1} e^{-ix/v} j_\ell(x)\,dx}\right|^2.
\end{equation}

Examples of the velocity-specific correction factors $f_{{\rm corr},\ell}(\xi|v)$ and the total correction factor $f_{{\rm corr},\ell}(\xi)$ are shown in Fig.~\ref{fig:corrfactor}.

\bibliography{main}

\begin{thebibliography}{55}%
\makeatletter
\providecommand \@ifxundefined [1]{%
 \@ifx{#1\undefined}
}%
\providecommand \@ifnum [1]{%
 \ifnum #1\expandafter \@firstoftwo
 \else \expandafter \@secondoftwo
 \fi
}%
\providecommand \@ifx [1]{%
 \ifx #1\expandafter \@firstoftwo
 \else \expandafter \@secondoftwo
 \fi
}%
\providecommand \natexlab [1]{#1}%
\providecommand \enquote  [1]{``#1''}%
\providecommand \bibnamefont  [1]{#1}%
\providecommand \bibfnamefont [1]{#1}%
\providecommand \citenamefont [1]{#1}%
\providecommand \href@noop [0]{\@secondoftwo}%
\providecommand \href [0]{\begingroup \@sanitize@url \@href}%
\providecommand \@href[1]{\@@startlink{#1}\@@href}%
\providecommand \@@href[1]{\endgroup#1\@@endlink}%
\providecommand \@sanitize@url [0]{\catcode `\\12\catcode `\$12\catcode `\&12\catcode `\#12\catcode `\^12\catcode `\_12\catcode `\%12\relax}%
\providecommand \@@startlink[1]{}%
\providecommand \@@endlink[0]{}%
\providecommand \url  [0]{\begingroup\@sanitize@url \@url }%
\providecommand \@url [1]{\endgroup\@href {#1}{\urlprefix }}%
\providecommand \urlprefix  [0]{URL }%
\providecommand \Eprint [0]{\href }%
\providecommand \doibase [0]{https://doi.org/}%
\providecommand \selectlanguage [0]{\@gobble}%
\providecommand \bibinfo  [0]{\@secondoftwo}%
\providecommand \bibfield  [0]{\@secondoftwo}%
\providecommand \translation [1]{[#1]}%
\providecommand \BibitemOpen [0]{}%
\providecommand \bibitemStop [0]{}%
\providecommand \bibitemNoStop [0]{.\EOS\space}%
\providecommand \EOS [0]{\spacefactor3000\relax}%
\providecommand \BibitemShut  [1]{\csname bibitem#1\endcsname}%
\let\auto@bib@innerbib\@empty
\bibitem [{\citenamefont {{Zel'dovich}}\ and\ \citenamefont {{Novikov}}(1967)}]{1967SvA....10..602Z}%
  \BibitemOpen
  \bibfield  {author} {\bibinfo {author} {\bibfnamefont {Y.~B.}\ \bibnamefont {{Zel'dovich}}}\ and\ \bibinfo {author} {\bibfnamefont {I.~D.}\ \bibnamefont {{Novikov}}},\ }\bibfield  {title} {\bibinfo {title} {{The Hypothesis of Cores Retarded during Expansion and the Hot Cosmological Model}},\ }\href@noop {} {\bibfield  {journal} {\bibinfo  {journal} {\sovast}\ }\textbf {\bibinfo {volume} {10}},\ \bibinfo {pages} {602} (\bibinfo {year} {1967})}\BibitemShut {NoStop}%
\bibitem [{\citenamefont {{Hawking}}(1971)}]{1971MNRAS.152...75H}%
  \BibitemOpen
  \bibfield  {author} {\bibinfo {author} {\bibfnamefont {S.}~\bibnamefont {{Hawking}}},\ }\bibfield  {title} {\bibinfo {title} {{Gravitationally collapsed objects of very low mass}},\ }\href {https://doi.org/10.1093/mnras/152.1.75} {\bibfield  {journal} {\bibinfo  {journal} {\mnras}\ }\textbf {\bibinfo {volume} {152}},\ \bibinfo {pages} {75} (\bibinfo {year} {1971})}\BibitemShut {NoStop}%
\bibitem [{\citenamefont {{Kashlinsky}}\ \emph {et~al.}(2019)\citenamefont {{Kashlinsky}}, \citenamefont {{Ali-Ha{\"\i}moud}}, \citenamefont {{Clesse}}, \citenamefont {{Garcia-Bellido}}, \citenamefont {{Amendola}}, \citenamefont {{Wyrzykowski}}, \citenamefont {{Annis}}, \citenamefont {{Arbey}}, \citenamefont {{Arendt}}, \citenamefont {{Atrio-Barandela}}, \citenamefont {{Bellomo}}, \citenamefont {{Belotskiy}}, \citenamefont {{Bernal}}, \citenamefont {{Bird}}, \citenamefont {{Bozza}}, \citenamefont {{Byrnes}}, \citenamefont {{Calore}}, \citenamefont {{Carr}}, \citenamefont {{Chluba}}, \citenamefont {{Cholis}}, \citenamefont {{Cieplak}}, \citenamefont {{Cole}}, \citenamefont {{Dalianis}}, \citenamefont {{Davis}}, \citenamefont {{Davis}}, \citenamefont {{de Luca}}, \citenamefont {{Dvorkin}}, \citenamefont {{Emparan}}, \citenamefont {{Ezquiaga}}, \citenamefont {{Fleury}}, \citenamefont {{Franciolini}}, \citenamefont {{Georg}}, \citenamefont {{Germani}}, \citenamefont {{Giudice}}, \citenamefont {{Goobar}},
  \citenamefont {{Hasinger}}, \citenamefont {{Hector}}, \citenamefont {{Hundertmark}}, \citenamefont {{Hutsi}}, \citenamefont {{Jansen}}, \citenamefont {{Kamionkowski}}, \citenamefont {{Kawasaki}}, \citenamefont {{Kazanas}}, \citenamefont {{Kehagias}}, \citenamefont {{Khlopov}}, \citenamefont {{Knebe}}, \citenamefont {{Kohri}}, \citenamefont {{Koushiappas}}, \citenamefont {{Kovetz}}, \citenamefont {{Kuhnel}}, \citenamefont {{MacGibbon}}, \citenamefont {{Marzola}}, \citenamefont {{Mediavilla}}, \citenamefont {{Meszaros}}, \citenamefont {{Mroz}}, \citenamefont {{Munoz}}, \citenamefont {{Musco}}, \citenamefont {{Nesseris}}, \citenamefont {{Ozsoy}}, \citenamefont {{Pani}}, \citenamefont {{Poulin}}, \citenamefont {{Raccanelli}}, \citenamefont {{Racco}}, \citenamefont {{Raidal}}, \citenamefont {{Ranc}}, \citenamefont {{Rattenbury}}, \citenamefont {{Rhodes}}, \citenamefont {{Ricotti}}, \citenamefont {{Riotto}}, \citenamefont {{Rubin}}, \citenamefont {{Rubio}}, \citenamefont {{Ruiz-Morales}}, \citenamefont
  {{Sasaki}}, \citenamefont {{Schnittman}}, \citenamefont {{Shvartzvald}}, \citenamefont {{Street}}, \citenamefont {{Takada}}, \citenamefont {{Takhistov}}, \citenamefont {{Tashiro}}, \citenamefont {{Tasinato}}, \citenamefont {{Tringas}}, \citenamefont {{Unal}}, \citenamefont {{Tada}}, \citenamefont {{Tsapras}}, \citenamefont {{Vaskonen}}, \citenamefont {{Veerm{\"a}e}}, \citenamefont {{Vidotto}}, \citenamefont {{Watson}}, \citenamefont {{Windhorst}}, \citenamefont {{Yokoyama}},\ and\ \citenamefont {{Young}}}]{2019BAAS...51c..51K}%
  \BibitemOpen
  \bibfield  {author} {\bibinfo {author} {\bibfnamefont {A.}~\bibnamefont {{Kashlinsky}}}, \bibinfo {author} {\bibfnamefont {Y.}~\bibnamefont {{Ali-Ha{\"\i}moud}}}, \bibinfo {author} {\bibfnamefont {S.}~\bibnamefont {{Clesse}}}, \bibinfo {author} {\bibfnamefont {J.}~\bibnamefont {{Garcia-Bellido}}}, \bibinfo {author} {\bibfnamefont {L.}~\bibnamefont {{Amendola}}}, \bibinfo {author} {\bibfnamefont {L.}~\bibnamefont {{Wyrzykowski}}}, \bibinfo {author} {\bibfnamefont {J.}~\bibnamefont {{Annis}}}, \bibinfo {author} {\bibfnamefont {A.}~\bibnamefont {{Arbey}}}, \bibinfo {author} {\bibfnamefont {R.~G.}\ \bibnamefont {{Arendt}}}, \bibinfo {author} {\bibfnamefont {F.}~\bibnamefont {{Atrio-Barandela}}}, \bibinfo {author} {\bibfnamefont {N.}~\bibnamefont {{Bellomo}}}, \bibinfo {author} {\bibfnamefont {K.}~\bibnamefont {{Belotskiy}}}, \bibinfo {author} {\bibfnamefont {J.~L.}\ \bibnamefont {{Bernal}}}, \bibinfo {author} {\bibfnamefont {S.}~\bibnamefont {{Bird}}}, \bibinfo {author} {\bibfnamefont {V.}~\bibnamefont
  {{Bozza}}}, \bibinfo {author} {\bibfnamefont {C.}~\bibnamefont {{Byrnes}}}, \bibinfo {author} {\bibfnamefont {F.}~\bibnamefont {{Calore}}}, \bibinfo {author} {\bibfnamefont {B.~J.}\ \bibnamefont {{Carr}}}, \bibinfo {author} {\bibfnamefont {J.}~\bibnamefont {{Chluba}}}, \bibinfo {author} {\bibfnamefont {I.}~\bibnamefont {{Cholis}}}, \bibinfo {author} {\bibfnamefont {A.}~\bibnamefont {{Cieplak}}}, \bibinfo {author} {\bibfnamefont {P.}~\bibnamefont {{Cole}}}, \bibinfo {author} {\bibfnamefont {I.}~\bibnamefont {{Dalianis}}}, \bibinfo {author} {\bibfnamefont {A.~C.}\ \bibnamefont {{Davis}}}, \bibinfo {author} {\bibfnamefont {T.}~\bibnamefont {{Davis}}}, \bibinfo {author} {\bibfnamefont {V.}~\bibnamefont {{de Luca}}}, \bibinfo {author} {\bibfnamefont {I.}~\bibnamefont {{Dvorkin}}}, \bibinfo {author} {\bibfnamefont {R.}~\bibnamefont {{Emparan}}}, \bibinfo {author} {\bibfnamefont {J.~M.}\ \bibnamefont {{Ezquiaga}}}, \bibinfo {author} {\bibfnamefont {P.}~\bibnamefont {{Fleury}}}, \bibinfo {author} {\bibfnamefont
  {G.}~\bibnamefont {{Franciolini}}}, \bibinfo {author} {\bibfnamefont {J.}~\bibnamefont {{Georg}}}, \bibinfo {author} {\bibfnamefont {C.}~\bibnamefont {{Germani}}}, \bibinfo {author} {\bibfnamefont {G.~F.}\ \bibnamefont {{Giudice}}}, \bibinfo {author} {\bibfnamefont {A.}~\bibnamefont {{Goobar}}}, \bibinfo {author} {\bibfnamefont {G.}~\bibnamefont {{Hasinger}}}, \bibinfo {author} {\bibfnamefont {A.}~\bibnamefont {{Hector}}}, \bibinfo {author} {\bibfnamefont {M.}~\bibnamefont {{Hundertmark}}}, \bibinfo {author} {\bibfnamefont {G.}~\bibnamefont {{Hutsi}}}, \bibinfo {author} {\bibfnamefont {R.}~\bibnamefont {{Jansen}}}, \bibinfo {author} {\bibfnamefont {M.}~\bibnamefont {{Kamionkowski}}}, \bibinfo {author} {\bibfnamefont {M.}~\bibnamefont {{Kawasaki}}}, \bibinfo {author} {\bibfnamefont {D.}~\bibnamefont {{Kazanas}}}, \bibinfo {author} {\bibfnamefont {A.}~\bibnamefont {{Kehagias}}}, \bibinfo {author} {\bibfnamefont {M.}~\bibnamefont {{Khlopov}}}, \bibinfo {author} {\bibfnamefont {A.}~\bibnamefont {{Knebe}}},
  \bibinfo {author} {\bibfnamefont {K.}~\bibnamefont {{Kohri}}}, \bibinfo {author} {\bibfnamefont {S.}~\bibnamefont {{Koushiappas}}}, \bibinfo {author} {\bibfnamefont {E.}~\bibnamefont {{Kovetz}}}, \bibinfo {author} {\bibfnamefont {F.}~\bibnamefont {{Kuhnel}}}, \bibinfo {author} {\bibfnamefont {J.}~\bibnamefont {{MacGibbon}}}, \bibinfo {author} {\bibfnamefont {L.}~\bibnamefont {{Marzola}}}, \bibinfo {author} {\bibfnamefont {E.}~\bibnamefont {{Mediavilla}}}, \bibinfo {author} {\bibfnamefont {P.}~\bibnamefont {{Meszaros}}}, \bibinfo {author} {\bibfnamefont {P.}~\bibnamefont {{Mroz}}}, \bibinfo {author} {\bibfnamefont {J.}~\bibnamefont {{Munoz}}}, \bibinfo {author} {\bibfnamefont {I.}~\bibnamefont {{Musco}}}, \bibinfo {author} {\bibfnamefont {S.}~\bibnamefont {{Nesseris}}}, \bibinfo {author} {\bibfnamefont {O.}~\bibnamefont {{Ozsoy}}}, \bibinfo {author} {\bibfnamefont {P.}~\bibnamefont {{Pani}}}, \bibinfo {author} {\bibfnamefont {V.}~\bibnamefont {{Poulin}}}, \bibinfo {author} {\bibfnamefont {A.}~\bibnamefont
  {{Raccanelli}}}, \bibinfo {author} {\bibfnamefont {D.}~\bibnamefont {{Racco}}}, \bibinfo {author} {\bibfnamefont {M.}~\bibnamefont {{Raidal}}}, \bibinfo {author} {\bibfnamefont {C.}~\bibnamefont {{Ranc}}}, \bibinfo {author} {\bibfnamefont {N.}~\bibnamefont {{Rattenbury}}}, \bibinfo {author} {\bibfnamefont {J.}~\bibnamefont {{Rhodes}}}, \bibinfo {author} {\bibfnamefont {M.}~\bibnamefont {{Ricotti}}}, \bibinfo {author} {\bibfnamefont {A.}~\bibnamefont {{Riotto}}}, \bibinfo {author} {\bibfnamefont {S.}~\bibnamefont {{Rubin}}}, \bibinfo {author} {\bibfnamefont {J.}~\bibnamefont {{Rubio}}}, \bibinfo {author} {\bibfnamefont {E.}~\bibnamefont {{Ruiz-Morales}}}, \bibinfo {author} {\bibfnamefont {M.}~\bibnamefont {{Sasaki}}}, \bibinfo {author} {\bibfnamefont {J.}~\bibnamefont {{Schnittman}}}, \bibinfo {author} {\bibfnamefont {Y.}~\bibnamefont {{Shvartzvald}}}, \bibinfo {author} {\bibfnamefont {R.}~\bibnamefont {{Street}}}, \bibinfo {author} {\bibfnamefont {M.}~\bibnamefont {{Takada}}}, \bibinfo {author}
  {\bibfnamefont {V.}~\bibnamefont {{Takhistov}}}, \bibinfo {author} {\bibfnamefont {H.}~\bibnamefont {{Tashiro}}}, \bibinfo {author} {\bibfnamefont {G.}~\bibnamefont {{Tasinato}}}, \bibinfo {author} {\bibfnamefont {G.}~\bibnamefont {{Tringas}}}, \bibinfo {author} {\bibfnamefont {C.}~\bibnamefont {{Unal}}}, \bibinfo {author} {\bibfnamefont {Y.}~\bibnamefont {{Tada}}}, \bibinfo {author} {\bibfnamefont {Y.}~\bibnamefont {{Tsapras}}}, \bibinfo {author} {\bibfnamefont {V.}~\bibnamefont {{Vaskonen}}}, \bibinfo {author} {\bibfnamefont {H.}~\bibnamefont {{Veerm{\"a}e}}}, \bibinfo {author} {\bibfnamefont {F.}~\bibnamefont {{Vidotto}}}, \bibinfo {author} {\bibfnamefont {S.}~\bibnamefont {{Watson}}}, \bibinfo {author} {\bibfnamefont {R.}~\bibnamefont {{Windhorst}}}, \bibinfo {author} {\bibfnamefont {S.}~\bibnamefont {{Yokoyama}}},\ and\ \bibinfo {author} {\bibfnamefont {S.}~\bibnamefont {{Young}}},\ }\bibfield  {title} {\bibinfo {title} {{Electromagnetic probes of primordial black holes as dark matter}},\ }\href
  {https://doi.org/10.48550/arXiv.1903.04424} {\bibfield  {journal} {\bibinfo  {journal} {\baas}\ }\textbf {\bibinfo {volume} {51}},\ \bibinfo {eid} {51} (\bibinfo {year} {2019})},\ \Eprint {https://arxiv.org/abs/1903.04424} {arXiv:1903.04424 [astro-ph.CO]} \BibitemShut {NoStop}%
\bibitem [{\citenamefont {{Green}}\ and\ \citenamefont {{Kavanagh}}(2021{\natexlab{a}})}]{2021JPhG...48d3001G}%
  \BibitemOpen
  \bibfield  {author} {\bibinfo {author} {\bibfnamefont {A.~M.}\ \bibnamefont {{Green}}}\ and\ \bibinfo {author} {\bibfnamefont {B.~J.}\ \bibnamefont {{Kavanagh}}},\ }\bibfield  {title} {\bibinfo {title} {{Primordial black holes as a dark matter candidate}},\ }\href {https://doi.org/10.1088/1361-6471/abc534} {\bibfield  {journal} {\bibinfo  {journal} {Journal of Physics G Nuclear Physics}\ }\textbf {\bibinfo {volume} {48}},\ \bibinfo {eid} {043001} (\bibinfo {year} {2021}{\natexlab{a}})},\ \Eprint {https://arxiv.org/abs/2007.10722} {arXiv:2007.10722 [astro-ph.CO]} \BibitemShut {NoStop}%
\bibitem [{\citenamefont {{Carr}}\ \emph {et~al.}(2021)\citenamefont {{Carr}}, \citenamefont {{Kohri}}, \citenamefont {{Sendouda}},\ and\ \citenamefont {{Yokoyama}}}]{2021RPPh...84k6902C}%
  \BibitemOpen
  \bibfield  {author} {\bibinfo {author} {\bibfnamefont {B.}~\bibnamefont {{Carr}}}, \bibinfo {author} {\bibfnamefont {K.}~\bibnamefont {{Kohri}}}, \bibinfo {author} {\bibfnamefont {Y.}~\bibnamefont {{Sendouda}}},\ and\ \bibinfo {author} {\bibfnamefont {J.}~\bibnamefont {{Yokoyama}}},\ }\bibfield  {title} {\bibinfo {title} {{Constraints on primordial black holes}},\ }\href {https://doi.org/10.1088/1361-6633/ac1e31} {\bibfield  {journal} {\bibinfo  {journal} {Reports on Progress in Physics}\ }\textbf {\bibinfo {volume} {84}},\ \bibinfo {eid} {116902} (\bibinfo {year} {2021})},\ \Eprint {https://arxiv.org/abs/2002.12778} {arXiv:2002.12778 [astro-ph.CO]} \BibitemShut {NoStop}%
\bibitem [{\citenamefont {{Bird}}\ \emph {et~al.}(2023)\citenamefont {{Bird}}, \citenamefont {{Albert}}, \citenamefont {{Dawson}}, \citenamefont {{Ali-Ha{\"\i}moud}}, \citenamefont {{Coogan}}, \citenamefont {{Drlica-Wagner}}, \citenamefont {{Feng}}, \citenamefont {{Inman}}, \citenamefont {{Inomata}}, \citenamefont {{Kovetz}}, \citenamefont {{Kusenko}}, \citenamefont {{Lehmann}}, \citenamefont {{Mu{\~n}oz}}, \citenamefont {{Singh}}, \citenamefont {{Takhistov}},\ and\ \citenamefont {{Tsai}}}]{2023PDU....4101231B}%
  \BibitemOpen
  \bibfield  {author} {\bibinfo {author} {\bibfnamefont {S.}~\bibnamefont {{Bird}}}, \bibinfo {author} {\bibfnamefont {A.}~\bibnamefont {{Albert}}}, \bibinfo {author} {\bibfnamefont {W.}~\bibnamefont {{Dawson}}}, \bibinfo {author} {\bibfnamefont {Y.}~\bibnamefont {{Ali-Ha{\"\i}moud}}}, \bibinfo {author} {\bibfnamefont {A.}~\bibnamefont {{Coogan}}}, \bibinfo {author} {\bibfnamefont {A.}~\bibnamefont {{Drlica-Wagner}}}, \bibinfo {author} {\bibfnamefont {Q.}~\bibnamefont {{Feng}}}, \bibinfo {author} {\bibfnamefont {D.}~\bibnamefont {{Inman}}}, \bibinfo {author} {\bibfnamefont {K.}~\bibnamefont {{Inomata}}}, \bibinfo {author} {\bibfnamefont {E.}~\bibnamefont {{Kovetz}}}, \bibinfo {author} {\bibfnamefont {A.}~\bibnamefont {{Kusenko}}}, \bibinfo {author} {\bibfnamefont {B.~V.}\ \bibnamefont {{Lehmann}}}, \bibinfo {author} {\bibfnamefont {J.~B.}\ \bibnamefont {{Mu{\~n}oz}}}, \bibinfo {author} {\bibfnamefont {R.}~\bibnamefont {{Singh}}}, \bibinfo {author} {\bibfnamefont {V.}~\bibnamefont {{Takhistov}}},\ and\
  \bibinfo {author} {\bibfnamefont {Y.-D.}\ \bibnamefont {{Tsai}}},\ }\bibfield  {title} {\bibinfo {title} {{Snowmass2021 Cosmic Frontier White Paper: Primordial black hole dark matter}},\ }\href {https://doi.org/10.1016/j.dark.2023.101231} {\bibfield  {journal} {\bibinfo  {journal} {Physics of the Dark Universe}\ }\textbf {\bibinfo {volume} {41}},\ \bibinfo {eid} {101231} (\bibinfo {year} {2023})},\ \Eprint {https://arxiv.org/abs/2203.08967} {arXiv:2203.08967 [hep-ph]} \BibitemShut {NoStop}%
\bibitem [{\citenamefont {{Carr}}\ \emph {et~al.}(2024)\citenamefont {{Carr}}, \citenamefont {{Clesse}}, \citenamefont {{Garc{\'\i}a-Bellido}}, \citenamefont {{Hawkins}},\ and\ \citenamefont {{K{\"u}hnel}}}]{2024PhR..1054....1C}%
  \BibitemOpen
  \bibfield  {author} {\bibinfo {author} {\bibfnamefont {B.~J.}\ \bibnamefont {{Carr}}}, \bibinfo {author} {\bibfnamefont {S.}~\bibnamefont {{Clesse}}}, \bibinfo {author} {\bibfnamefont {J.}~\bibnamefont {{Garc{\'\i}a-Bellido}}}, \bibinfo {author} {\bibfnamefont {M.~R.~S.}\ \bibnamefont {{Hawkins}}},\ and\ \bibinfo {author} {\bibfnamefont {F.}~\bibnamefont {{K{\"u}hnel}}},\ }\bibfield  {title} {\bibinfo {title} {{Observational evidence for primordial black holes: A positivist perspective}},\ }\href {https://doi.org/10.1016/j.physrep.2023.11.005} {\bibfield  {journal} {\bibinfo  {journal} {\physrep}\ }\textbf {\bibinfo {volume} {1054}},\ \bibinfo {pages} {1} (\bibinfo {year} {2024})},\ \Eprint {https://arxiv.org/abs/2306.03903} {arXiv:2306.03903 [astro-ph.CO]} \BibitemShut {NoStop}%
\bibitem [{\citenamefont {{Green}}(2024)}]{2024NuPhB100316494G}%
  \BibitemOpen
  \bibfield  {author} {\bibinfo {author} {\bibfnamefont {A.~M.}\ \bibnamefont {{Green}}},\ }\bibfield  {title} {\bibinfo {title} {{Primordial black holes as a dark matter candidate - a brief overview}},\ }\href {https://doi.org/10.1016/j.nuclphysb.2024.116494} {\bibfield  {journal} {\bibinfo  {journal} {Nuclear Physics B}\ }\textbf {\bibinfo {volume} {1003}},\ \bibinfo {eid} {116494} (\bibinfo {year} {2024})},\ \Eprint {https://arxiv.org/abs/2402.15211} {arXiv:2402.15211 [astro-ph.CO]} \BibitemShut {NoStop}%
\bibitem [{\citenamefont {{Green}}\ and\ \citenamefont {{Kavanagh}}(2021{\natexlab{b}})}]{green2020}%
  \BibitemOpen
  \bibfield  {author} {\bibinfo {author} {\bibfnamefont {A.~M.}\ \bibnamefont {{Green}}}\ and\ \bibinfo {author} {\bibfnamefont {B.~J.}\ \bibnamefont {{Kavanagh}}},\ }\bibfield  {title} {\bibinfo {title} {{Primordial black holes as a dark matter candidate}},\ }\href {https://doi.org/10.1088/1361-6471/abc534} {\bibfield  {journal} {\bibinfo  {journal} {Journal of Physics G Nuclear Physics}\ }\textbf {\bibinfo {volume} {48}},\ \bibinfo {eid} {043001} (\bibinfo {year} {2021}{\natexlab{b}})},\ \Eprint {https://arxiv.org/abs/2007.10722} {arXiv:2007.10722 [astro-ph.CO]} \BibitemShut {NoStop}%
\bibitem [{\citenamefont {{Griest}}\ \emph {et~al.}(2014)\citenamefont {{Griest}}, \citenamefont {{Cieplak}},\ and\ \citenamefont {{Lehner}}}]{2014ApJ...786..158G}%
  \BibitemOpen
  \bibfield  {author} {\bibinfo {author} {\bibfnamefont {K.}~\bibnamefont {{Griest}}}, \bibinfo {author} {\bibfnamefont {A.~M.}\ \bibnamefont {{Cieplak}}},\ and\ \bibinfo {author} {\bibfnamefont {M.~J.}\ \bibnamefont {{Lehner}}},\ }\bibfield  {title} {\bibinfo {title} {{Experimental Limits on Primordial Black Hole Dark Matter from the First 2 yr of Kepler Data}},\ }\href {https://doi.org/10.1088/0004-637X/786/2/158} {\bibfield  {journal} {\bibinfo  {journal} {\apj}\ }\textbf {\bibinfo {volume} {786}},\ \bibinfo {eid} {158} (\bibinfo {year} {2014})},\ \Eprint {https://arxiv.org/abs/1307.5798} {arXiv:1307.5798 [astro-ph.CO]} \BibitemShut {NoStop}%
\bibitem [{\citenamefont {{Niikura}}\ \emph {et~al.}(2019)\citenamefont {{Niikura}}, \citenamefont {{Takada}}, \citenamefont {{Yasuda}}, \citenamefont {{Lupton}}, \citenamefont {{Sumi}}, \citenamefont {{More}}, \citenamefont {{Kurita}}, \citenamefont {{Sugiyama}}, \citenamefont {{More}}, \citenamefont {{Oguri}},\ and\ \citenamefont {{Chiba}}}]{2019NatAs...3..524N}%
  \BibitemOpen
  \bibfield  {author} {\bibinfo {author} {\bibfnamefont {H.}~\bibnamefont {{Niikura}}}, \bibinfo {author} {\bibfnamefont {M.}~\bibnamefont {{Takada}}}, \bibinfo {author} {\bibfnamefont {N.}~\bibnamefont {{Yasuda}}}, \bibinfo {author} {\bibfnamefont {R.~H.}\ \bibnamefont {{Lupton}}}, \bibinfo {author} {\bibfnamefont {T.}~\bibnamefont {{Sumi}}}, \bibinfo {author} {\bibfnamefont {S.}~\bibnamefont {{More}}}, \bibinfo {author} {\bibfnamefont {T.}~\bibnamefont {{Kurita}}}, \bibinfo {author} {\bibfnamefont {S.}~\bibnamefont {{Sugiyama}}}, \bibinfo {author} {\bibfnamefont {A.}~\bibnamefont {{More}}}, \bibinfo {author} {\bibfnamefont {M.}~\bibnamefont {{Oguri}}},\ and\ \bibinfo {author} {\bibfnamefont {M.}~\bibnamefont {{Chiba}}},\ }\bibfield  {title} {\bibinfo {title} {{Microlensing constraints on primordial black holes with Subaru/HSC Andromeda observations}},\ }\href {https://doi.org/10.1038/s41550-019-0723-1} {\bibfield  {journal} {\bibinfo  {journal} {Nature Astronomy}\ }\textbf {\bibinfo {volume} {3}},\ \bibinfo
  {pages} {524} (\bibinfo {year} {2019})},\ \Eprint {https://arxiv.org/abs/1701.02151} {arXiv:1701.02151 [astro-ph.CO]} \BibitemShut {NoStop}%
\bibitem [{\citenamefont {{Smyth}}\ \emph {et~al.}(2020)\citenamefont {{Smyth}}, \citenamefont {{Profumo}}, \citenamefont {{English}}, \citenamefont {{Jeltema}}, \citenamefont {{McKinnon}},\ and\ \citenamefont {{Guhathakurta}}}]{2020PhRvD.101f3005S}%
  \BibitemOpen
  \bibfield  {author} {\bibinfo {author} {\bibfnamefont {N.}~\bibnamefont {{Smyth}}}, \bibinfo {author} {\bibfnamefont {S.}~\bibnamefont {{Profumo}}}, \bibinfo {author} {\bibfnamefont {S.}~\bibnamefont {{English}}}, \bibinfo {author} {\bibfnamefont {T.}~\bibnamefont {{Jeltema}}}, \bibinfo {author} {\bibfnamefont {K.}~\bibnamefont {{McKinnon}}},\ and\ \bibinfo {author} {\bibfnamefont {P.}~\bibnamefont {{Guhathakurta}}},\ }\bibfield  {title} {\bibinfo {title} {{Updated constraints on asteroid-mass primordial black holes as dark matter}},\ }\href {https://doi.org/10.1103/PhysRevD.101.063005} {\bibfield  {journal} {\bibinfo  {journal} {\prd}\ }\textbf {\bibinfo {volume} {101}},\ \bibinfo {eid} {063005} (\bibinfo {year} {2020})},\ \Eprint {https://arxiv.org/abs/1910.01285} {arXiv:1910.01285 [astro-ph.CO]} \BibitemShut {NoStop}%
\bibitem [{\citenamefont {{Hawking}}(1975)}]{1975CMaPh..43..199H}%
  \BibitemOpen
  \bibfield  {author} {\bibinfo {author} {\bibfnamefont {S.~W.}\ \bibnamefont {{Hawking}}},\ }\bibfield  {title} {\bibinfo {title} {{Particle creation by black holes}},\ }\href {https://doi.org/10.1007/BF02345020} {\bibfield  {journal} {\bibinfo  {journal} {Communications in Mathematical Physics}\ }\textbf {\bibinfo {volume} {43}},\ \bibinfo {pages} {199} (\bibinfo {year} {1975})}\BibitemShut {NoStop}%
\bibitem [{\citenamefont {{Carr}}\ \emph {et~al.}(2016)\citenamefont {{Carr}}, \citenamefont {{Kohri}}, \citenamefont {{Sendouda}},\ and\ \citenamefont {{Yokoyama}}}]{2016PhRvD..94d4029C}%
  \BibitemOpen
  \bibfield  {author} {\bibinfo {author} {\bibfnamefont {B.~J.}\ \bibnamefont {{Carr}}}, \bibinfo {author} {\bibfnamefont {K.}~\bibnamefont {{Kohri}}}, \bibinfo {author} {\bibfnamefont {Y.}~\bibnamefont {{Sendouda}}},\ and\ \bibinfo {author} {\bibfnamefont {J.}~\bibnamefont {{Yokoyama}}},\ }\bibfield  {title} {\bibinfo {title} {{Constraints on primordial black holes from the Galactic gamma-ray background}},\ }\href {https://doi.org/10.1103/PhysRevD.94.044029} {\bibfield  {journal} {\bibinfo  {journal} {\prd}\ }\textbf {\bibinfo {volume} {94}},\ \bibinfo {eid} {044029} (\bibinfo {year} {2016})},\ \Eprint {https://arxiv.org/abs/1604.05349} {arXiv:1604.05349 [astro-ph.CO]} \BibitemShut {NoStop}%
\bibitem [{\citenamefont {{Boudaud}}\ and\ \citenamefont {{Cirelli}}(2019)}]{2019PhRvL.122d1104B}%
  \BibitemOpen
  \bibfield  {author} {\bibinfo {author} {\bibfnamefont {M.}~\bibnamefont {{Boudaud}}}\ and\ \bibinfo {author} {\bibfnamefont {M.}~\bibnamefont {{Cirelli}}},\ }\bibfield  {title} {\bibinfo {title} {{Voyager 1 e$^{{\ensuremath{\pm}}}$ Further Constrain Primordial Black Holes as Dark Matter}},\ }\href {https://doi.org/10.1103/PhysRevLett.122.041104} {\bibfield  {journal} {\bibinfo  {journal} {\prl}\ }\textbf {\bibinfo {volume} {122}},\ \bibinfo {eid} {041104} (\bibinfo {year} {2019})},\ \Eprint {https://arxiv.org/abs/1807.03075} {arXiv:1807.03075 [astro-ph.HE]} \BibitemShut {NoStop}%
\bibitem [{\citenamefont {{DeRocco}}\ and\ \citenamefont {{Graham}}(2019)}]{2019PhRvL.123y1102D}%
  \BibitemOpen
  \bibfield  {author} {\bibinfo {author} {\bibfnamefont {W.}~\bibnamefont {{DeRocco}}}\ and\ \bibinfo {author} {\bibfnamefont {P.~W.}\ \bibnamefont {{Graham}}},\ }\bibfield  {title} {\bibinfo {title} {{Constraining Primordial Black Hole Abundance with the Galactic 511 keV Line}},\ }\href {https://doi.org/10.1103/PhysRevLett.123.251102} {\bibfield  {journal} {\bibinfo  {journal} {\prl}\ }\textbf {\bibinfo {volume} {123}},\ \bibinfo {eid} {251102} (\bibinfo {year} {2019})},\ \Eprint {https://arxiv.org/abs/1906.07740} {arXiv:1906.07740 [astro-ph.CO]} \BibitemShut {NoStop}%
\bibitem [{\citenamefont {{Laha}}(2019)}]{2019PhRvL.123y1101L}%
  \BibitemOpen
  \bibfield  {author} {\bibinfo {author} {\bibfnamefont {R.}~\bibnamefont {{Laha}}},\ }\bibfield  {title} {\bibinfo {title} {{Primordial Black Holes as a Dark Matter Candidate Are Severely Constrained by the Galactic Center 511 keV {\ensuremath{\gamma}} -Ray Line}},\ }\href {https://doi.org/10.1103/PhysRevLett.123.251101} {\bibfield  {journal} {\bibinfo  {journal} {\prl}\ }\textbf {\bibinfo {volume} {123}},\ \bibinfo {eid} {251101} (\bibinfo {year} {2019})},\ \Eprint {https://arxiv.org/abs/1906.09994} {arXiv:1906.09994 [astro-ph.HE]} \BibitemShut {NoStop}%
\bibitem [{\citenamefont {{Roncadelli}}\ \emph {et~al.}(2009)\citenamefont {{Roncadelli}}, \citenamefont {{Treves}},\ and\ \citenamefont {{Turolla}}}]{2009arXiv0901.1093R}%
  \BibitemOpen
  \bibfield  {author} {\bibinfo {author} {\bibfnamefont {M.}~\bibnamefont {{Roncadelli}}}, \bibinfo {author} {\bibfnamefont {A.}~\bibnamefont {{Treves}}},\ and\ \bibinfo {author} {\bibfnamefont {R.}~\bibnamefont {{Turolla}}},\ }\bibfield  {title} {\bibinfo {title} {{Primordial black holes are again on the limelight}},\ }\href {https://doi.org/10.48550/arXiv.0901.1093} {\bibfield  {journal} {\bibinfo  {journal} {arXiv e-prints}\ ,\ \bibinfo {eid} {arXiv:0901.1093}} (\bibinfo {year} {2009})},\ \Eprint {https://arxiv.org/abs/0901.1093} {arXiv:0901.1093 [astro-ph.CO]} \BibitemShut {NoStop}%
\bibitem [{\citenamefont {{G{\'e}nolini}}\ \emph {et~al.}(2020)\citenamefont {{G{\'e}nolini}}, \citenamefont {{Serpico}},\ and\ \citenamefont {{Tinyakov}}}]{2020PhRvD.102h3004G}%
  \BibitemOpen
  \bibfield  {author} {\bibinfo {author} {\bibfnamefont {Y.}~\bibnamefont {{G{\'e}nolini}}}, \bibinfo {author} {\bibfnamefont {P.~D.}\ \bibnamefont {{Serpico}}},\ and\ \bibinfo {author} {\bibfnamefont {P.}~\bibnamefont {{Tinyakov}}},\ }\bibfield  {title} {\bibinfo {title} {{Revisiting primordial black hole capture into neutron stars}},\ }\href {https://doi.org/10.1103/PhysRevD.102.083004} {\bibfield  {journal} {\bibinfo  {journal} {\prd}\ }\textbf {\bibinfo {volume} {102}},\ \bibinfo {eid} {083004} (\bibinfo {year} {2020})},\ \Eprint {https://arxiv.org/abs/2006.16975} {arXiv:2006.16975 [astro-ph.HE]} \BibitemShut {NoStop}%
\bibitem [{\citenamefont {{Oncins}}\ \emph {et~al.}(2022)\citenamefont {{Oncins}}, \citenamefont {{Miralda-Escud{\'e}}}, \citenamefont {{Guti{\'e}rrez}},\ and\ \citenamefont {{Gil-Pons}}}]{2022MNRAS.517...28O}%
  \BibitemOpen
  \bibfield  {author} {\bibinfo {author} {\bibfnamefont {M.}~\bibnamefont {{Oncins}}}, \bibinfo {author} {\bibfnamefont {J.}~\bibnamefont {{Miralda-Escud{\'e}}}}, \bibinfo {author} {\bibfnamefont {J.~L.}\ \bibnamefont {{Guti{\'e}rrez}}},\ and\ \bibinfo {author} {\bibfnamefont {P.}~\bibnamefont {{Gil-Pons}}},\ }\bibfield  {title} {\bibinfo {title} {{Primordial black holes capture by stars and induced collapse to low-mass stellar black holes}},\ }\href {https://doi.org/10.1093/mnras/stac2647} {\bibfield  {journal} {\bibinfo  {journal} {\mnras}\ }\textbf {\bibinfo {volume} {517}},\ \bibinfo {pages} {28} (\bibinfo {year} {2022})},\ \Eprint {https://arxiv.org/abs/2205.13003} {arXiv:2205.13003 [astro-ph.GA]} \BibitemShut {NoStop}%
\bibitem [{\citenamefont {{Esser}}\ and\ \citenamefont {{Tinyakov}}(2023)}]{2023PhRvD.107j3052E}%
  \BibitemOpen
  \bibfield  {author} {\bibinfo {author} {\bibfnamefont {N.}~\bibnamefont {{Esser}}}\ and\ \bibinfo {author} {\bibfnamefont {P.}~\bibnamefont {{Tinyakov}}},\ }\bibfield  {title} {\bibinfo {title} {{Constraints on primordial black holes from observation of stars in dwarf galaxies}},\ }\href {https://doi.org/10.1103/PhysRevD.107.103052} {\bibfield  {journal} {\bibinfo  {journal} {\prd}\ }\textbf {\bibinfo {volume} {107}},\ \bibinfo {eid} {103052} (\bibinfo {year} {2023})},\ \Eprint {https://arxiv.org/abs/2207.07412} {arXiv:2207.07412 [astro-ph.HE]} \BibitemShut {NoStop}%
\bibitem [{\citenamefont {{Tinyakov}}(2024)}]{2024arXiv240603114T}%
  \BibitemOpen
  \bibfield  {author} {\bibinfo {author} {\bibfnamefont {P.}~\bibnamefont {{Tinyakov}}},\ }\bibfield  {title} {\bibinfo {title} {{Primordial black holes: the asteroid mass window}},\ }\href {https://doi.org/10.48550/arXiv.2406.03114} {\bibfield  {journal} {\bibinfo  {journal} {arXiv e-prints}\ ,\ \bibinfo {eid} {arXiv:2406.03114}} (\bibinfo {year} {2024})},\ \Eprint {https://arxiv.org/abs/2406.03114} {arXiv:2406.03114 [astro-ph.CO]} \BibitemShut {NoStop}%
\bibitem [{\citenamefont {{Bai}}\ and\ \citenamefont {{Orlofsky}}(2019)}]{2019PhRvD..99l3019B}%
  \BibitemOpen
  \bibfield  {author} {\bibinfo {author} {\bibfnamefont {Y.}~\bibnamefont {{Bai}}}\ and\ \bibinfo {author} {\bibfnamefont {N.}~\bibnamefont {{Orlofsky}}},\ }\bibfield  {title} {\bibinfo {title} {{Microlensing of x-ray pulsars: A method to detect primordial black hole dark matter}},\ }\href {https://doi.org/10.1103/PhysRevD.99.123019} {\bibfield  {journal} {\bibinfo  {journal} {\prd}\ }\textbf {\bibinfo {volume} {99}},\ \bibinfo {eid} {123019} (\bibinfo {year} {2019})},\ \Eprint {https://arxiv.org/abs/1812.01427} {arXiv:1812.01427 [astro-ph.HE]} \BibitemShut {NoStop}%
\bibitem [{\citenamefont {{Tamta}}\ \emph {et~al.}(2024)\citenamefont {{Tamta}}, \citenamefont {{Raj}},\ and\ \citenamefont {{Sharma}}}]{2024arXiv240520365T}%
  \BibitemOpen
  \bibfield  {author} {\bibinfo {author} {\bibfnamefont {M.}~\bibnamefont {{Tamta}}}, \bibinfo {author} {\bibfnamefont {N.}~\bibnamefont {{Raj}}},\ and\ \bibinfo {author} {\bibfnamefont {P.}~\bibnamefont {{Sharma}}},\ }\bibfield  {title} {\bibinfo {title} {{Breaking into the window of primordial black hole dark matter with x-ray microlensing}},\ }\href {https://doi.org/10.48550/arXiv.2405.20365} {\bibfield  {journal} {\bibinfo  {journal} {arXiv e-prints}\ ,\ \bibinfo {eid} {arXiv:2405.20365}} (\bibinfo {year} {2024})},\ \Eprint {https://arxiv.org/abs/2405.20365} {arXiv:2405.20365 [astro-ph.HE]} \BibitemShut {NoStop}%
\bibitem [{\citenamefont {{Nemiroff}}\ and\ \citenamefont {{Gould}}(1995)}]{1995ApJ...452L.111N}%
  \BibitemOpen
  \bibfield  {author} {\bibinfo {author} {\bibfnamefont {R.~J.}\ \bibnamefont {{Nemiroff}}}\ and\ \bibinfo {author} {\bibfnamefont {A.}~\bibnamefont {{Gould}}},\ }\bibfield  {title} {\bibinfo {title} {{Probing for MACHOs of Mass 10 -15 M$_{sun}$ to 10 -7 M$_{sun}$ with Gamma-Ray Burst Parallax Spacecraft}},\ }\href {https://doi.org/10.1086/309722} {\bibfield  {journal} {\bibinfo  {journal} {\apjl}\ }\textbf {\bibinfo {volume} {452}},\ \bibinfo {pages} {L111} (\bibinfo {year} {1995})},\ \Eprint {https://arxiv.org/abs/astro-ph/9505019} {arXiv:astro-ph/9505019 [astro-ph]} \BibitemShut {NoStop}%
\bibitem [{\citenamefont {{Marani}}\ \emph {et~al.}(1999)\citenamefont {{Marani}}, \citenamefont {{Nemiroff}}, \citenamefont {{Norris}}, \citenamefont {{Hurley}},\ and\ \citenamefont {{Bonnell}}}]{1999ApJ...512L..13M}%
  \BibitemOpen
  \bibfield  {author} {\bibinfo {author} {\bibfnamefont {G.~F.}\ \bibnamefont {{Marani}}}, \bibinfo {author} {\bibfnamefont {R.~J.}\ \bibnamefont {{Nemiroff}}}, \bibinfo {author} {\bibfnamefont {J.~P.}\ \bibnamefont {{Norris}}}, \bibinfo {author} {\bibfnamefont {K.}~\bibnamefont {{Hurley}}},\ and\ \bibinfo {author} {\bibfnamefont {J.~T.}\ \bibnamefont {{Bonnell}}},\ }\bibfield  {title} {\bibinfo {title} {{Gravitationally Lensed Gamma-Ray Bursts as Probes of Dark Compact Objects}},\ }\href {https://doi.org/10.1086/311868} {\bibfield  {journal} {\bibinfo  {journal} {\apjl}\ }\textbf {\bibinfo {volume} {512}},\ \bibinfo {pages} {L13} (\bibinfo {year} {1999})},\ \Eprint {https://arxiv.org/abs/astro-ph/9810391} {arXiv:astro-ph/9810391 [astro-ph]} \BibitemShut {NoStop}%
\bibitem [{\citenamefont {{Ray}}\ \emph {et~al.}(2021)\citenamefont {{Ray}}, \citenamefont {{Laha}}, \citenamefont {{Mu{\~n}oz}},\ and\ \citenamefont {{Caputo}}}]{2021PhRvD.104b3516R}%
  \BibitemOpen
  \bibfield  {author} {\bibinfo {author} {\bibfnamefont {A.}~\bibnamefont {{Ray}}}, \bibinfo {author} {\bibfnamefont {R.}~\bibnamefont {{Laha}}}, \bibinfo {author} {\bibfnamefont {J.~B.}\ \bibnamefont {{Mu{\~n}oz}}},\ and\ \bibinfo {author} {\bibfnamefont {R.}~\bibnamefont {{Caputo}}},\ }\bibfield  {title} {\bibinfo {title} {{Near future MeV telescopes can discover asteroid-mass primordial black hole dark matter}},\ }\href {https://doi.org/10.1103/PhysRevD.104.023516} {\bibfield  {journal} {\bibinfo  {journal} {\prd}\ }\textbf {\bibinfo {volume} {104}},\ \bibinfo {eid} {023516} (\bibinfo {year} {2021})},\ \Eprint {https://arxiv.org/abs/2102.06714} {arXiv:2102.06714 [astro-ph.CO]} \BibitemShut {NoStop}%
\bibitem [{\citenamefont {{Auffinger}}(2023)}]{2023PrPNP.13104040A}%
  \BibitemOpen
  \bibfield  {author} {\bibinfo {author} {\bibfnamefont {J.}~\bibnamefont {{Auffinger}}},\ }\bibfield  {title} {\bibinfo {title} {{Primordial black hole constraints with Hawking radiation-A review}},\ }\href {https://doi.org/10.1016/j.ppnp.2023.104040} {\bibfield  {journal} {\bibinfo  {journal} {Progress in Particle and Nuclear Physics}\ }\textbf {\bibinfo {volume} {131}},\ \bibinfo {eid} {104040} (\bibinfo {year} {2023})},\ \Eprint {https://arxiv.org/abs/2206.02672} {arXiv:2206.02672 [astro-ph.CO]} \BibitemShut {NoStop}%
\bibitem [{\citenamefont {{Page}}(1976)}]{1976PhRvD..13..198P}%
  \BibitemOpen
  \bibfield  {author} {\bibinfo {author} {\bibfnamefont {D.~N.}\ \bibnamefont {{Page}}},\ }\bibfield  {title} {\bibinfo {title} {{Particle emission rates from a black hole: Massless particles from an uncharged, nonrotating hole}},\ }\href {https://doi.org/10.1103/PhysRevD.13.198} {\bibfield  {journal} {\bibinfo  {journal} {\prd}\ }\textbf {\bibinfo {volume} {13}},\ \bibinfo {pages} {198} (\bibinfo {year} {1976})}\BibitemShut {NoStop}%
\bibitem [{\citenamefont {{Page}}(1977)}]{1977PhRvD..16.2402P}%
  \BibitemOpen
  \bibfield  {author} {\bibinfo {author} {\bibfnamefont {D.~N.}\ \bibnamefont {{Page}}},\ }\bibfield  {title} {\bibinfo {title} {{Particle emission rates from a black hole. III. Charged leptons from a nonrotating hole}},\ }\href {https://doi.org/10.1103/PhysRevD.16.2402} {\bibfield  {journal} {\bibinfo  {journal} {\prd}\ }\textbf {\bibinfo {volume} {16}},\ \bibinfo {pages} {2402} (\bibinfo {year} {1977})}\BibitemShut {NoStop}%
\bibitem [{\citenamefont {{Arbey}}\ and\ \citenamefont {{Auffinger}}(2019)}]{2019EPJC...79..693A}%
  \BibitemOpen
  \bibfield  {author} {\bibinfo {author} {\bibfnamefont {A.}~\bibnamefont {{Arbey}}}\ and\ \bibinfo {author} {\bibfnamefont {J.}~\bibnamefont {{Auffinger}}},\ }\bibfield  {title} {\bibinfo {title} {{BlackHawk: a public code for calculating the Hawking evaporation spectra of any black hole distribution}},\ }\href {https://doi.org/10.1140/epjc/s10052-019-7161-1} {\bibfield  {journal} {\bibinfo  {journal} {European Physical Journal C}\ }\textbf {\bibinfo {volume} {79}},\ \bibinfo {eid} {693} (\bibinfo {year} {2019})},\ \Eprint {https://arxiv.org/abs/1905.04268} {arXiv:1905.04268 [gr-qc]} \BibitemShut {NoStop}%
\bibitem [{\citenamefont {{Arbey}}\ \emph {et~al.}(2021{\natexlab{a}})\citenamefont {{Arbey}}, \citenamefont {{Auffinger}}, \citenamefont {{Geiller}}, \citenamefont {{Livine}},\ and\ \citenamefont {{Sartini}}}]{2021PhRvD.103j4010A}%
  \BibitemOpen
  \bibfield  {author} {\bibinfo {author} {\bibfnamefont {A.}~\bibnamefont {{Arbey}}}, \bibinfo {author} {\bibfnamefont {J.}~\bibnamefont {{Auffinger}}}, \bibinfo {author} {\bibfnamefont {M.}~\bibnamefont {{Geiller}}}, \bibinfo {author} {\bibfnamefont {E.~R.}\ \bibnamefont {{Livine}}},\ and\ \bibinfo {author} {\bibfnamefont {F.}~\bibnamefont {{Sartini}}},\ }\bibfield  {title} {\bibinfo {title} {{Hawking radiation by spherically-symmetric static black holes for all spins: Teukolsky equations and potentials}},\ }\href {https://doi.org/10.1103/PhysRevD.103.104010} {\bibfield  {journal} {\bibinfo  {journal} {\prd}\ }\textbf {\bibinfo {volume} {103}},\ \bibinfo {eid} {104010} (\bibinfo {year} {2021}{\natexlab{a}})},\ \Eprint {https://arxiv.org/abs/2101.02951} {arXiv:2101.02951 [gr-qc]} \BibitemShut {NoStop}%
\bibitem [{\citenamefont {{Arbey}}\ and\ \citenamefont {{Auffinger}}(2021)}]{2021EPJC...81..910A}%
  \BibitemOpen
  \bibfield  {author} {\bibinfo {author} {\bibfnamefont {A.}~\bibnamefont {{Arbey}}}\ and\ \bibinfo {author} {\bibfnamefont {J.}~\bibnamefont {{Auffinger}}},\ }\bibfield  {title} {\bibinfo {title} {{Physics beyond the standard model with BlackHawk v2.0}},\ }\href {https://doi.org/10.1140/epjc/s10052-021-09702-8} {\bibfield  {journal} {\bibinfo  {journal} {European Physical Journal C}\ }\textbf {\bibinfo {volume} {81}},\ \bibinfo {eid} {910} (\bibinfo {year} {2021})},\ \Eprint {https://arxiv.org/abs/2108.02737} {arXiv:2108.02737 [gr-qc]} \BibitemShut {NoStop}%
\bibitem [{\citenamefont {{Knipp}}\ and\ \citenamefont {{Uhlenbeck}}(1936)}]{1936Phy.....3..425K}%
  \BibitemOpen
  \bibfield  {author} {\bibinfo {author} {\bibfnamefont {J.~K.}\ \bibnamefont {{Knipp}}}\ and\ \bibinfo {author} {\bibfnamefont {G.~E.}\ \bibnamefont {{Uhlenbeck}}},\ }\bibfield  {title} {\bibinfo {title} {{Emission of gamma radiation during the beta decay of nuclei}},\ }\href {https://doi.org/10.1016/S0031-8914(36)80008-1} {\bibfield  {journal} {\bibinfo  {journal} {Physica}\ }\textbf {\bibinfo {volume} {3}},\ \bibinfo {pages} {425} (\bibinfo {year} {1936})}\BibitemShut {NoStop}%
\bibitem [{\citenamefont {{Altarelli}}\ and\ \citenamefont {{Parisi}}(1977)}]{1977NuPhB.126..298A}%
  \BibitemOpen
  \bibfield  {author} {\bibinfo {author} {\bibfnamefont {G.}~\bibnamefont {{Altarelli}}}\ and\ \bibinfo {author} {\bibfnamefont {G.}~\bibnamefont {{Parisi}}},\ }\bibfield  {title} {\bibinfo {title} {{Asymptotic freedom in parton language}},\ }\href {https://doi.org/10.1016/0550-3213(77)90384-4} {\bibfield  {journal} {\bibinfo  {journal} {Nuclear Physics B}\ }\textbf {\bibinfo {volume} {126}},\ \bibinfo {pages} {298} (\bibinfo {year} {1977})}\BibitemShut {NoStop}%
\bibitem [{\citenamefont {{Coogan}}\ \emph {et~al.}(2020)\citenamefont {{Coogan}}, \citenamefont {{Morrison}},\ and\ \citenamefont {{Profumo}}}]{2020JCAP...01..056C}%
  \BibitemOpen
  \bibfield  {author} {\bibinfo {author} {\bibfnamefont {A.}~\bibnamefont {{Coogan}}}, \bibinfo {author} {\bibfnamefont {L.}~\bibnamefont {{Morrison}}},\ and\ \bibinfo {author} {\bibfnamefont {S.}~\bibnamefont {{Profumo}}},\ }\bibfield  {title} {\bibinfo {title} {{Hazma: a python toolkit for studying indirect detection of sub-GeV dark matter}},\ }\href {https://doi.org/10.1088/1475-7516/2020/01/056} {\bibfield  {journal} {\bibinfo  {journal} {\jcap}\ }\textbf {\bibinfo {volume} {2020}},\ \bibinfo {eid} {056} (\bibinfo {year} {2020})},\ \Eprint {https://arxiv.org/abs/1907.11846} {arXiv:1907.11846 [hep-ph]} \BibitemShut {NoStop}%
\bibitem [{\citenamefont {{Page}}\ \emph {et~al.}(2008)\citenamefont {{Page}}, \citenamefont {{Carr}},\ and\ \citenamefont {{MacGibbon}}}]{2008PhRvD..78f4044P}%
  \BibitemOpen
  \bibfield  {author} {\bibinfo {author} {\bibfnamefont {D.~N.}\ \bibnamefont {{Page}}}, \bibinfo {author} {\bibfnamefont {B.~J.}\ \bibnamefont {{Carr}}},\ and\ \bibinfo {author} {\bibfnamefont {J.~H.}\ \bibnamefont {{MacGibbon}}},\ }\bibfield  {title} {\bibinfo {title} {{Bremsstrahlung effects around evaporating black holes}},\ }\href {https://doi.org/10.1103/PhysRevD.78.064044} {\bibfield  {journal} {\bibinfo  {journal} {\prd}\ }\textbf {\bibinfo {volume} {78}},\ \bibinfo {eid} {064044} (\bibinfo {year} {2008})},\ \Eprint {https://arxiv.org/abs/0709.2381} {arXiv:0709.2381 [astro-ph]} \BibitemShut {NoStop}%
\bibitem [{\citenamefont {{Coogan}}\ \emph {et~al.}(2021)\citenamefont {{Coogan}}, \citenamefont {{Morrison}},\ and\ \citenamefont {{Profumo}}}]{coogan2021}%
  \BibitemOpen
  \bibfield  {author} {\bibinfo {author} {\bibfnamefont {A.}~\bibnamefont {{Coogan}}}, \bibinfo {author} {\bibfnamefont {L.}~\bibnamefont {{Morrison}}},\ and\ \bibinfo {author} {\bibfnamefont {S.}~\bibnamefont {{Profumo}}},\ }\bibfield  {title} {\bibinfo {title} {{Direct Detection of Hawking Radiation from Asteroid-Mass Primordial Black Holes}},\ }\href {https://doi.org/10.1103/PhysRevLett.126.171101} {\bibfield  {journal} {\bibinfo  {journal} {\prl}\ }\textbf {\bibinfo {volume} {126}},\ \bibinfo {eid} {171101} (\bibinfo {year} {2021})},\ \Eprint {https://arxiv.org/abs/2010.04797} {arXiv:2010.04797 [astro-ph.CO]} \BibitemShut {NoStop}%
\bibitem [{\citenamefont {{Ballesteros}}\ \emph {et~al.}(2020)\citenamefont {{Ballesteros}}, \citenamefont {{Coronado-Bl{\'a}zquez}},\ and\ \citenamefont {{Gaggero}}}]{2020PhLB..80835624B}%
  \BibitemOpen
  \bibfield  {author} {\bibinfo {author} {\bibfnamefont {G.}~\bibnamefont {{Ballesteros}}}, \bibinfo {author} {\bibfnamefont {J.}~\bibnamefont {{Coronado-Bl{\'a}zquez}}},\ and\ \bibinfo {author} {\bibfnamefont {D.}~\bibnamefont {{Gaggero}}},\ }\bibfield  {title} {\bibinfo {title} {{X-ray and gamma-ray limits on the primordial black hole abundance from Hawking radiation}},\ }\href {https://doi.org/10.1016/j.physletb.2020.135624} {\bibfield  {journal} {\bibinfo  {journal} {Physics Letters B}\ }\textbf {\bibinfo {volume} {808}},\ \bibinfo {eid} {135624} (\bibinfo {year} {2020})},\ \Eprint {https://arxiv.org/abs/1906.10113} {arXiv:1906.10113 [astro-ph.CO]} \BibitemShut {NoStop}%
\bibitem [{\citenamefont {{Silva}}\ \emph {et~al.}(2023)\citenamefont {{Silva}}, \citenamefont {{Vasquez}}, \citenamefont {{Koivu}}, \citenamefont {{Das}},\ and\ \citenamefont {{Hirata}}}]{Paper1}%
  \BibitemOpen
  \bibfield  {author} {\bibinfo {author} {\bibfnamefont {M.}~\bibnamefont {{Silva}}}, \bibinfo {author} {\bibfnamefont {G.}~\bibnamefont {{Vasquez}}}, \bibinfo {author} {\bibfnamefont {E.}~\bibnamefont {{Koivu}}}, \bibinfo {author} {\bibfnamefont {A.}~\bibnamefont {{Das}}},\ and\ \bibinfo {author} {\bibfnamefont {C.~M.}\ \bibnamefont {{Hirata}}},\ }\bibfield  {title} {\bibinfo {title} {{Corrections to Hawking radiation from asteroid mass primordial black holes: Formalism of dissipative interactions in quantum electrodynamics}},\ }\href {https://doi.org/10.1103/PhysRevD.107.045004} {\bibfield  {journal} {\bibinfo  {journal} {\prd}\ }\textbf {\bibinfo {volume} {107}},\ \bibinfo {eid} {045004} (\bibinfo {year} {2023})},\ \Eprint {https://arxiv.org/abs/2210.01914} {arXiv:2210.01914 [gr-qc]} \BibitemShut {NoStop}%
\bibitem [{\citenamefont {{Vasquez}}\ \emph {et~al.}(2024)\citenamefont {{Vasquez}}, \citenamefont {{Kushan}}, \citenamefont {{Silva}}, \citenamefont {{Koivu}}, \citenamefont {{Das}},\ and\ \citenamefont {{Hirata}}}]{2024arXiv240709724V}%
  \BibitemOpen
  \bibfield  {author} {\bibinfo {author} {\bibfnamefont {G.}~\bibnamefont {{Vasquez}}}, \bibinfo {author} {\bibfnamefont {J.}~\bibnamefont {{Kushan}}}, \bibinfo {author} {\bibfnamefont {M.}~\bibnamefont {{Silva}}}, \bibinfo {author} {\bibfnamefont {E.}~\bibnamefont {{Koivu}}}, \bibinfo {author} {\bibfnamefont {A.}~\bibnamefont {{Das}}},\ and\ \bibinfo {author} {\bibfnamefont {C.~M.}\ \bibnamefont {{Hirata}}},\ }\bibfield  {title} {\bibinfo {title} {{Corrections to Hawking radiation from asteroid-mass primordial black holes: description of the stochastic charge effect in quantum electrodynamics}},\ }\href {https://doi.org/10.48550/arXiv.2407.09724} {\bibfield  {journal} {\bibinfo  {journal} {arXiv e-prints}\ ,\ \bibinfo {eid} {arXiv:2407.09724}} (\bibinfo {year} {2024})},\ \Eprint {https://arxiv.org/abs/2407.09724} {arXiv:2407.09724 [astro-ph.CO]} \BibitemShut {NoStop}%
\bibitem [{\citenamefont {{Chandrasekhar}}(1976)}]{1976RSPSA.348...39C}%
  \BibitemOpen
  \bibfield  {author} {\bibinfo {author} {\bibfnamefont {S.}~\bibnamefont {{Chandrasekhar}}},\ }\bibfield  {title} {\bibinfo {title} {{On a Transformation of Teukolsky's Equation and the Electromagnetic Perturbations of the Kerr Black Hole}},\ }\href {https://doi.org/10.1098/rspa.1976.0022} {\bibfield  {journal} {\bibinfo  {journal} {Proceedings of the Royal Society of London Series A}\ }\textbf {\bibinfo {volume} {348}},\ \bibinfo {pages} {39} (\bibinfo {year} {1976})}\BibitemShut {NoStop}%
\bibitem [{\citenamefont {{Arbey}}\ \emph {et~al.}(2021{\natexlab{b}})\citenamefont {{Arbey}}, \citenamefont {{Auffinger}}, \citenamefont {{Geiller}}, \citenamefont {{Livine}},\ and\ \citenamefont {{Sartini}}}]{2021PhRvD.104h4016A}%
  \BibitemOpen
  \bibfield  {author} {\bibinfo {author} {\bibfnamefont {A.}~\bibnamefont {{Arbey}}}, \bibinfo {author} {\bibfnamefont {J.}~\bibnamefont {{Auffinger}}}, \bibinfo {author} {\bibfnamefont {M.}~\bibnamefont {{Geiller}}}, \bibinfo {author} {\bibfnamefont {E.~R.}\ \bibnamefont {{Livine}}},\ and\ \bibinfo {author} {\bibfnamefont {F.}~\bibnamefont {{Sartini}}},\ }\bibfield  {title} {\bibinfo {title} {{Hawking radiation by spherically-symmetric static black holes for all spins. II. Numerical emission rates, analytical limits, and new constraints}},\ }\href {https://doi.org/10.1103/PhysRevD.104.084016} {\bibfield  {journal} {\bibinfo  {journal} {\prd}\ }\textbf {\bibinfo {volume} {104}},\ \bibinfo {eid} {084016} (\bibinfo {year} {2021}{\natexlab{b}})},\ \Eprint {https://arxiv.org/abs/2107.03293} {arXiv:2107.03293 [gr-qc]} \BibitemShut {NoStop}%
\bibitem [{\citenamefont {{Misner}}\ \emph {et~al.}(1973)\citenamefont {{Misner}}, \citenamefont {{Thorne}},\ and\ \citenamefont {{Wheeler}}}]{1973grav.book.....M}%
  \BibitemOpen
  \bibfield  {author} {\bibinfo {author} {\bibfnamefont {C.~W.}\ \bibnamefont {{Misner}}}, \bibinfo {author} {\bibfnamefont {K.~S.}\ \bibnamefont {{Thorne}}},\ and\ \bibinfo {author} {\bibfnamefont {J.~A.}\ \bibnamefont {{Wheeler}}},\ }\href@noop {} {\emph {\bibinfo {title} {{Gravitation}}}}\ (\bibinfo {year} {1973})\BibitemShut {NoStop}%
\bibitem [{\citenamefont {{Teukolsky}}(1973)}]{1973ApJ...185..635T}%
  \BibitemOpen
  \bibfield  {author} {\bibinfo {author} {\bibfnamefont {S.~A.}\ \bibnamefont {{Teukolsky}}},\ }\bibfield  {title} {\bibinfo {title} {{Perturbations of a Rotating Black Hole. I. Fundamental Equations for Gravitational, Electromagnetic, and Neutrino-Field Perturbations}},\ }\href {https://doi.org/10.1086/152444} {\bibfield  {journal} {\bibinfo  {journal} {\apj}\ }\textbf {\bibinfo {volume} {185}},\ \bibinfo {pages} {635} (\bibinfo {year} {1973})}\BibitemShut {NoStop}%
\bibitem [{\citenamefont {Meurer}\ \emph {et~al.}(2017)\citenamefont {Meurer}, \citenamefont {Smith}, \citenamefont {Paprocki}, \citenamefont {\v{C}ert\'{i}k}, \citenamefont {Kirpichev}, \citenamefont {Rocklin}, \citenamefont {Kumar}, \citenamefont {Ivanov}, \citenamefont {Moore}, \citenamefont {Singh}, \citenamefont {Rathnayake}, \citenamefont {Vig}, \citenamefont {Granger}, \citenamefont {Muller}, \citenamefont {Bonazzi}, \citenamefont {Gupta}, \citenamefont {Vats}, \citenamefont {Johansson}, \citenamefont {Pedregosa}, \citenamefont {Curry}, \citenamefont {Terrel}, \citenamefont {Rou\v{c}ka}, \citenamefont {Saboo}, \citenamefont {Fernando}, \citenamefont {Kulal}, \citenamefont {Cimrman},\ and\ \citenamefont {Scopatz}}]{SymPy}%
  \BibitemOpen
  \bibfield  {author} {\bibinfo {author} {\bibfnamefont {A.}~\bibnamefont {Meurer}}, \bibinfo {author} {\bibfnamefont {C.~P.}\ \bibnamefont {Smith}}, \bibinfo {author} {\bibfnamefont {M.}~\bibnamefont {Paprocki}}, \bibinfo {author} {\bibfnamefont {O.}~\bibnamefont {\v{C}ert\'{i}k}}, \bibinfo {author} {\bibfnamefont {S.~B.}\ \bibnamefont {Kirpichev}}, \bibinfo {author} {\bibfnamefont {M.}~\bibnamefont {Rocklin}}, \bibinfo {author} {\bibfnamefont {A.}~\bibnamefont {Kumar}}, \bibinfo {author} {\bibfnamefont {S.}~\bibnamefont {Ivanov}}, \bibinfo {author} {\bibfnamefont {J.~K.}\ \bibnamefont {Moore}}, \bibinfo {author} {\bibfnamefont {S.}~\bibnamefont {Singh}}, \bibinfo {author} {\bibfnamefont {T.}~\bibnamefont {Rathnayake}}, \bibinfo {author} {\bibfnamefont {S.}~\bibnamefont {Vig}}, \bibinfo {author} {\bibfnamefont {B.~E.}\ \bibnamefont {Granger}}, \bibinfo {author} {\bibfnamefont {R.~P.}\ \bibnamefont {Muller}}, \bibinfo {author} {\bibfnamefont {F.}~\bibnamefont {Bonazzi}}, \bibinfo {author} {\bibfnamefont
  {H.}~\bibnamefont {Gupta}}, \bibinfo {author} {\bibfnamefont {S.}~\bibnamefont {Vats}}, \bibinfo {author} {\bibfnamefont {F.}~\bibnamefont {Johansson}}, \bibinfo {author} {\bibfnamefont {F.}~\bibnamefont {Pedregosa}}, \bibinfo {author} {\bibfnamefont {M.~J.}\ \bibnamefont {Curry}}, \bibinfo {author} {\bibfnamefont {A.~R.}\ \bibnamefont {Terrel}}, \bibinfo {author} {\bibfnamefont {v.}~\bibnamefont {Rou\v{c}ka}}, \bibinfo {author} {\bibfnamefont {A.}~\bibnamefont {Saboo}}, \bibinfo {author} {\bibfnamefont {I.}~\bibnamefont {Fernando}}, \bibinfo {author} {\bibfnamefont {S.}~\bibnamefont {Kulal}}, \bibinfo {author} {\bibfnamefont {R.}~\bibnamefont {Cimrman}},\ and\ \bibinfo {author} {\bibfnamefont {A.}~\bibnamefont {Scopatz}},\ }\bibfield  {title} {\bibinfo {title} {Sympy: symbolic computing in python},\ }\href {https://doi.org/10.7717/peerj-cs.103} {\bibfield  {journal} {\bibinfo  {journal} {PeerJ Computer Science}\ }\textbf {\bibinfo {volume} {3}},\ \bibinfo {pages} {e103} (\bibinfo {year}
  {2017})}\BibitemShut {NoStop}%
\bibitem [{\citenamefont {{Saha}}\ and\ \citenamefont {{Laha}}(2022)}]{2022PhRvD.105j3026S}%
  \BibitemOpen
  \bibfield  {author} {\bibinfo {author} {\bibfnamefont {A.~K.}\ \bibnamefont {{Saha}}}\ and\ \bibinfo {author} {\bibfnamefont {R.}~\bibnamefont {{Laha}}},\ }\bibfield  {title} {\bibinfo {title} {{Sensitivities on nonspinning and spinning primordial black hole dark matter with global 21-cm troughs}},\ }\href {https://doi.org/10.1103/PhysRevD.105.103026} {\bibfield  {journal} {\bibinfo  {journal} {\prd}\ }\textbf {\bibinfo {volume} {105}},\ \bibinfo {eid} {103026} (\bibinfo {year} {2022})},\ \Eprint {https://arxiv.org/abs/2112.10794} {arXiv:2112.10794 [astro-ph.CO]} \BibitemShut {NoStop}%
\bibitem [{\citenamefont {{Caputo}}\ \emph {et~al.}(2023)\citenamefont {{Caputo}}, \citenamefont {{Negro}}, \citenamefont {{Regis}},\ and\ \citenamefont {{Taoso}}}]{2023JCAP...02..006C}%
  \BibitemOpen
  \bibfield  {author} {\bibinfo {author} {\bibfnamefont {A.}~\bibnamefont {{Caputo}}}, \bibinfo {author} {\bibfnamefont {M.}~\bibnamefont {{Negro}}}, \bibinfo {author} {\bibfnamefont {M.}~\bibnamefont {{Regis}}},\ and\ \bibinfo {author} {\bibfnamefont {M.}~\bibnamefont {{Taoso}}},\ }\bibfield  {title} {\bibinfo {title} {{Dark matter prospects with COSI: ALPs, PBHs and sub-GeV dark matter}},\ }\href {https://doi.org/10.1088/1475-7516/2023/02/006} {\bibfield  {journal} {\bibinfo  {journal} {\jcap}\ }\textbf {\bibinfo {volume} {2023}},\ \bibinfo {eid} {006} (\bibinfo {year} {2023})},\ \Eprint {https://arxiv.org/abs/2210.09310} {arXiv:2210.09310 [hep-ph]} \BibitemShut {NoStop}%
\bibitem [{\citenamefont {{Ohio~Supercomputer~Center}}(2018)}]{Pitzer2018}%
  \BibitemOpen
  \bibfield  {author} {\bibinfo {author} {\bibnamefont {{Ohio~Supercomputer~Center}}},\ }\href {http://osc.edu/ark:/19495/hpc56htp} {\bibinfo {title} {Pitzer supercomputer}} (\bibinfo {year} {2018})\BibitemShut {NoStop}%
\bibitem [{\citenamefont {{Bloch}}(1936)}]{1936PhRv...50..272B}%
  \BibitemOpen
  \bibfield  {author} {\bibinfo {author} {\bibfnamefont {F.}~\bibnamefont {{Bloch}}},\ }\bibfield  {title} {\bibinfo {title} {{On the Continuous {\ensuremath{\gamma}}-Radiation Accompanying the {\ensuremath{\beta}}-Decay}},\ }\href {https://doi.org/10.1103/PhysRev.50.272} {\bibfield  {journal} {\bibinfo  {journal} {Physical Review}\ }\textbf {\bibinfo {volume} {50}},\ \bibinfo {pages} {272} (\bibinfo {year} {1936})}\BibitemShut {NoStop}%
\bibitem [{\citenamefont {{Cardoso}}\ \emph {et~al.}(2003)\citenamefont {{Cardoso}}, \citenamefont {{Lemos}},\ and\ \citenamefont {{Yoshida}}}]{2003PhRvD..68h4011C}%
  \BibitemOpen
  \bibfield  {author} {\bibinfo {author} {\bibfnamefont {V.}~\bibnamefont {{Cardoso}}}, \bibinfo {author} {\bibfnamefont {J.~P.}\ \bibnamefont {{Lemos}}},\ and\ \bibinfo {author} {\bibfnamefont {S.}~\bibnamefont {{Yoshida}}},\ }\bibfield  {title} {\bibinfo {title} {{Electromagnetic radiation from collisions at almost the speed of light: An extremely relativistic charged particle falling into a Schwarzschild black hole}},\ }\href {https://doi.org/10.1103/PhysRevD.68.084011} {\bibfield  {journal} {\bibinfo  {journal} {\prd}\ }\textbf {\bibinfo {volume} {68}},\ \bibinfo {eid} {084011} (\bibinfo {year} {2003})},\ \Eprint {https://arxiv.org/abs/gr-qc/0307104} {arXiv:gr-qc/0307104 [gr-qc]} \BibitemShut {NoStop}%
\bibitem [{\citenamefont {{Jackson}}(1998)}]{1998clel.book.....J}%
  \BibitemOpen
  \bibfield  {author} {\bibinfo {author} {\bibfnamefont {J.~D.}\ \bibnamefont {{Jackson}}},\ }\href@noop {} {\emph {\bibinfo {title} {{Classical Electrodynamics, 3rd Edition}}}}\ (\bibinfo {year} {1998})\BibitemShut {NoStop}%
\bibitem [{\citenamefont {{Abramowitz}}\ and\ \citenamefont {{Stegun}}(1972)}]{1972hmfw.book.....A}%
  \BibitemOpen
  \bibfield  {author} {\bibinfo {author} {\bibfnamefont {M.}~\bibnamefont {{Abramowitz}}}\ and\ \bibinfo {author} {\bibfnamefont {I.~A.}\ \bibnamefont {{Stegun}}},\ }\href@noop {} {\emph {\bibinfo {title} {{Handbook of Mathematical Functions}}}}\ (\bibinfo {year} {1972})\BibitemShut {NoStop}%
\bibitem [{\citenamefont {{Sudakov}}(1956)}]{Sudakov56}%
  \BibitemOpen
  \bibfield  {author} {\bibinfo {author} {\bibfnamefont {V.~V.}\ \bibnamefont {{Sudakov}}},\ }\href@noop {} {\bibfield  {journal} {\bibinfo  {journal} {{Zh. Eksp. Teor. Fiz.}}\ }\textbf {\bibinfo {volume} {30}},\ \bibinfo {pages} {87} (\bibinfo {year} {1956})}\BibitemShut {NoStop}%
\bibitem [{\citenamefont {{Abrikosov}}(1956)}]{Abrikosov56}%
  \BibitemOpen
  \bibfield  {author} {\bibinfo {author} {\bibfnamefont {A.~A.}\ \bibnamefont {{Abrikosov}}},\ }\href@noop {} {\bibfield  {journal} {\bibinfo  {journal} {{Zh. Eksp. Teor. Fiz.}}\ }\textbf {\bibinfo {volume} {30}},\ \bibinfo {pages} {96} (\bibinfo {year} {1956})}\BibitemShut {NoStop}%
\end{thebibliography}%

\end{document}